\documentclass[pdflatex,sn-mathphys-ay]{sn-jnl}

\usepackage{graphicx}%
\graphicspath{ {figures/} }
\usepackage{multirow}%
\usepackage{amsmath,amssymb,amsfonts}%
\usepackage{amsthm}%
\usepackage{mathrsfs}%
\usepackage[title]{appendix}%
\usepackage{xcolor}%
\usepackage{textcomp}%
\usepackage{manyfoot}%
\usepackage{booktabs}%
\usepackage{algorithm}%
\usepackage{algorithmicx}%
\usepackage{algpseudocode}%
\usepackage{listings}%
\usepackage{array}

\begin{document}

\title[History Matters: Damage-Mediated Amplification of Brain Deformation and Injury Risk under Repeated Head Impacts
]{History Matters: Damage-Mediated Amplification of Brain Deformation and Injury Risk under Repeated Head Impacts
}


\author[1]{\fnm{Carson} \sur{Cooper}}\email{Carson.Cooper@lsu.edu}
\equalcont{Current email: ccoope@umich.edu}

\author[2]{\fnm{Anu} \sur{Tripathi}}\email{tripathia@rmu.edu}

\author[1]{\fnm{Genevieve} \sur{Palardy}}\email{gpalardy@lsu.edu}

\author*[3]{\fnm{Kshitiz} \sur{Upadhyay}}\email{kshitizu@umn.edu}


\affil[1]{\orgdiv{Department of Mechanical and Industrial Engineering}, \orgname{Louisiana State University}, \orgaddress{\city{Baton Rouge}, \state{LA}, \country{USA}}}

\affil[2]{\orgdiv{Department of Engineering}, \orgname{Robert Morris University}, \orgaddress{\city{Moon Township}, \state{PA}, \country{USA}}}

\affil*[3]{\orgdiv{Department of Aerospace Engineering and Mechanics}, \orgname{University of Minnesota}, \orgaddress{\city{Minneapolis}, \state{MN}, \country{USA}}}

\abstract{
Computational head models are traditionally applied to isolated impact events, leaving the effects of repeated head loading largely unexplored. An Ogden--Roxburgh Mullins damage formulation was implemented in a high-fidelity finite element--based head model to represent the loading-history-dependent stress softening observed during cyclic deformation of brain tissue. Repeated-loading histories derived from mixed martial arts head-impact data were applied and compared with damage-free hyperelastic (HE) and linear visco-hyperelastic (LVHE) model variants. Under five identical single-axis loading cycles, Mullins-type softening progressively increased strain and strain rate metrics relative to the HE model. Mullins-based injury probabilities progressively exceeded strain-based HE predictions and diverged from unchanged kinematics-based predictions, indicating that neglecting prior softening may underestimate tissue-level injury risk. In a randomized twenty-cycle multiaxial sequence, loading cycles of similar kinematic intensity produced different deformation and injury-risk estimates depending on the accumulated softening history. The HE and LVHE models predicted higher injury probabilities during the initial cycles, whereas the Mullins-based model generally produced the largest estimates during later cycles and the highest probability of at least one injury over the full sequence. Regional deformation amplification depended jointly on loading direction and prior softening, with no general direction-independent hierarchy among brain substructures. Gyral elements exhibited higher cumulative maximum principal strain than sulcal elements, whereas sulcal elements exhibited greater amplification relative to their initial responses. These findings demonstrate that short-term damage-mediated softening can substantially amplify tissue deformation and injury-risk estimates beyond those predicted by conventional damage-free head models under the same prescribed loading histories. Further experimental characterization of cyclic brain-tissue softening is needed to support improved models of repeated head loading and traumatic brain injury.}


\maketitle

\section{Introduction}\label{chap:intro}
Traumatic brain injury (TBI) is a major public health concern in the United States, accounting for 214,110 hospitalizations in 2020 \citep{CDC_TBI_Data} and 67,842 deaths in 2024 \citep{CDC_TBI_Data_death}. 
These injuries are prevalent in both civilian and military populations, often caused by “closed-head” impacts (no penetration of the skull) resulting from motor-vehicle accidents, falls, contact sports, abuse, and blasts \citep{Meaney2014}.
Up to 75\% of reported TBI incidents result in mild TBI \citep{CDC_2003}, defined by less than thirty minutes of loss of consciousness, and less than one day of alteration of the victim’s mental state \citep{Brasure2012PostacuteTBI}.
The ability to predict severity of TBI for a given kinematic loading has wide-reaching benefits for protection and prevention of head injuries.
Computational head models are valuable tools to quickly predict injury outcomes for a variety of head loadings, and have seen extensive use in recent years \citep{Taylor2006, Takhounts2008, Cloots2008, Takhounts2013, Wright2013, Mao2013, Zhang2013, Ji2015, Ghajari2017, Ganpule2018, Fagan2020, Duckworth2021, Alshareef2021, Upadhyay2022, Tripathi2026}.
These models offer detailed predictions of the full-field mechanical response (i.e., stress, strain, and strain rate) of the brain, and offer potential to establish a crucial link between mechanical stimuli and resulting biological damage and injury. 

The mechanics of brain injury have been studied across tissue and organ scales, with the central goal of relating externally imposed impact forces and head kinematics to local mechanical fields in the brain and, ultimately, to neurological injury outcomes.  Two commonly studied outcomes are diffuse axonal injury (DAI) and mild TBI (mTBI), including concussion. DAI is characterized by widespread disruption of axonal fiber function and is commonly associated with large tissue deformation, particularly shear deformation, during inertial loading \citep{Cloots2011}. Although concussion, the injury of interest in this study, typically does not involve visible macroscopic tissue damage, mild axonal deformation and disruption are believed to play important roles in the underlying pathophysiology \citep{Giza2014}. Tissue-level studies have further established connections between mechanical deformation, axonal injury, and functional impairment in the context of both DAI and concussion \citep{Margulies1992, Smith2000, Bain2000, Elkin2007, Meaney2011}. Computational head models leverage these deformation--injury relationships to estimate TBI risk by predicting full-field mechanical measures, such as strain and strain rate, under prescribed head-loading conditions \citep{Zhou1995, Kleiven2007, Takhounts2008, Takhounts2013, Gabler2018a}.

However, computational studies often consider individual head impacts as isolated events. This simplification is consequential because many TBI-relevant exposure scenarios involve repeated head impacts, especially in sports contexts such as American football \citep{Greenwald2008}, soccer, and mixed martial arts \citep{OKeeffe2020}. The limitation is evident even in studies motivated by the long-term consequences of head trauma. For example, \citet{Ghajari2017} used a computational head model to examine brain deformation in impact scenarios relevant to chronic traumatic encephalopathy, including an American football impact, a fall with occipital impact, and a motorcycle accident. Although these cases provided important insight into how different impact scenarios produce distinct patterns of brain deformation, each loading event was treated as an independent simulation rather than as part of a successive loading history with evolving tissue properties. Similarly, parametric studies examining the effects of loading conditions \citep{Carlsen2021, Upadhyay2024, Takhounts2013, Gabler2018a}, model geometry \citep{Cloots2008}, fluid modeling techniques \citep{Duckworth2021}, and meshing \citep{Zhou2025} have largely focused on isolated impact events. Even studies focused on sports-related head loading \citep{OKeeffe2020, Zuidema2023}, where repeated impact exposure is most likely, typically analyze impacts independently. As a result, the mechanical consequences of successive head loadings remain an open question in computational injury biomechanics. This omission may partly explain the imperfect correspondence between model-predicted injury metrics and real-world injury outcomes, since nominally similar kinematic inputs or peak tissue deformations may produce injury in some cases but not others depending on prior exposure history \citep{Carlsen2021, OKeeffe2020}.

Studies have indicated that repeated head impacts can uniquely contribute to the onset and escalation of brain injury.
\cite{Longhi2005} subjected mice to repeated concussive injuries at 3-, 5-, and 7-day intervals and observed significant cognitive impairment in mice injured at 3-day intervals, but not in those injured at longer time intervals. 
Additionally, repetitive mild TBI spaced over years or decades has been linked to neurodegenerative disorders such as Chronic Traumatic Encephalopathy (CTE) \citep{McKee2013}.
The gradual onset of CTE involves a complex interplay between mechanical response and biochemical processes that remains an active area of research \citep{Nol2019, Ghajari2017, McKee2013}.
Even subconcussive impacts may accumulate over time and contribute to the unexpected onset of concussion, despite producing no immediate clinical symptoms \citep{Ji2022}. 
Regarding repeated loading scenarios, \cite{Meaney2011} asserted that the tissue response in repeated injuries is not merely a superposition of the responses of the individual impacts, supporting medical observations of compounding risk in repeated head loading \citep{Longhi2005, Ji2022, Giza2014}. This issue is further complicated by the prevalence of mild TBI and the tendency for some events to go unreported or untreated, both of which may increase the likelihood of repeated exposure before full recovery.

A possible mechanical contributor to this compounding response is loading-history-dependent alteration of brain-tissue properties. As an extremely compliant tissue, brain can undergo large deformations that may produce temporary or permanent changes in mechanical response \citep{Budday2020}. Such effects may be represented using a damage mechanics framework, in which an internal variable evolves with the material loading history and modifies the constitutive response. Degradation of brain mechanical properties has been observed in cyclic and failure-inducing loading of brain and similar tissues \citep{Franceschini2006, Bain2000, Budday2020, DeRooij2016, Begonia2021}, motivating several damage-based constitutive descriptions. \citet{Nol2019} modeled continuum damage in brain tissue using an isotropic scalar damage variable, an exponential-type damage evolution law, and a nonlocal strain-energy-density formulation. \citet{Sumelka2017} proposed a fractional-calculus-based damage model capable of capturing history dependence, primary load-path softening, permanent failure, and rate-dependent damage onset, with later extensions to anisotropic damage onset \citep{Voyiadjis2019}. In the context of short-term cyclic loading, \citet{Franceschini2006} observed cycle-to-cycle softening in brain tissue under quasi-static uniaxial tension and compression. This behavior resembles the Mullins effect, a stress-softening phenomenon first characterized in filled rubbers by \citet{Mullins1947}. \citet{Ogden1998} proposed a pseudoelastic damage model for Mullins-type softening, which \citet{Franceschini2006} calibrated to cyclic white-matter data and found to capture the observed softening response. Although the physical mechanisms underlying Mullins-type softening in brain tissue remain unclear, proposed explanations include microstructural damage \citep{Gerber346700, Li2016} and poroelastic fluid redistribution \citep{Budday2020}. Because tissue deformation directly influences brain injury, short-term Mullins-type softening may alter local deformation and injury-risk predictions during successive head loadings over time scales of seconds to minutes, before substantial recovery of the softened response occurs \citep{Budday2020}.

Despite these developments, the organ-scale consequences of damage evolution in repeated head loading remain largely unexplored. Existing damage-based brain models have either used idealized geometries, simplified loading cases, non-human tissue properties, or constitutive descriptions that are difficult to implement directly in full three-dimensional head-impact simulations. For example, \citet{Nol2019} simulated a small two-dimensional section to examine damage evolution near the grey--white matter interface and the influence of sulcal geometry, using a damage law calibrated to the \citet{Bain2000} data on guinea pig optic nerves. \citet{Gerber346700} implemented the Ogden--Roxburgh model \citep{Ogden1998} in a cyclic loading simulation of a small representative brain volume. Although these studies provided important insight into local damage evolution, their simplified geometries and idealized loading conditions limit their ability to reveal how damage-mediated softening affects deformation throughout the full human brain under realistic head kinematics. This limitation is especially important because the brain is a geometrically complex, heterogeneous organ composed of distinct tissue structures, curved cortical folds, deep grey and white matter regions, ventricles, membranes, and brainstem structures, all of which may experience different deformation histories depending on loading direction and magnitude. As a result, simplified local models cannot fully capture the spatial redistribution of deformation, regional accumulation of damage, or potential differences between gyral and sulcal responses during successive impacts. The short-term softening effects observed by \citet{Franceschini2006} therefore remain unexplored at the full-head scale. Because the Ogden--Roxburgh model directly captures Mullins-type stress softening, has been calibrated to human brain tissue cyclic loading data, and is readily implemented in finite element simulations, it provides a suitable framework for investigating how short-term repeated head loading may alter organ-scale deformation and injury-risk predictions.

This study examines how short-term Mullins-type softening of brain tissue may alter organ-scale deformation and injury-risk predictions during successive head loadings. To this end, a high-fidelity finite element (FE)–based human head model is developed and equipped with the Ogden--Roxburgh Mullins damage formulation, parameterized using cyclic human brain tissue data, to represent history-dependent stress softening in the brain. Sequences of head-loading histories drawn from real-world sports-related kinematic data are applied to investigate deformation amplification under repeated loading. Simulation results are evaluated using tissue-deformation-based injury metrics and established injury risk functions. In controlled single-axis simulations, damage-based predictions are compared against a corresponding damage-free hyperelastic model to isolate the mechanical effect of Mullins softening. In multiaxial simulations, predictions are compared against both damage-free hyperelastic and visco-hyperelastic models to assess the consequences of damage evolution relative to established intact-tissue modeling approaches. In addition to global injury measures, regional analyses are performed to assess spatially heterogeneous deformation and damage evolution across brain structures, including differences between gyral and sulcal locations. Together, these simulations provide a first step toward understanding how short-term damage-mediated softening may influence brain deformation and injury-risk prediction in repetitive head loading scenarios.

The paper is organized as follows. 
Section~\ref{chap:methods} presents the Mullins damage constitutive model, FE–based head model development, loading conditions, and analysis framework.
Section~\ref{chap:results} presents results for single-axis and multiaxial repeated loading simulations, including global injury metrics, regional deformation trends, and gyral--sulcal comparisons.
Section~\ref{chap:discussion} discusses the implications of damage-mediated deformation amplification for repeated head loading and injury-risk prediction.
Section~\ref{chap:conclusion} summarizes the main findings, limitations, and future directions.

\section{Methods}\label{chap:methods}
\begin{figure}[hbt!]
    \centering    
    \includegraphics[width=4.997in]{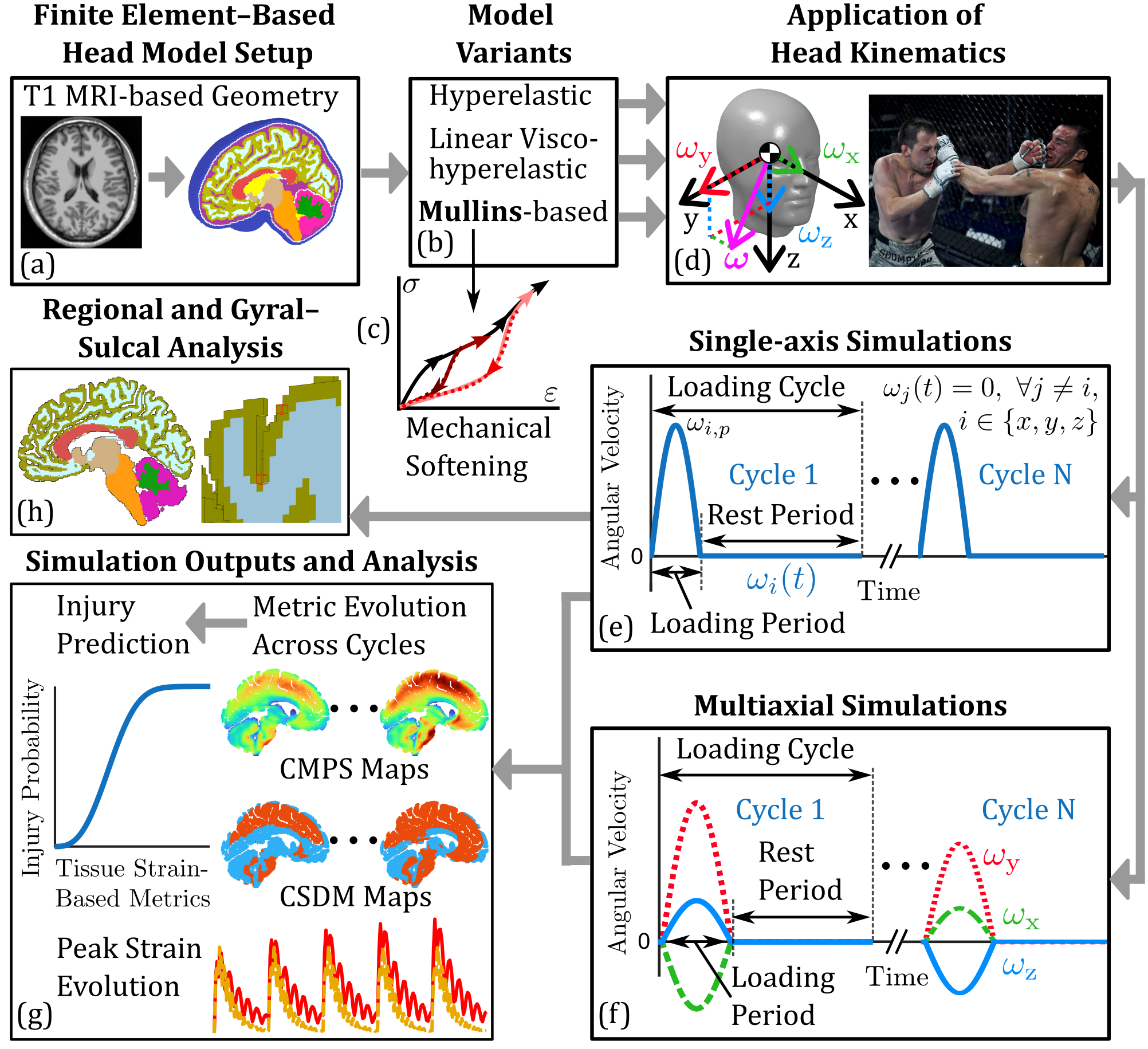}
    \caption{Overview of the computational framework for modeling repeated head loading and evaluating damage-mediated changes in brain deformation and injury risk. (a) Magnetic resonance imaging (MRI) scans are used to generate the finite element head geometry. (b) Brain-bulk constitutive variants: hyperelastic (HE), linear viscohyperelastic (LVHE), and Mullins-based models. (c) Schematic stress--strain response illustrating Mullins-type softening. (d) Head coordinate system and rotational velocity components used to prescribe mixed martial arts (MMA)-derived head kinematics; the MMA image is adapted from a U.S. Air Force photograph by Airman 1st Class Riley Johnson (public domain, Wikimedia Commons). Representative loading histories are shown for (e) identical single-axis loading cycles and (f) randomized multiaxial loading cycles. (g) Simulation outputs include tissue strain and strain rate metrics, together with injury-risk predictions. (h) Regional and gyral--sulcal analyses use the retained anatomical segmentation and paired cortical sampling locations}
    \label{fig:overview}
\end{figure}

Figure \ref{fig:overview} provides an overview of the computational workflow employed in this study. Beginning with subject-specific magnetic resonance imaging (MRI) scans, a finite element-based head model is developed and brain tissue constitutive model variants are subjected to repeated head loading conditions. Simulation outputs, including brain deformation, injury metrics, and regional responses, are subsequently analyzed to examine the effects of short-term mechanical softening.

\subsection{Constitutive Modeling of Brain Tissue Damage Applicable to Repetitive Mechanical Loading}

\label{sec:constitutivemodeling}
To capture Mullins-type softening as observed in brain tissue, the pseudoelastic damage model developed by \citet{Ogden1998} (i.e., the Ogden--Roxburgh model) is used. This model is a history-dependent modification of classical hyperelasticity. The strain energy density $W$ in the material is decomposed additively into isochoric (i.e., distortional) and volumetric components \citep{Ogden1998, Holzapfel2000},
\begin{equation}
    W(\eta,\mathbf{F})=\underbrace{\eta W_{\mathrm{dev}}(\mathbf{F}^{*})
    +\phi(\eta)}_{W_{\mathrm{dev}}(\eta,\mathbf{F}^{*})}
    +W_{\mathrm{vol}}(J),
\end{equation}
where \(\mathbf{F}\) is the deformation gradient, \(J=\det(\mathbf{F})\), and
\begin{equation}
\mathbf{F}^{*}=J^{-1/3}\mathbf{F}
\end{equation}
is the isochoric component of the deformation gradient. Here, $W_{\mathrm{dev}}(\eta,\mathbf{F}^{*})$ is the pseudo strain energy density associated with the distortional response and depends on the scalar Mullins damage variable $\eta$, the strain energy density associated with the intact material response $W_{\mathrm{dev}}(\mathbf{F}^{*})$, and the damage function $\phi(\eta)$ \citep{Ogden1998}. The damage variable $\eta$ allows load--unload paths to diverge during successive loading as stress softening evolves (i.e., as damage accumulates). $W_{\mathrm{vol}}(J)$ is the volumetric energy density.
Damage is assumed to affect only the isochoric/deviatoric response, leaving the volumetric response unchanged.

The damage variable $\eta$ is defined as  
\begin{equation}
\label{eq:eta}
\eta=1-\frac{1}{r}\operatorname{erf}\left[ \frac{1}{m}\left(W_{\mathrm{dev,max}}(t)-W_{\mathrm{dev}}(t)\right)\right] \ ,
\end{equation}
where $W_{\mathrm{dev,max}}(t)=\underset{0 \le \tau \le t}{\max}W_{\mathrm{dev}}(\tau)$ is the maximum strain energy density attained by the material during its loading history (i.e., from time $\tau=0$ to $t$, $t$ being the current time), $W_{\mathrm{dev}}(t)$ is the strain energy density corresponding to the current deformation state, $\operatorname{erf}(\cdot)$ is the Gaussian error function, and $r$ ($r>1$) and $m$ are strictly positive material parameters \citep{Ogden1998, abaqus2004, Franceschini2006}. Inclusion of the maximum previously attained strain energy imposes history dependence, such that damage is retained even after the material is unloaded. In the initial undeformed state, $W_\mathrm{dev,max}=W_\mathrm{dev}=0$, and therefore, $\eta=1$. During unloading and reloading below a previously attained maximum strain energy, \(\eta<1\), reducing the isochoric stress response. When the material is loaded beyond its previous maximum strain energy, \(W_{\mathrm{dev,max}}\) is updated and the response follows the primary hyperelastic loading path.

The parameter $r$ is dimensionless and controls the maximum amount of Mullins-type softening permitted by the model. Because $r^{-1}$ scales the bounded error function $\operatorname{erf}(\cdot)\in(-1,1)$, the damage variable has a lower bound, 
\begin{equation}
\eta_{\mathrm{min}}= 1-\frac{1}{r},
\end{equation}
which is approached as $(W_{\mathrm{dev,max}}-W_{\mathrm{dev}})\to\infty$. Thus, for large values of $r$, $\eta_{\mathrm{min}}$ remains close to unity and the material response remains nearly intact even after large-amplitude cyclic deformation. In contrast, for \(r\approx 1\), \(\eta_{\min}\) approaches zero, allowing substantial stress softening to accumulate. The parameter $m$, with units of strain energy density, controls the energy scale over which softening evolves. Larger values of $m$ require larger changes in strain energy before appreciable softening occurs, whereas smaller values of $m$ allow \(\eta\) to decrease at lower deformation levels.

An isotropic material response is assumed following \cite{Budday2020}, allowing the strain energy density function to be recast in terms of the principal invariants of $\mathbf{B}$, defined as \citep{Ericksen1954}
\begin{equation}
\begin{aligned}
    I_{1}&=\operatorname{tr}(\mathbf{B}) \ , \\
    I_{2}&= \frac{1}{2}\left(I_{1}^2-\operatorname{tr}(\mathbf{B}^{2})\right) \ , \\
    I_{3}&= \det(\mathbf{B}) \ ,
\end{aligned}
\end{equation}
where $\mathbf{B}$ is the left Cauchy-Green deformation tensor, with $\mathbf{B}=\mathbf{F}\cdot\mathbf{F}^{\mathrm{T}}$. Corrections to the invariants are made to isolate distortional deformation, with
\begin{equation}
\begin{aligned}
    I_{1}^{*}&=J^{-2/3}I_{1}=\operatorname{tr}(\mathbf{B}^{*}) \ , \\
    I_{2}^{*}&= J^{-4/3}I_{2}=\frac{1}{2}\left(I_{1}^{*2}-\operatorname{tr}(\mathbf{B}^{*2})\right) \ ,
\end{aligned}
\end{equation}
such that \(W_{\mathrm{dev}}=W_{\mathrm{dev}}(I_{1}^{*},I_{2}^{*})\) and \(W_{\mathrm{vol}}=W_{\mathrm{vol}}(J)\). Here, $\mathbf{B}^*$ is the corrected left Cauchy-Green deformation tensor based on the deviatoric part of the deformation gradient; \(\mathbf{B}^*=\mathbf{F}^{*}\cdot\mathbf{\left(F^{*}\right)}^{\mathrm{T}}\) \citep{Holzapfel2000}.

A neo-Hookean functional form for the strain energy density is used in this work, a well-known constitutive model to capture the behavior of soft nonlinear solids, including brain tissue \citep{Holzapfel2000, DeRooij2016, Madhukar2019}. The isochoric and volumetric strain energy densities are given by
\begin{equation}
\label{eq:NHdev}
    W_{\mathrm{dev,neo-Hookean}}=C_{10}(I_{1}^{*}-3) \ ,
\end{equation}
\begin{equation}
\label{eq:NHvol}
    W_{\mathrm{vol,neo-Hookean}}=\frac{1}{D_{1}}(J-1)^{2} \ .
\end{equation}
The constants $C_{10}=\frac{\mu}{2}$ and $D_{1}=\frac{2}{K}$ are material parameters related to the shear modulus ($\mu$) and bulk modulus ($K$), respectively.

The Cauchy stress in the material is given by
\begin{equation}
\begin{aligned}
\label{eq:cauchystress}
\boldsymbol{\sigma}&=\underbrace{\eta\operatorname{dev}\left(\frac{2}{J}\frac{\partial W_{\mathrm{dev}}(\mathbf{B}^*)}{\partial\mathbf{B}^{*}}\cdot\mathbf{B}^{*}\right)}_{\boldsymbol{\sigma}_{\mathrm{dev}}} &&+\underbrace{\frac{2}{J}\frac{\partial W_{\mathrm{vol}}(\mathbf{B})}{\partial\mathbf{B}}\cdot\mathbf{B}}_{\boldsymbol{\sigma}_{\mathrm{vol}}} \\
&= \eta\left(\frac{2}{J}\operatorname{dev}(C_{10}\mathbf{B}^{*})\right)&&+\frac{2}{D_{1}}(J-1)\mathbf{I} \ ,
\end{aligned}
\end{equation}
where $\mathbf{I}$ is the second-order identity tensor and the deviatoric operator is defined as \(\mathrm{dev}(\cdot)=(\cdot)-\frac{1}{3}\operatorname{tr}(\cdot)\mathbf{I}\), with \(\operatorname{tr}(\cdot)\) as the trace of the tensor. The first term in Eq.~(\ref{eq:cauchystress}) is the stress due to distortional deformation, while the second term is due to volumetric deformation. In the limiting case of severe Mullins-type softening, $\eta \to 0$, the isochoric stress contribution vanishes and the material offers negligible resistance to distortional deformation. 

\cite{Franceschini2006} observed Mullins effect in human brain tissue under cyclic loading. Successive cycles followed progressively softened loading paths, and the response was well captured by the Ogden--Roxburgh model. The Mullins damage parameters \(r\) and \(m\), calibrated from these cyclic human brain tissue experiments, are given in Table~\ref{tab:mullins} and are used directly to describe damage-mediated softening of the brain tissue in the present FE-based head injury model, described in Sec.~\ref{sec:FEmodeldevelopment}.
\begin{table}[hbt!]
\label{tab:mullins}
\caption{Mullins damage constitutive model parameters from \cite{Franceschini2006}}
\begin{tabular}{lcc}
\toprule
   Parameter & $r$ (-) & $m$ (kPa) \\
 \midrule
    Value & 1.2 & 0.05 \\
\bottomrule
\end{tabular}
\end{table}

Together with the intact hyperelastic parameters \(C_{10}\) and \(D_{1}\), or equivalently \(\mu\) and \(K\), these four parameters define the Mullins damage--based constitutive response of the brain tissue. The shear and bulk moduli are taken as \(\mu=6.44~\mathrm{kPa}\) and \(K=1.46~\mathrm{GPa}\), respectively \citep{Alshareef2021}. All other properties relevant to non-brain components of the computational head model are given in Sec.~\ref{subsec:matmodelsforFE}.

\subsection{Finite Element--Based Head Injury Model Development}
\label{sec:FEmodeldevelopment}
A finite element (FE)--based computational head model is developed to study the effects of repeated head impacts on brain tissue deformation and associated injury risk. Briefly, patient-specific medical scans are processed to generate an anatomically accurate head model, and appropriate material models are assigned to the segmented structures. The computational model is validated against experimental data from sub-injurious head motion. Kinematic loading histories representative of real-world, potentially injurious head impacts are then applied to investigate the tissue-level biomechanical response and injury outcomes, including regional variability associated with the brain's complex geometry.

\subsubsection{Image Processing and Segmentation} 
The Neuroimaging Tools and Resources Collaboratory (NITRC) Brain Biomechanics Imaging Resources (BBIR) database was utilized for FE head model generation in this work \citep{Bayly2021}. This database provides subject-specific, high-resolution brain MRI, magnetic resonance elastography (MRE), and tagged MRI data for computational model development and validation. Specifically, the present study processed T1-weighted MRI scans (resolution: 1 x 1 x 1 mm$^3$) of a 31-year-old male subject using the open-source FreeSurfer software \citep{Fischl2012}.
Briefly, features like the skull, skin, and meninges are removed with a watershed skull-stripping algorithm \citep{Sgonne2004}, and the brain is then segmented into distinct tissue structures (i.e., grey matter, corpus callosum, ventricles, white matter, etc.) according to the Destrieux atlas \citep{Fischl2004}. 

\begin{figure}[hbt!]
    \centering    
    \includegraphics[width=4.997in]{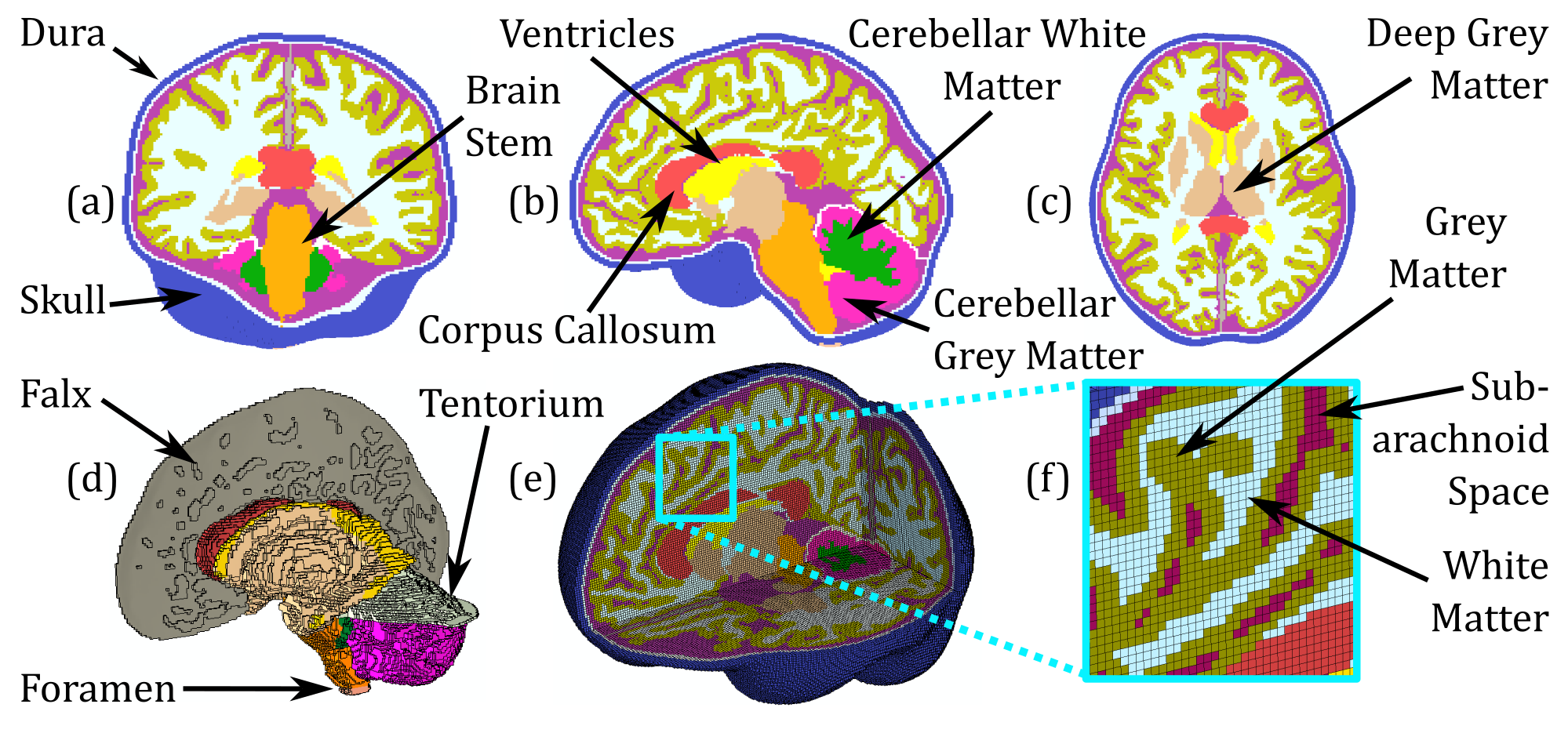}
    \caption{Finite element (FE)--based head model of the present study: (a) coronal view, (b) sagittal view, (c) axial view, (d) inner structures, (e) isometric cut-view, and (f) magnified view of mesh. The parenchymal brain substructures, including grey matter, white matter, deep grey matter, corpus callosum, cerebellar grey matter, cerebellar white matter, and brain stem, are assigned a common ``brain bulk" material response in the FE simulations}
    \label{fig:femodel}
\end{figure}

Additional structures are then added to the initially segmented volume. The corpus callosum, a structure that physically and functionally links the right and left hemispheres of the brain, is semi-automatically segmented from the original T1-weighted MRI volume using the ITK-SNAP software \citep{py06nimg} via a bubble-growth classification method~\citep{Shiino2017}. The falx and tentorium are generated in the 3D Slicer software~\citep{Kikinis2014}. The falx is a thin membrane that physically separates the right and left hemispheres of the brain. This is created by dilating desired regions of the left and right hemispheres and considering the overlap between these two growths to be the falx, following the method described by \cite{Glaister2017a, Glaister2017b}. 
The tentorium provides a barrier between the cerebellum and the posterior region of the brain, and is manually constructed in 3D Slicer by a similar growth-overlap method. 
The subarachnoid space and intracranial cerebrospinal fluid are combined into a single label, following \cite{Alshareef2021, Upadhyay2022, Ghajari2017}.
Lastly, the dura and skull are created by dilating a heavily smoothed cortical surface (i.e., a brainmask), creating a simplified skull that encapsulates the brain, as is commonly done in computational head modeling literature \citep{Taylor2006, Takhounts2008, Cloots2008, Takhounts2013, Wright2013, Ghajari2017, Ganpule2018, Fagan2020, Duckworth2021, Alshareef2021, Tripathi2026}. 
The full 3-D model is shown in Fig.~\ref{fig:femodel}.

In total, the head is segmented into fourteen regions: the grey matter, white matter, deep grey matter, corpus callosum, cerebellar grey matter, cerebellar white matter, brain stem, subarachnoid space/cerebrospinal fluid, ventricles, dura, skull, falx, tentorium, and foramen.
Although these anatomical regions are retained in the segmented model, the seven parenchymal brain substructures (i.e., grey matter, white matter, deep grey matter, corpus callosum, cerebellar grey and white matter, and brain stem) are consolidated into a single ``brain bulk" label and assigned a common set of material properties in the FE simulations. This simplification is necessary because region-specific experimental data for Mullins-type damage in brain tissue are not currently available. Nevertheless, retaining the regional segmentation enables analysis of region-wise biomechanical response and damage evolution arising from the brain's complex three-dimensional geometry. As regional Mullins damage properties become available, they can be incorporated into the present head model to enable spatially heterogeneous constitutive descriptions of brain tissue.
 
\subsubsection{Meshing, Material Models, and Setup}
\label{subsec:matmodelsforFE}
The open-source toolbox GibbonCode \citep{Moerman2018} is used to generate a hexahedral FE mesh in MATLAB R2025a (Mathworks, Inc.), where each voxel in the segmented head volume is converted one-to-one into a brick FE element. This voxelated mesh maintains the high (1 x 1 x 1 mm$^3$) resolution of the MRI scans and preserves boundaries between the brain’s structures. 
Other techniques, such as the use of smoothed conformal elements \citep{Zhou2025}, 
have seen use, but carry increased computational cost and are specialized toward more intense loadings (such as blast waves \citep{Zhang2013}) than those considered here or in the study of localized phenomena like cerebrospinal fluid cavitation in the brain’s sulci \citep{Fagan2020}. 
The voxelated meshing technique used here has been found to perform similarly to conformal meshes that mimic the smoothness of interfaces in the brain \citep{Zhou2025}. The final meshed model of this work contains 1,948,979 nodes and 1,888,586 hexahedral elements.

In terms of the constitutive material models adopted in this work, the Mullins damage model described in Sec.~\ref{sec:constitutivemodeling} is employed to describe the bulk brain tissue, including grey matter, white matter, deep grey matter, corpus callosum, cerebellum, brain stem. This FE-based head injury model, in which the brain bulk is described using the Mullins damage formulation, serves as the primary model of the present study and is hereafter referred to as the \textit{Mullins-based} model. The falx, tentorium, dura, and foramen are treated as linear elastic solids \citep{Upadhyay2022b}. The ventricles are modeled as a soft, nearly incompressible linear elastic solid to mimic their fluidic behavior \citep{Takhounts2008}. The subarachnoid space/cerebrospinal fluid region is modeled as a soft hyperelastic solid with Prony-series based linear viscoelasticity~\citep{Takhounts2008, Mao2013}, described by 
\begin{equation}
\label{eq:visco}
\mu(t)=\mu_{0}-\mu_{0}\sum^{n}_{i=1}g_{i}\left(1-\exp{\left(-t/\tau_{i}\right)}\right) \ ,
\end{equation}
where $\mu(t)$ is the time-dependent shear modulus and $\mu_{0}$ is the instantaneous shear modulus. The parameters $g_i$ and $\tau_i$ are fractional shear contributions and characteristic decay times of branches $i=1,2,\ldots,n$ in a generalized Maxwell model \citep{Alshareef2021, DeRooij2016}. Finally, the skull is modeled as a rigid body as it is orders of magnitude stiffer than brain tissue \citep{Budday2020} and experiences negligible deformation in typical closed head impacts \citep{Takhounts2013}. 
Skull rigidity is a common assumption in the head modeling literature \citep{Cloots2008, Wright2013, Ghajari2017, Alshareef2021, Tripathi2026}. Material properties for all anatomical regions in the primary Mullins-based head model are given in Table~\ref{tab:props}.

The Mullins-based computational head injury model is validated by comparing its predicted spatiotemporal strain fields under sub-injurious head rotation against subject-specific tagged MRI measurements from the NITRC BBIR database \citep{Knutsen2020}. The database provides both the experimentally measured tagged MRI strain fields and the corresponding head rotational kinematics for the same human subject used to generate the FE head geometry, thereby enabling subject-specific model validation.
In tagged MRI, tag lines are imposed over the brain \citep{Knutsen2014} and deform with the brain tissue during controlled head rotation.
Strain is estimated from the displacement of the tag lines through the image sequence, providing full-field spatiotemporal strain measurements. The corresponding experimentally measured head rotation kinematics are prescribed as loading inputs in the validation simulations using the Abaqus/Explicit (2023, Dassault Systèmes) solver to reproduce the tagged MRI loading conditions. Specifically, to impose skull motion, a reference point is created at the head model's center of mass, and a rigid-body constraint is applied to the skull such that the skull and reference point move together. The prescribed head kinematic history is applied to this reference point; as the skull moves, inertial forces accelerate the brain, commonly described as inertial loading \citep{Wright2013}. Validation results are provided in the Supplementary Material Section~S1. All simulations in this work are run using the Abaqus/Explicit solver with double precision on 24 CPUs using Louisiana State University's High Performance Computing (HPC) resources.

\renewcommand{\thefootnote}{\alph{footnote}}

\begin{table}
\caption{Selected material properties for anatomical regions in the primary Mullins damage-based FE head model}
\label{tab:props}
\begin{tabular}{c>{\centering\arraybackslash}m{1in}cc}
\multicolumn{4}{c}{\textbf{Mullins-Based Head Model}} \\
\toprule
     Anatomical region(s) & Material Type & Material Properties & Ref.\\
     \addlinespace[0.8mm]
     Brain Bulk\footnotemark[1] & Neo-Hookean Hyperelastic with Mullins Damage & \begin{tabular}{@{}l@{}}
     $\rho$ = 1.13 g cm$^{-3}$ \\ $\mu$ = 6.44 kPa \\ $r$ = 1.2 \\ $m$ = 0.05 kPa \\ $K$ = 1.46 GPa \end{tabular} & \begin{tabular}{@{}c@{}} \cite{Alshareef2021} \\ \cite{Franceschini2006} \end{tabular} \\  
\midrule
Subarachnoid Space & Neo-Hookean Hyperelastic with Linear Viscoelasticity (1st-order) & \begin{tabular}{@{}l@{}} $\rho$ = 1.04 g cm$^{-3}$ \\ $\mu_{0}$ = 0.5 kPa \\ $g_{1}$ = 0.8 \\ $\tau_{1}$ = 1.25 ms \\ $K$ = 2.19 GPa \end{tabular}& \cite{Alshareef2021}\\
\midrule
\begin{tabular}{c}Falx, Tentorium,\\ Dura, Foramen\end{tabular}& Linear Elastic &\begin{tabular}{@{}l@{}} $\rho$ = 1.13 g cm$^{-3}$ \\ $E$ = 31.5 MPa \\ $\nu$ = 0.45 \end{tabular}& \cite{Mao2013}\\
\midrule
Ventricles& Linear Elastic & \begin{tabular}{@{}l@{}}$\rho$ = 1.00 g cm$^{-3}$ \\ $E$ = 1 MPa \\ $\nu$ = 0.4999 \end{tabular} & \cite{Takhounts2008} \\
\bottomrule
\end{tabular}

\textsuperscript{a}Brain Bulk includes grey matter, white matter, deep grey matter, cerebellar grey matter, cerebellar white matter, corpus callosum, and brain stem 
\end{table}

To isolate the effect of Mullins damage on brain tissue response, two additional FE-based head injury models are developed for comparison. These models are identical to the primary Mullins-based model in geometry, mesh, boundary conditions, loading procedure, and non-brain material properties, but replace the Mullins damage formulation for the brain bulk with alternative damage-free constitutive descriptions. In the first comparison model, hereafter referred to as the hyperelastic (HE) model, the brain bulk is modeled as an intact neo-Hookean hyperelastic solid; this corresponds to the hyperelastic response used in the Mullins-based model without the damage evolution variable \(\eta\). In the second comparison model, hereafter referred to as the linear visco-hyperelastic (LVHE) model, the brain bulk is modeled as a neo-Hookean hyperelastic solid with Prony-series-based linear viscoelasticity. Material properties for the HE and LVHE models are given in Table~\ref{tab:props2}. Explicit analytical equations for the HE and LVHE constitutive models are provided in Supplementary Section~S2.

\begin{table}
\caption{Selected material properties for anatomical regions in the comparison hyperelastic (HE) and linear visco-hyperelastic (LVHE) FE head models}
\label{tab:props2}
\begin{tabular}{c>{\centering\arraybackslash}m{1in}cc}
\multicolumn{4}{c}{\textbf{Hyperelastic Head Model --- No Damage (HE)}} \\
\toprule
     Anatomical region(s) & Material Type & Material Properties & Ref.\\
     \addlinespace[0.8mm]
Brain Bulk\footnotemark[1] & Neo-Hookean Hyperelastic &\begin{tabular}{@{}l@{}} $\rho$ = 1.13 g cm$^{-3}$\\ $\mu$ = 6.44 kPa \\ $K$ = 1.46 GPa \end{tabular}& \cite{Alshareef2021}\\
\addlinespace[5mm]
\multicolumn{4}{c}{\textbf{Visco-hyperelastic Head Model --- No Damage (LVHE)}} \\
\toprule
     Anatomical region(s) & Material Type & Material Properties & Ref.\\
     \addlinespace[0.8mm]
Brain Bulk\footnotemark[1] & Neo-Hookean Hyperelastic with Linear Viscoelasticity (2nd-order) & \begin{tabular}{@{}l@{}} $\rho$ = 1.13 g cm$^{-3}$ \\ $\mu_{0}$ = 6.44 kPa \\ $g_{1},g_{2}$ = 0.74, 0.048 \\ $\tau_{1},\tau_{2}$ = 1.0, 29.6 ms \\ $K$ = 1.46 GPa \end{tabular} & \cite{Alshareef2021}\\
\bottomrule
\end{tabular}
\footnotetext{Note: Anatomical regions not listed here (i.e., Subarachnoid Space, Falx, Tentorium, Dura, Foramen, Ventricles) are modeled using the properties shown in Table \ref{tab:props}}
\smallskip
\textsuperscript{a}Brain Bulk includes grey matter, white matter, deep grey matter, cerebellar grey matter, cerebellar white matter, corpus callosum, and brain stem 
\end{table}

\subsection{Analysis Framework for Repeated Head Loading Simulations}
\label{subsec:analysis_framework}

The Mullins-based computational head injury model is subjected to sequences of loading cycles with conditions drawn from the literature, and brain deformation within each cycle is analyzed to evaluate cycle-to-cycle changes as mechanical damage accumulates. Where appropriate, these results are compared with those obtained from damage-free HE and/or LVHE head models under identical loading conditions. Two loading modes are considered: (1) single-axis rotation, in which head rotation is prescribed about one axis only during each loading cycle; (2) multiaxial rotation, in which rotational components about all three axes are present. Single-axis rotation simulations are used to isolate the effects of Mullins damage on brain deformation (i.e., strain and strain rate) by prescribing identical loading cycles within one rotational plane. Multiaxial rotation simulations are conducted to examine damage progression under more general repeated head loading, representative of a real-world impact scenario, where multiple regions of the brain may be strained by successive impacts varying in both intensity and directionality.

\begin{figure}[hbt!]
    \centering    
    \includegraphics[width=2.466in]{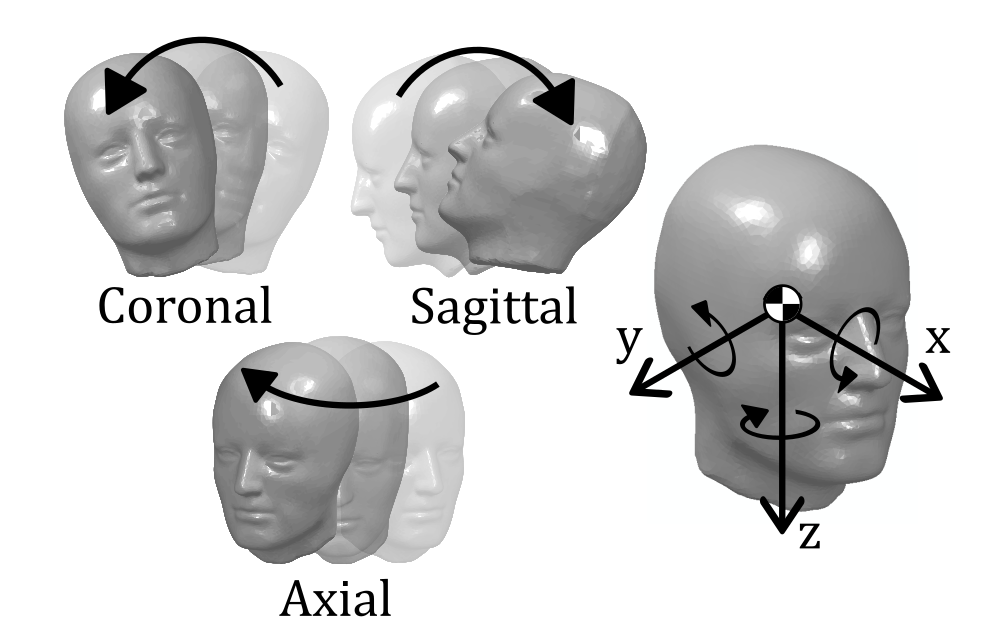}
    \caption{Head coordinate system and key rotational planes}
    \label{fig:coord}
\end{figure}

Simulation outputs are analyzed using strain-based injury predictors, injury risk functions (IRFs), and spatial analyses of regional and gyral/sulcal response. Strain-rate-based metrics are additionally evaluated in the single-axis simulations to examine how Mullins damage affects deformation-rate response during controlled repeated loading.
The head coordinate system follows the SAE J211 standard~\citep{J211/1_202208} and is presented in Fig.~\ref{fig:coord}.
All data and results hereafter adhere to this coordinate system. 

\subsubsection{Loading Conditions and Simulation Cases}
\label{subsec:loadingconditions}

Repeated head loading is prescribed through the angular velocity history of the skull. Each loading cycle consists of an active loading pulse followed by a rest period. During the active loading pulse, the skull is accelerated and decelerated according to a half-sinusoidal angular velocity profile, consistent with previous computational head-injury studies \citep{Takhounts2008, Cloots2008, Upadhyay2022, Upadhyay2024}. For loading cycle \(n\), each angular velocity component is prescribed as
\begin{equation}
\label{eq:kinematics_idealized}
\omega_i^{(n)}(t)
=
\omega_{i,p}^{(n)}
\sin\left[
\frac{\pi (t-t_n)}{\Delta t_n}
\right],
\qquad
t_n \leq t \leq t_n+\Delta t_n,
\end{equation}
where \(i=x,y,z\), \(\omega_{i,p}^{(n)}\) is the peak angular velocity of component \(i\) in cycle \(n\), \(\Delta t_n\) is the pulse duration, and \(t_n\) is the start time of the loading pulse. The skull is therefore at rest at the beginning and end of each loading pulse. The corresponding angular acceleration history is obtained from the time derivative of Eq.~\eqref{eq:kinematics_idealized}. During the rest period following the loading pulse, the prescribed skull angular velocity is zero.

Two classes of repeated loading simulations are considered: single-axis and multiaxial loading. In the single-axis simulations, head rotation is restricted to one anatomical axis at a time, corresponding to coronal, sagittal, or axial rotation as defined by the coordinate system in Fig.~\ref{fig:coord}. Thus, only one angular velocity component is nonzero in a given single-axis simulation,
\begin{equation}
\label{eq:singleaxis}
    \begin{aligned}
    \boldsymbol{\omega}_{\mathrm{coronal}}(t)&=[\omega_{x}(t), && 0, && 0] ^{\mathrm{T}} \ , \\
    \boldsymbol{\omega}_{\mathrm{sagittal}}(t)&=[0, && \omega_{y}(t), && 0] ^{\mathrm{T}} \ , \\
    \boldsymbol{\omega}_{\mathrm{axial}}(t)&=[0, && 0, && \omega_{z}(t)] ^{\mathrm{T}} \ .
    \end{aligned}
\end{equation}

\begin{figure}[hbt!]
    \centering    
    \includegraphics[width=4.997in]{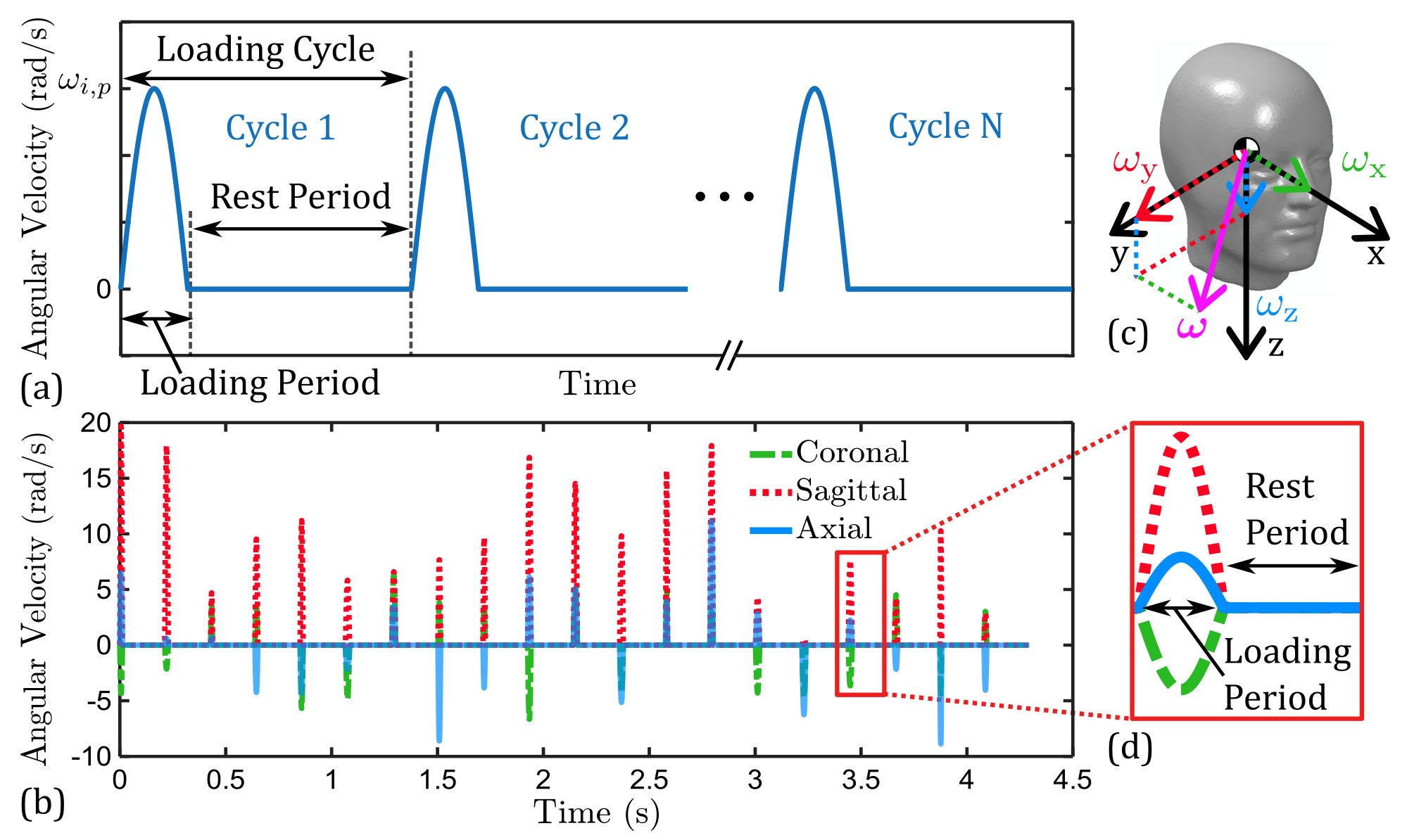}
    \caption{Description of head simulation kinematic history: (a) representative single-axis simulation history comprised of $N$ loading cycles, with peak angular velocity $\omega_{i,p}$ for component direction $i$, (b) multiaxial loading history for $N=20$ cycles, (c) schematic of general head rotation for multiaxial simulations, and (d) magnified view of a multiaxial loading cycle}
    \label{fig:history}
\end{figure}

The same half-sinusoidal pulse is repeated from cycle to cycle in each single-axis case, allowing the effect of Mullins damage on brain deformation to be isolated under controlled, identical repeated loading. In the multiaxial simulation, all three angular velocity components may be nonzero within a given cycle, and the peak angular velocity components and pulse duration are allowed to vary from cycle to cycle. This produces a more general loading history in which successive impacts vary in both intensity and directionality, as expected in real-world repeated head impact scenarios. Representative single-axis and multiaxial loading histories, along with the general head rotation schematic, are shown in Fig.~\ref{fig:history}.

The peak angular velocity components and pulse durations used in the simulations are derived from literature data representative of sports-related head impacts. Specifically, head kinematic data are taken from the mixed martial arts (MMA) dataset of \citet{OKeeffe2020}, in which head impacts were recorded using instrumented mouthguards. This dataset was chosen because MMA athletes may experience multiple head impacts over short time intervals, with a typical MMA round lasting 5 min, providing a relevant scenario for investigating the effects of short-term mechanical softening under repeated loading. \citet{Laksari2020} reported the mean and standard deviation of the peak angular velocity components measured by \citet{OKeeffe2020}, as listed in Table~\ref{tab:kinematicsvalues}. Because the \citet{OKeeffe2020} dataset does not report the duration of individual impacts, pulse-duration statistics are taken from the work of \citet{Adamec2020}, who reported impulse durations for punches. Peak angular velocity is used because it is strongly correlated with injury risk for short-duration impacts, defined as \(\Delta t<36\) ms by \citet{Gabler2018a}, which are characteristic of punching impacts \citep{Carlsen2021, Gabler2018a, Laksari2020}. Pulse duration is prescribed to ensure that the loading conditions remain in the short-duration regime associated with sports-related head impacts, including those in American football, rugby, and MMA \citep{Gabler2018a}.

\begin{table}[hbt!]
\caption{Angular kinematic data representative of mixed martial arts head impacts}
\label{tab:kinematicsvalues}
\begin{tabular}{crrc}
\toprule
\multicolumn{4}{c}{Peak Angular Velocity (rad/s)} \\
\midrule
    Direction & Average & Standard Deviation & Reference \\
    Coronal & 15.0 & 3.6 & \cite{Laksari2020} \\
    Sagittal & 36.0 & 26.0 & \cite{Laksari2020} \\
    Axial & 19.8 & 14.0 & \cite{Laksari2020} \\
    \midrule
    \multicolumn{4}{c}{Impact Duration (ms)} \\
    \midrule
        & Average & Standard Deviation & Reference \\
    Duration & 15 & 4 & \cite{Adamec2020} \\
    \bottomrule
\end{tabular}
\end{table}

The statistical data in Table~\ref{tab:kinematicsvalues} are used differently for the single-axis and multiaxial simulations. For the single-axis simulations, three loading cases are considered, one for each anatomical rotation direction. Each case consists of five identical loading cycles, as shown schematically in Fig.~\ref{fig:history}(a). The peak angular velocity in each direction is set equal to the corresponding average value in Table~\ref{tab:kinematicsvalues}, and the pulse duration is set equal to the average impact duration. The standard deviations are not used in the single-axis cases, because these simulations are intended to isolate the cycle-to-cycle effects of Mullins damage under controlled repeated loading. Identical single-axis loading histories are applied to the Mullins-based and HE models for direct comparison. This comparison isolates the effect of Mullins damage relative to the same underlying intact neo-Hookean hyperelastic response, without introducing additional rate-dependent effects from viscoelasticity.

For the multiaxial simulation, a statistical sampling approach is used to generate a twenty-cycle loading history. For each cycle, the peak angular velocity magnitudes \(\omega_{x,p}\), \(\omega_{y,p}\), and \(\omega_{z,p}\) are sampled from a multivariate normal distribution constructed using the mean and standard deviation values in Table~\ref{tab:kinematicsvalues}, with zero covariance assumed between components. The pulse duration \(\Delta t\) is sampled independently from a univariate normal distribution constructed using the mean and standard deviation of punch impulse duration reported in Table~\ref{tab:kinematicsvalues}. To represent the directionality of realistic head rotations, randomly selected positive or negative signs are assigned to the coronal and axial components, \(\omega_{x,p}\) and \(\omega_{z,p}\), with equal probability, allowing the head to rotate toward either side in these planes. The sagittal component, \(\omega_{y,p}\), is kept positive to represent whiplash-like extension motion, such as that associated with uppercut-type punches, while avoiding downward sagittal rotation that is less representative of the MMA loading scenario considered here. Because the objective of the multiaxial simulation is to examine cumulative deformation amplification over successive loadings rather than individual severe impacts, the sampled angular velocity values are reduced by 75\%, i.e., multiplied by 0.25. This scaling preserves the statistical cycle-to-cycle variation and loading directionality of the \citet{Laksari2020} dataset while preventing immediate saturation of injury metrics. Twenty such randomly generated loading cycles define the multiaxial loading history shown in Fig.~\ref{fig:history}(b,d). Identical multiaxial loading histories are applied to the Mullins-based, HE, and LVHE models to enable direct comparison of their repeated-loading responses. The HE model provides a damage-free counterpart to the Mullins-based model, while the LVHE model provides an additional benchmark against a commonly used intact, rate-dependent brain-tissue constitutive description.

A rest period is inserted after each active loading pulse in both the single-axis and multiaxial simulations. During this rest period, the skull is stationary, but the brain may continue to deform as shear waves propagate through the tissue. Because shear waves propagate relatively slowly in brain tissue (\(\sim 1~\mathrm{m/s}\)), deformation continues after skull motion ceases \citep{Ganpule2018}. Several studies have shown that brain strain fields can continue to evolve beyond the active loading phase, even after the skull has come to rest; therefore, a post-loading window is needed to capture the full evolution of brain dynamics and the maximum strains experienced by the brain. \citet{Takhounts2008} recommended a 32 ms post-loading window, \citet{Carlsen2021} used a 20 ms period, and \citet{Ji2022} used a 70 ms post-impact window. In the present study, a 200 ms rest period is used between successive loading pulses to ensure that brain strains largely subside before the next cycle is applied. This rest period was sufficient for strains to fall below 10\% of their peak value, thereby reducing artificial cycle-to-cycle strain increases due to residual dynamic deformation or pre-straining. Although real head impacts in MMA may be separated by several seconds, the 200 ms rest period provides a computationally efficient approximation that allows dynamic strains to decay while retaining the short-term Mullins-type softening effects considered in this study. This treatment is consistent with experimental observations that brain-tissue stress softening can persist over short time scales, with complete stiffness recovery reported after approximately 60 min \citep{Budday2020}.

\subsubsection{Biomechanical Analysis of the Effects of Repeated Loading}
\label{subsec:analysismeth}

\paragraph{Tissue-strain- and kinematics-based injury metrics and risk prediction}
Simulation outputs contain full-field logarithmic strain, also called left Hencky strain, defined as
\begin{equation}
    \boldsymbol{\varepsilon}^{L}=\ln(\mathbf{V}),
\end{equation}
where \(\mathbf{V}\) is the left stretch tensor, with \(\mathbf{V}=\sqrt{\mathbf{B}}\). Logarithmic strain is commonly used to track large deformations of hyperelastic solids \citep{Rudnicki2015,Chandrashekar2026} and biological tissues \citep{Cloots2008,Carlsen2021}. In this work, logarithmic strain is extracted at the integration point of each element. Because the brain mesh uses reduced-integration hexahedral elements, specifically C3D8R elements in Abaqus/Explicit, each element contains one integration point and therefore one strain tensor at each output time. The temporal resolution of the simulation output used for analysis is approximately \(5~\mathrm{ms}\).

The analysis primarily uses two tissue-strain-based injury predictor variables, 95th percentile cumulative maximum principal strain (CMPS95) and cumulative strain damage measure with a 0.15 strain threshold (CSDM15), and one kinematics-based injury predictor, the universal brain injury criterion (UBrIC) \citep{Gabler2018a}. These quantities are used to evaluate cycle-wise changes in tissue deformation and injury risk during repeated loading.

CMPS95 is computed from the maximum principal strain (MPS), defined as the maximum eigenvalue ($\lambda_{\mathrm{max}}$) of \(\boldsymbol{\varepsilon}^{L}\) at each element and time point:
\begin{equation}
    \mathrm{MPS}_{e}(t)
    =
    \lambda_{\max}
    \left[
    \boldsymbol{\varepsilon}^{L}_{e}(t)
    \right],
\end{equation}
where \(e\) denotes a brain-bulk element. Since MPS is a spatiotemporal field, the instantaneous 95th percentile maximum principal strain, denoted MPS95, is used to characterize the temporal evolution of peak brain strain:
\begin{equation}
\label{eq:mps95}
    \mathrm{MPS95}(t)
    =
    P_{95}
    \left(
    \left\{
    \mathrm{MPS}_{e}(t)
    \right\}_{e\in\mathcal{B}}
    \right),
\end{equation}
where \(P_{95}(\cdot)\) denotes the 95th percentile and \(\mathcal{B}\) is the set of all brain-bulk elements. To evaluate the peak strain experienced by each element during a loading cycle, cumulative maximum principal strain (CMPS) is defined as
\begin{equation}
    \mathrm{CMPS}_{e}^{(n)}
    =
    \max_{t\in T_n}
    \mathrm{MPS}_{e}(t),
\end{equation}
where \(T_n=[t_n,t_{n+1})\) denotes the complete time interval of loading cycle \(n\), including both the active loading pulse and the subsequent rest period. Thus, CMPS is a scalar field in which each element stores its maximum MPS value over the duration of a given loading cycle. CMPS95 is then defined as
\begin{equation}
    \mathrm{CMPS95}^{(n)}
    =
    P_{95}
    \left(
    \left\{
    \mathrm{CMPS}_{e}^{(n)}
    \right\}_{e\in\mathcal{B}}
    \right),
\end{equation}
which yields a single cycle-wise value for a given loading pulse. CMPS95 is a standard measure for characterizing the extent of brain deformation and injury risk \citep{Takhounts2008,Gabler2018a,Upadhyay2022,Upadhyay2024}.

CSDM is a volumetric measure computed from the CMPS field and is defined as the fraction of brain-bulk volume whose CMPS exceeds a prescribed strain threshold. In this work, CSDM15 is used, corresponding to a CMPS threshold value of 0.15:
\begin{equation}
    \mathrm{CSDM15}^{(n)}
    =
    \frac{
    \sum\limits_{e\in\mathcal{B}}
    V_e\,
    H
    \left(
    \mathrm{CMPS}_{e}^{(n)}-0.15
    \right)
    }{
    \sum\limits_{e\in\mathcal{B}} V_e
    },
\end{equation}
where \(V_e\) is the volume of element \(e\), and \(H(\cdot)\) is the Heaviside step function. A strain threshold of 0.15 is commonly used in studies of TBI and mild TBI (mTBI) \citep{Carlsen2021,Wu2022}. Like CMPS95, CSDM15 is computed separately for each loading cycle to enable cycle-by-cycle evaluation of changes in strain-based injury risk.

In addition to these tissue-strain-based measures, UBrIC is used as a kinematics-based injury predictor. Unlike CMPS95 and CSDM15, UBrIC depends only on the prescribed rotational input kinematics, specifically peak angular velocities and peak angular accelerations \citep{Gabler2018a,Wu2022}. Mathematically, UBrIC is defined as
\begin{equation}
    \label{eq:ubric}
    \mathrm{UBrIC}
    =
    \left\{
    \sum_{i}
    \left[
    \omega_{i}^{*}
    +
    \left(
    \alpha_{i}^{*}-\omega_{i}^{*}
    \right)
    \exp
    \left(
    \frac{-\alpha_{i}^{*}}{\omega_{i}^{*}}
    \right)
    \right]^{2}
    \right\}^{1/2},
\end{equation}
where \(\omega_{i}^{*}=\omega_{i}/\omega_{i,\mathrm{crit}}\) are the normalized peak angular velocity components and \(\alpha_{i}^{*}=\alpha_{i}/\alpha_{i,\mathrm{crit}}\) are the normalized peak angular acceleration components, with \(i=x,y,z\) \citep{Gabler2018a}. The critical angular velocities, \(\omega_{i,\mathrm{crit}}\), and accelerations, \(\alpha_{i,\mathrm{crit}}\), are empirically determined values given in \citet{Gabler2018a}.

Strain-based injury predictors, including CMPS95 and CSDM15, and kinematics-based predictors, including UBrIC, have been correlated with probability of brain injury \citep{Takhounts2008,Wu2022} according to the abbreviated injury scale (AIS)~\citep{aaam_ais1998}. Here, AIS2, corresponding to mTBI or concussion, is the injury outcome of interest, consistent with similar investigations of concussive loading \citep{Carlsen2021,Wu2022}. For strain-based predictors, \citet{Wu2022} represented injury probability using a Weibull cumulative distribution function,
\begin{equation}
    \label{eq:IRFprob}
    P_\mathrm{inj}
    =
    1-\exp{\left[
    -\exp{\left(
    \frac{\ln(x)-a_1}{a_2}
    \right)}
    \right]},
\end{equation}
where \(x\) denotes the strain-based injury predictor (i.e., CMPS95 or CSDM15), and \(a_1\) and \(a_2\) are empirically determined coefficients corresponding to Weibull scale \(e^{a_1}\) and shape \(1/a_2\), respectively. In the present study, CMPS95 and CSDM15 values computed from the simulations are substituted directly into their corresponding IRFs to estimate AIS2 injury probability. For UBrIC, \citet{Wu2022} derived a kinematics-based IRF from a linear regression relating UBrIC (\(y\)) to CMPS95 (\(x\)), expressed as \(y=b_1x+b_2\). Using the resulting published UBrIC-based IRF coefficients, AIS2 injury probabilities are computed directly from UBrIC values in the present study.

For cycle-wise cumulative measures, including CMPS95 and CSDM15, an element's history is taken to be the duration of the corresponding loading cycle only; strain from previous cycles is not included directly in the metric calculation. This definition enables cycle-by-cycle comparison of injury predictors while allowing material history effects to enter through the evolving Mullins damage state. In the single-axis simulations, cycle-to-cycle changes in injury predictors and injury probabilities are compared between the Mullins-based and HE models. In the multiaxial simulations, injury predictors and injury probabilities are compared among the Mullins-based, HE, and LVHE models. Injury probabilities predicted from simulated tissue deformation and from prescribed kinematics are expected to diverge as damage accumulates, because UBrIC does not account for the evolving mechanical state of the brain tissue.

\paragraph{Strain-rate-based injury metrics}
Recently, strain-rate-based metrics have seen growing interest as supplementary indicators of TBI \citep{Kleiven2007,Elkin2007,Carlsen2021,Nakarmi2025}. Here, logarithmic strain rate is used, defined as the symmetric part of the velocity gradient tensor:
\begin{equation}
    \dot{\boldsymbol{\varepsilon}}^{L}
    =
    \operatorname{sym}
    \left(
    \frac{\partial\mathbf{v}}{\partial\mathbf{x}}
    \right),
\end{equation}
where \(\mathbf{v}\) is the particle velocity. The maximum principal strain rate (MPSR) is defined as the maximum eigenvalue ($\lambda_\mathrm{max}$) of \(\dot{\boldsymbol{\varepsilon}}^{L}\):
\begin{equation}
    \mathrm{MPSR}_{e}(t)
    =
    \lambda_{\max}
    \left[
    \dot{\boldsymbol{\varepsilon}}^{L}_{e}(t)
    \right].
\end{equation}
Analogous to the strain-based metrics, cumulative maximum principal strain rate (CMPSR) is defined for each element as the maximum MPSR attained over a given loading cycle:
\begin{equation}
    \mathrm{CMPSR}_{e}^{(n)}
    =
    \max_{t\in T_n}
    \mathrm{MPSR}_{e}(t).
\end{equation}
The 95th percentile cumulative maximum principal strain rate, CMPSR95, is then defined as
\begin{equation}
    \mathrm{CMPSR95}^{(n)}
    =
    P_{95}
    \left(
    \left\{
    \mathrm{CMPSR}_{e}^{(n)}
    \right\}_{e\in\mathcal{B}}
    \right).
\end{equation}

Cumulative strain rate damage measure (CSRDM) is defined as the volume fraction of brain-bulk elements whose CMPSR exceeds a prescribed threshold. Commonly used thresholds for studying concussion are \(40~\mathrm{s}^{-1}\), denoted CSRDM40 \citep{Hajiaghamemar2021,Carlsen2021}, and \(50~\mathrm{s}^{-1}\), denoted CSRDM50 \citep{Zhan2022,Nakarmi2025}. These quantities are computed as
\begin{equation}
    \begin{split}
    \mathrm{CSRDM40}^{(n)}
    &=
    \frac{
    \sum\limits_{e\in\mathcal{B}}
    V_e\,
    H
    \left(
    \mathrm{CMPSR}_{e}^{(n)}-40~\mathrm{s}^{-1}
    \right)
    }{
    \sum\limits_{e\in\mathcal{B}} V_e
    }, \\
    \mathrm{CSRDM50}^{(n)}
    &=
    \frac{
    \sum\limits_{e\in\mathcal{B}}
    V_e\,
    H
    \left(
    \mathrm{CMPSR}_{e}^{(n)}-50~\mathrm{s}^{-1}
    \right)
    }{
    \sum\limits_{e\in\mathcal{B}} V_e
    }.
    \end{split}
\end{equation}

These strain-rate-based metrics are computed for the single-axis simulations to provide insight into strain-rate evolution during successive loading. Because the IRFs used in this study are based on strain-based and kinematics-based predictors, strain-rate metrics are not converted to injury probabilities. Instead, strain-rate-based quantities are used as supplementary mechanical measures to assess how damage alters deformation-rate response from cycle to cycle.

\paragraph{Geometry-dependent regional and gyral--sulcal analyses}
A regional analysis is conducted to examine the interplay between the complex geometry of the brain, the directionality of applied loading, and the evolution of damage. The seven segmented parenchymal brain regions shown in Fig.~\ref{fig:femodel} are investigated: grey matter, white matter, deep grey matter, corpus callosum, cerebellar grey matter, cerebellar white matter, and brain stem. Although these regions are assigned a common brain-bulk material response in the FE simulations because region-specific Mullins damage parameters are unavailable, their anatomical labels are retained for post-processing. Region-wise CMPS95 values are computed for each cycle of the three single-axis simulations. From these regional CMPS95-versus-cycle curves, an exponential curve fit of the form
\begin{equation}
    \mathrm{CMPS95}(n)
    =
    A
    \left(
    b-\exp[-k(n-d)]
    \right)
    \label{eq:exp_k}
\end{equation}
is applied using the Curve Fitting Toolbox in MATLAB R2025a to capture the asymptotic behavior. Here, the fitting parameters \(A\) and \(b\) control the magnitude of the asymptotic value, and \(d\) is an arbitrary translation of the curve along the abscissa. The coefficient \(k\), here referred to as the regional damage evolution coefficient, captures the shape of the CMPS95 curve, with higher values indicating a more rapid approach to the asymptote. This coefficient is used to compare regional behavior across the three loading directions.

Lastly, deformation and damage response in the cerebral sulci and gyri are compared. Strain localization in the sulci is commonly observed in computational head models \citep{Ghajari2017,Nol2019,Fagan2020,Tripathi2026}. Indicators of neurodegeneration are also often present in the sulci \citep{McKee2013}. To investigate tissue response at gyral and sulcal loci, 75 gyri--sulci element pairs are manually identified in the grey matter of the computational head model. These elements are chosen to be approximately spatially uniform across the brain. The cycle-wise cumulative maximum principal strain attained by these elements, i.e., CMPS, is obtained, and the sample sets are compared using a paired \(t\)-test to identify statistically significant differences in mechanical response.

\section{Results}\label{chap:results}

This section presents the simulations results of this study and examines how Mullins-type damage alters brain deformation and injury-risk predictions during repeated head loading. The first subsection considers controlled single-axis simulations, where the Mullins-based model is compared with the HE model to isolate the effects of damage-mediated softening. The second subsection presents multiaxial loading results, where the Mullins-based model is compared with both HE and LVHE models under the same randomized twenty-cycle loading history. The final subsection examines spatial variability in the repeated-loading response through region-wise analysis and gyral--sulcal comparisons.

\subsection{Single-Axis Simulations Reveal Damage-Mediated Deformation Amplification under Controlled Repeated Loading}
As detailed in Sec.~\ref{subsec:analysis_framework}, three single-axis simulations are conducted, corresponding to coronal, sagittal, and axial rotation, with each simulation consisting of five identical loading cycles comprising a 15 ms active loading pulse and a 200 ms rest period. MPS95 histories, cycle-wise deformation measures, and cycle-wise injury probabilities are analyzed for the three simulations.

\begin{figure}[hbt!]
    \centering    
    \includegraphics[width=3.846in]{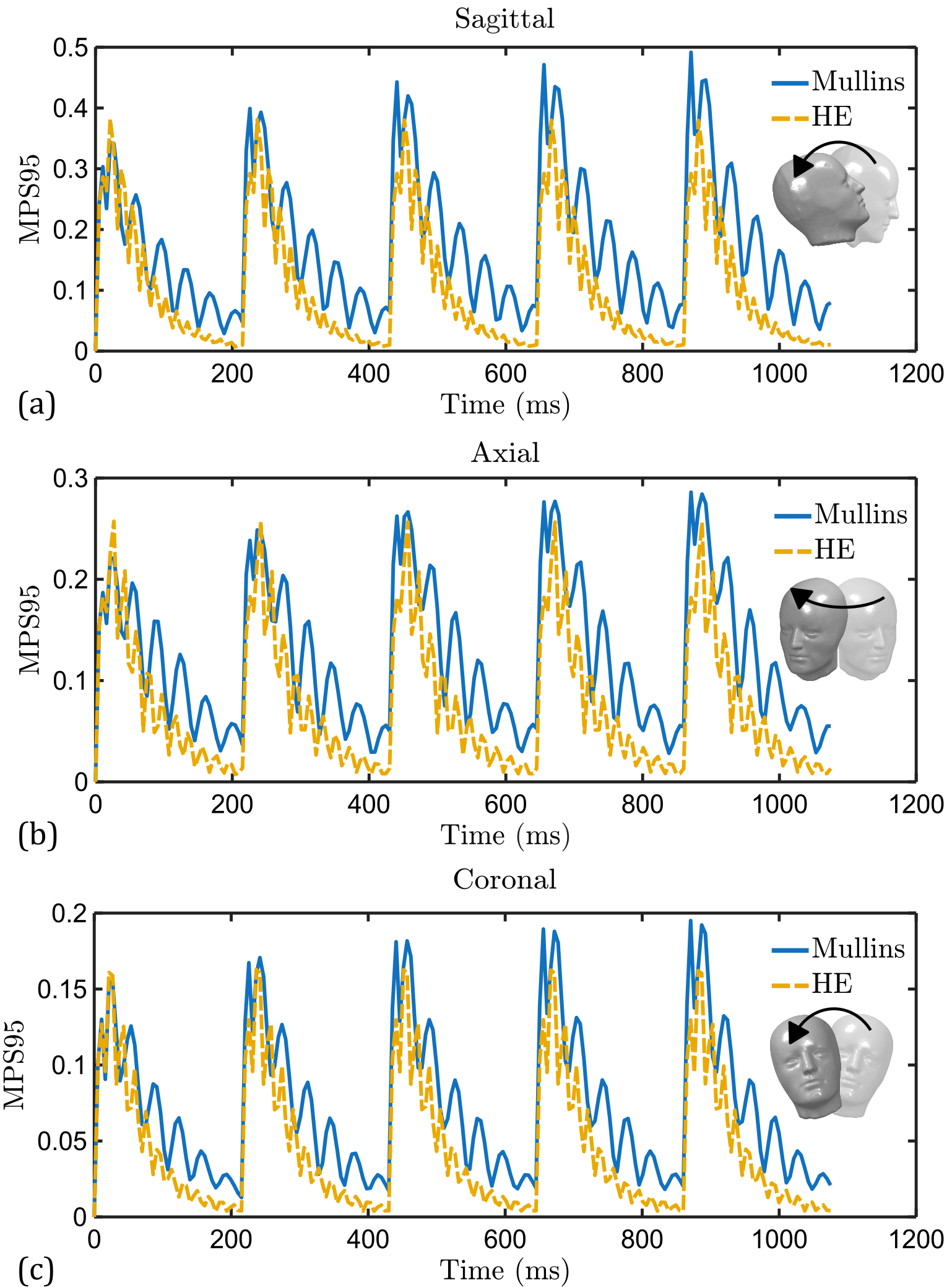}
    \caption{95th-percentile maximum principal strain (MPS95) histories for three single-axis rotation simulations in: (a) sagittal plane, (b) axial plane, and (c) coronal plane. Results of Mullins damage-based model and purely hyperelastic (HE) model are shown}
    \label{fig:singlemps}
\end{figure}

\paragraph{Temporal evolution of MPS95}
Figure~\ref{fig:singlemps} shows the MPS95 time histories for repeated loading in the three rotation directions, for the Mullins-based and the HE-based computational head models. 
MPS95, distinct from CMPS95, is an instantaneous, non-cumulative measure of peak brain strain that varies in time.

Within cycles, a consistent pattern is observed: an initial spike in strains during the loading period, followed by a secondary peak that may exceed the strains of the initial spike, and a subsequent gradual oscillatory decrease to strains around 10\% of the peak. 
Both models (Mullins and HE) decay to a small residual strain during the rest period, although the decay is more rapid for the HE model. 
The Mullins-based model exhibits higher strain amplitudes and an apparently longer oscillation period during the rest period; the origin of this difference is not examined further here.
The maximum MPS95 value within a loading cycle is sometimes reached after the skull comes to rest (i.e., after the completion of the 15 ms loading pulse), which is consistent with findings from \cite{Carlsen2021}.

Across all three head rotation direction cases, cycle-to-cycle variation in peak strains in the HE-based model remains within 2\% of the first cycle peak value. 
Cycle-to-cycle variations in HE model values are attributed to pre-strain effects, as some non-zero strains are present at the end of each rest period.
The Mullins-based model, on the other hand, predicts a significant, monotonic increase in peak strains from cycle to cycle.
Because the prescribed peak angular velocity is greatest for sagittal rotation (see the MMA head kinematics dataset of Table \ref{tab:kinematicsvalues}), the sagittal case produces the largest MPS95 value, reaching 0.49 during Cycle 5 in the Mullins-based model.
Were the three peak angular velocities equal, the axial loading would be expected to produce the largest deformations \citep{Gabler2018a}.

\begin{figure}[t!]
    \centering    
    \includegraphics[width=4.997in]{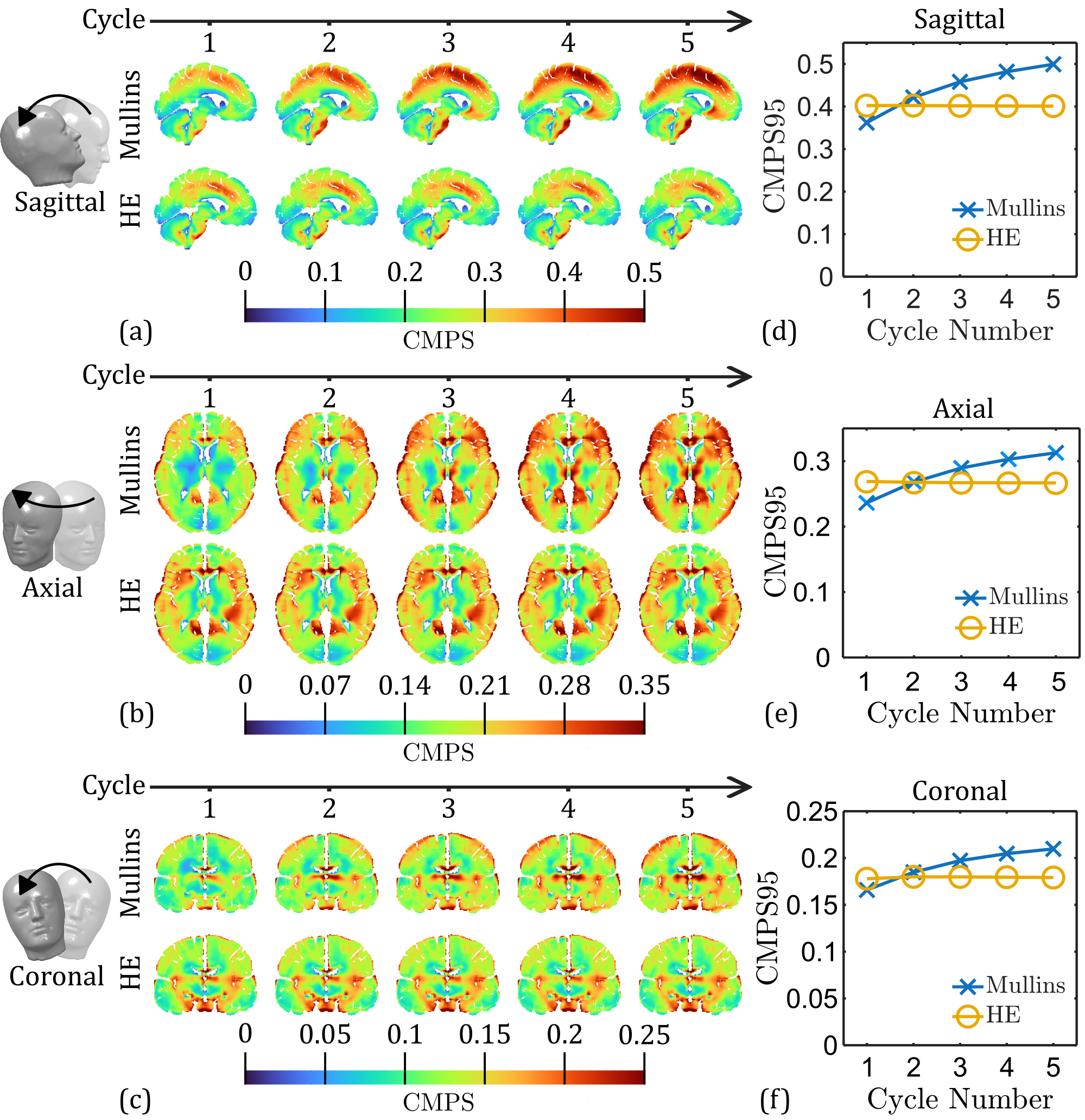}
    \caption{Evolution of cumulative maximum principal strain (CMPS) fields and 95th-percentile cumulative maximum principal strain (CMPS95) under repeated single-axis rotation. Left column: cycle-wise CMPS fields for (a) sagittal, (b) axial, and (c) coronal rotation. Within each left-column panel, Cycles 1--5 are shown from left to right, with results from the Mullins-based model in the top row and the hyperelastic (HE) model in the bottom row. The representative 2-D slices are located at \(y=5~\mathrm{mm}\), \(z=0~\mathrm{mm}\), and \(x=0~\mathrm{mm}\) for the sagittal, axial, and coronal cases, respectively, measured relative to the center of the mesh bounding box. Right column: evolution of CMPS95 with loading cycle for the Mullins-based and HE models under (d) sagittal, (e) axial, and (f) coronal rotation}
    \label{fig:singlecmps}
\end{figure}

\paragraph{Cycle-wise evolution of CMPS and CMPS95}
To evaluate spatial evolution of softening damage in brain over repeated cycles, the left column of Fig.~\ref{fig:singlecmps} presents the CMPS fields for successive loading cycles in the three rotation direction cases. 
Slice views are taken at distances $x=0$ mm, $y=5$ mm, and $z=0$ mm from the center of the mesh's bounding box for the coronal, sagittal, and axial simulations respectively. 
From the figure, CMPS fields, computed across each cycle separately, reveal visible increases in peak strains in the Mullins-based model. 
This amplification is spatially localized, suggesting a heterogeneous influence of Mullins softening on the brain deformation field. The strongest cycle-to-cycle deformation amplification occurs near the cortical sulci and brain stem under sagittal rotation, near the cortical sulci and along the longitudinal fissure under axial rotation, and within the corpus callosum and deep grey matter under coronal rotation.
As expected, the CMPS fields predicted by the HE model remain largely unchanged from cycle to cycle.

For a quantitative analysis of CMPS evolution, the right column of Fig.~\ref{fig:singlecmps} shows CMPS95---a scalar measure denoting the 95th-percentile value of CMPS---versus cycle number for each of the three rotation directions. For the Mullins-based model, CMPS95 increased most strongly from Cycle 1--2 in all three simulations, followed by progressively smaller cycle-to-cycle increases. For example, under sagittal loading, CMPS95 increased by 0.059 from Cycle 1--2 but by only 0.037 from Cycle 2--3, with similar trends observed under axial and coronal loading. The diminishing increments suggest an asymptotic approach toward a plateau under continued identical loading, consistent with progressive saturation of the bounded Mullins softening response. Owing to the functional form of the damage variable $\eta$ in Eq. (\ref{eq:eta}), initial softening is more rapid, with damage evolution becoming gradual at larger deformations. In contrast, the HE model exhibited little cycle-to-cycle variation, with CMPS95 remaining within 0.3\%, 0.8\%, and 1.1\% of the first-cycle value for the sagittal, axial, and coronal simulations, respectively.

\paragraph{Cycle-wise evolution of CSDM15}

\begin{figure}[hbt!]
    \centering    
    \includegraphics[width=5.001in]{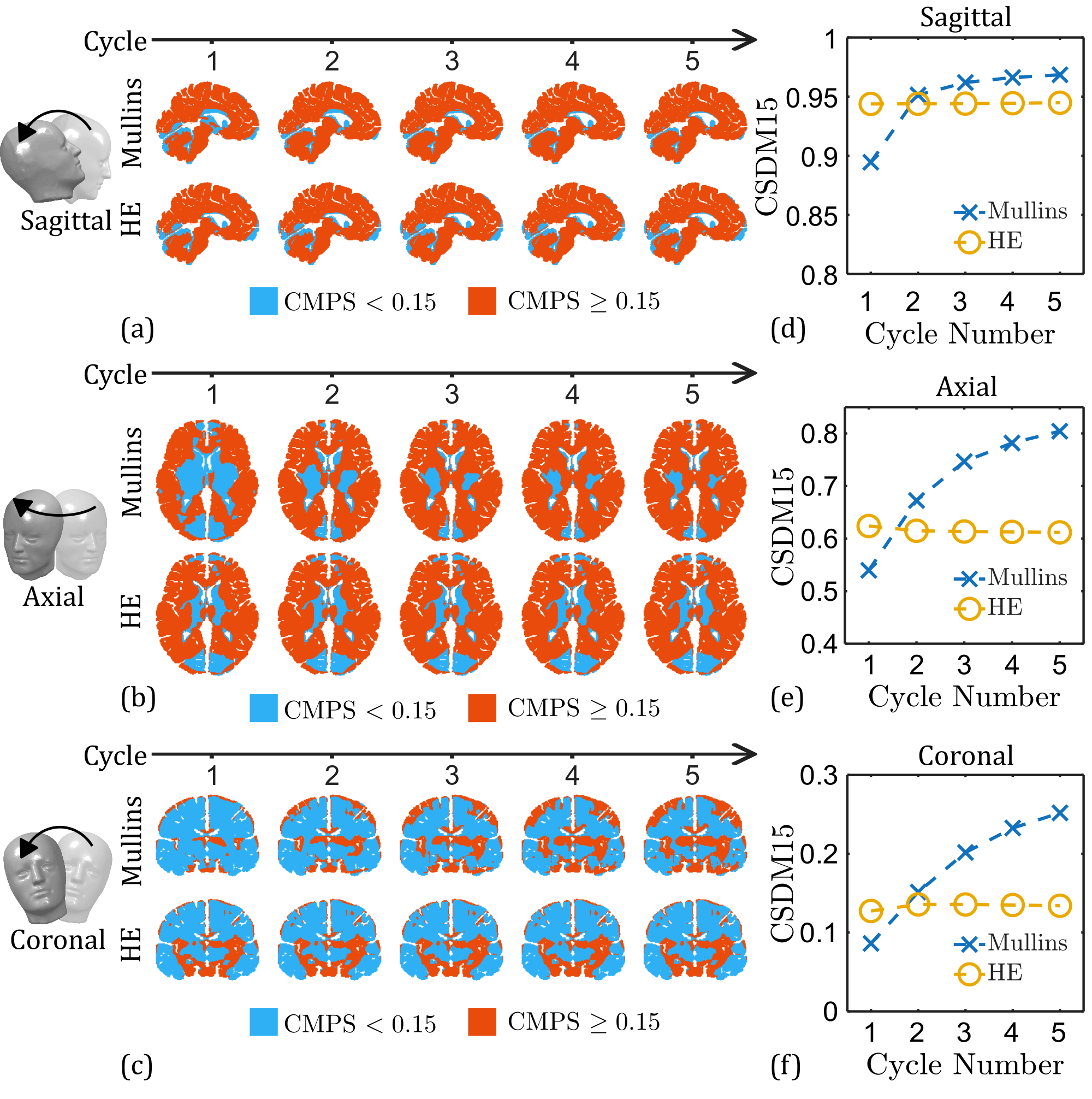}
    \caption{Evolution of the cumulative strain damage measure with a 0.15 strain threshold (CSDM15) under repeated single-axis rotation. Left column: cycle-wise threshold maps showing brain-bulk elements with cumulative maximum principal strain (CMPS) below 0.15 in blue and greater than or equal to 0.15 in red for (a) sagittal, (b) axial, and (c) coronal rotation. Within each left-column panel, Cycles 1--5 are shown from left to right, with results from the Mullins-based model in the top row and the hyperelastic (HE) model in the bottom row. The representative 2-D slices are located at \(y=5~\mathrm{mm}\), \(z=0~\mathrm{mm}\), and \(x=0~\mathrm{mm}\) for the sagittal, axial, and coronal cases, respectively, measured relative to the center of the mesh bounding box. Right column: evolution of CSDM15 with loading cycle for the Mullins-based and HE models under (d) sagittal, (e) axial, and (f) coronal rotation}
    \label{fig:singlecsdm}
\end{figure}

The left column of Fig.~\ref{fig:singlecsdm} shows the spatial evolution of brain-bulk regions whose cycle-wise CMPS meets or exceeds the 0.15 threshold, indicated in red, for the three single-axis rotation cases. Under sagittal rotation, this region encompasses large portions of the cortex, cerebellum, and brain stem, accounting for approximately 89\% of the brain-bulk volume after the first loading cycle in the Mullins-based model. The above-threshold region expands as damage-mediated softening accumulates, exceeding 97\% of the brain-bulk volume by Cycle 5. These volume fractions correspond to the CSDM15 values shown in Fig.~\ref{fig:singlecsdm}(d). Similar cycle-to-cycle amplification is observed under axial and coronal rotation, although the spatial distributions of tissue exceeding the threshold differ. Under axial rotation, the above-threshold region encompasses most of the brain, except for portions near the ventricles and posterior cortex. Under coronal rotation, the above-threshold regions are concentrated primarily within the corpus callosum and outer cortex and occupy a comparatively smaller fraction of the brain-bulk volume. Across all three loading directions, the progressively expanding regions indicate that repeated loading causes a larger volume of tissue to experience potentially injurious strain levels. In contrast, the corresponding above-threshold regions predicted by the HE model remain nearly unchanged from cycle to cycle.

The right column of Fig.~\ref{fig:singlecsdm} quantifies the cycle-wise evolution of CSDM15 for the three rotation directions. For the Mullins-based model, the largest increase occurs between Cycles 1 and 2 in all three cases. Under sagittal rotation, CSDM15 increases by 0.057 between Cycles 1 and 2 and by 0.011 between Cycles 2 and 3, with similarly diminishing cycle-to-cycle increases observed under axial and coronal rotation. By contrast, CSDM15 predicted by the HE model remains nearly constant, with total ranges of variation limited to 0.1\%, 2.0\%, and 6.5\% of the corresponding first-cycle values for the sagittal, axial, and coronal simulations, respectively.

\paragraph{Cycle-wise evolution of strain rate metrics}
Table~\ref{tab:strainrate} compares the strain rate metrics predicted by the Mullins-based and HE models during Cycles 1 and 5 of the three single-axis simulations. Intermediate cycles are not shown because their qualitative cycle-to-cycle trends are consistent with those observed for the tissue-strain-based injury metrics.

All three strain rate metrics increase between Cycles 1 and 5 for the Mullins-based model. CMPSR95 increases by 17.5\%, 11.9\%, and 21.1\% under coronal, sagittal, and axial rotation, respectively. Over the same interval, CSRDM40 increases by 100.0\%, 13.6\%, and 75.0\%, while CSRDM50 increases by 166.7\%, 30.6\%, and 107.4\%, respectively. The comparatively large relative increases in CSRDM40 and CSRDM50 under coronal rotation partly reflect their low Cycle 1 baseline values. In contrast, the corresponding HE-model metrics exhibit only minor cycle-to-cycle variation, with Cycle 1--5 changes of approximately 0.9\%, 1.7\%, and 3.1\% for CMPSR95, CSRDM40, and CSRDM50, respectively. These results demonstrate that Mullins softening alters not only the magnitude of tissue strain rate but also the volume of tissue experiencing elevated strain rates during repeated loading.

\begin{sidewaystable}[p]
\centering
\caption{Cycle-wise evolution of the 95th-percentile cumulative maximum principal strain rate (CMPSR95) and cumulative strain rate damage measures with thresholds of \(40~\mathrm{s}^{-1}\) (CSRDM40) and \(50~\mathrm{s}^{-1}\) (CSRDM50) for the Mullins-based and hyperelastic (HE) models under single-axis rotation}
\label{tab:strainrate}

\begin{tabular}{llcccccc}
\toprule
& &
\multicolumn{3}{c}{Cycle 1} &
\multicolumn{3}{c}{Cycle 5} \\
\cmidrule(lr){3-5}
\cmidrule(lr){6-8}
Model
& Direction
& CMPSR95 (\(\mathrm{s}^{-1}\))
& CSRDM40
& CSRDM50
& CMPSR95 (\(\mathrm{s}^{-1}\))
& CSRDM40
& CSRDM50 \\
\midrule
\multirow{3}{*}{Mullins-based model}
& Coronal  & 46.9  & 0.11 & 0.03 & 55.1  & 0.22 & 0.08 \\
& Sagittal & 122.8 & 0.81 & 0.62 & 137.4 & 0.92 & 0.81 \\
& Axial    & 75.3  & 0.44 & 0.27 & 91.2  & 0.77 & 0.56 \\
\midrule
\multirow{3}{*}{HE model}
& Coronal  & 43.7 & 0.09 & 0.02 & 44.1 & 0.09 & 0.02 \\
& Sagittal & 108.7 & 0.95 & 0.87 & 108.9 & 0.95 & 0.87 \\
& Axial    & 68.0 & 0.58 & 0.32 & 67.4 & 0.57 & 0.31 \\
\bottomrule
\end{tabular}
\end{sidewaystable}

Previous work from \cite{Elkin2007} indicates that heightened strain rates exacerbate the onset of tissue failure and injury, suggesting the increase in strain rate metrics observed here may alter the biomechanical response of brain tissue under repeated loading. These results further motivate that improved characterization of rate dependent damage onset in brain tissue may be a crucial future direction to improve injury prediction in successive loading scenarios. 

\begin{figure}[hbt!]
    \centering    
    \includegraphics[width=5.092in]{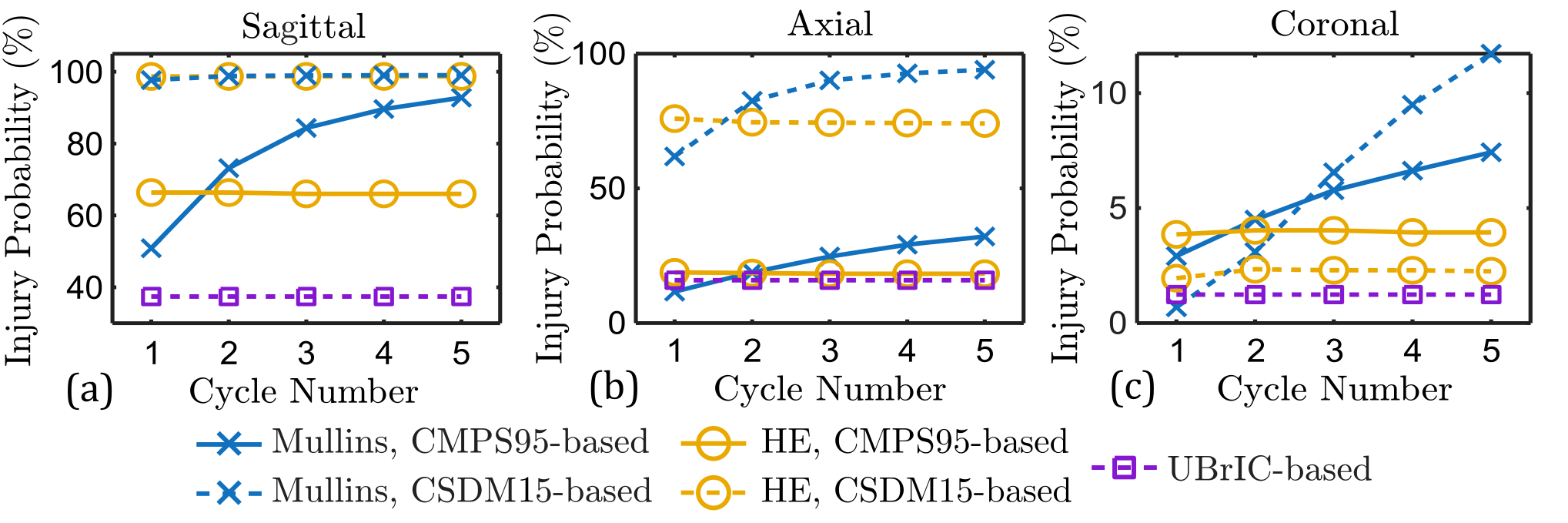}
    \caption{Cycle-wise AIS2 injury probability estimates for single-axis rotation in the (a) sagittal, (b) axial, and (c) coronal directions. Predictions based on the 95th-percentile cumulative maximum principal strain (CMPS95) and cumulative strain damage measure with a 0.15 strain threshold (CSDM15) are shown for the Mullins-based and hyperelastic (HE) models. Kinematics-based predictions obtained using the universal brain injury criterion (UBrIC) are also shown. All injury probabilities are calculated using the injury risk functions developed by \citet{Wu2022}}
    \label{fig:singleinjury}
\end{figure}

\paragraph{Cycle-wise estimation of AIS2 injury probability}
Figure~\ref{fig:singleinjury} presents cycle-wise estimates of AIS2 (i.e., mTBI or concussion) injury probability obtained from the CMPS95-, CSDM15-, and UBrIC-based injury risk functions (IRFs) developed by \citet{Wu2022}. Under sagittal and axial rotation, the CSDM15-based IRF predicts higher injury probabilities than the CMPS95-based IRF for both the Mullins-based and HE models across all loading cycles. Under coronal rotation, this ordering does not hold at the initially low predicted probabilities (\(<6\%\)): the CMPS95-based probabilities initially exceed the corresponding CSDM15-based values for both models. With successive loading cycles, however, the CSDM15-based probability surpasses the CMPS95-based probability for the Mullins-based model.

Across all three rotation directions, the injury probabilities predicted by the Mullins-based model increase with cycle number as damage-mediated softening progressively amplifies brain deformation. In the sagittal case, the CSDM15-based probability is already close to saturation during the first loading cycle and therefore exhibits only a small additional increase over subsequent cycles. The UBrIC values are calculated using Eq.~\eqref{eq:ubric} from the prescribed single-axis kinematic histories defined in Eq.~\eqref{eq:singleaxis}. Because the loading pulse is identical in every cycle of a given simulation, the UBrIC-based injury probability remains constant from cycle to cycle. Predictions from the HE model likewise remain nearly unchanged because the model contains no loading-history-dependent alteration of the brain-bulk material response. Consequently, as repeated loading progresses, the HE-model and UBrIC-based estimates increasingly underestimate the tissue-strain-based injury probabilities predicted by the Mullins-based model. These results indicate that damage-free constitutive models and kinematics-only injury criteria may substantially underpredict injury risk when short-term repeated loading produces progressive softening of brain tissue.

The largest cycle-to-cycle increase in Mullins-based injury probability generally occurs between Cycles 1 and 2, consistent with the corresponding CMPS95 and CSDM15 trends. This behavior reflects both the rapid initial evolution of Mullins softening and the nonlinear Weibull cumulative distribution functions used in the IRFs (see Eq.~(\ref{eq:IRFprob})). As CMPS95 and CSDM15 increase over successive cycles, the Weibull mappings initially produce comparatively large increases in predicted injury probability, followed by progressively smaller increases. Because the bounded Mullins response causes the strain-based predictor variables themselves to approach cycle-wise plateaus, the corresponding injury probabilities also approach finite limiting values determined by those plateaued predictor values, rather than necessarily reaching 100\%.

Together, the single-axis simulations demonstrate that Mullins-type softening produces progressive, spatially localized amplification of brain deformation under otherwise identical repeated loading. These changes increase CMPS95, CSDM15, strain rate metrics, and the corresponding tissue-strain-based injury probabilities, whereas the HE-model response and UBrIC-based predictions remain nearly unchanged. The resulting divergence between tissue-strain- and kinematics-based injury estimates demonstrates that identical head kinematics need not produce identical tissue-level injury-risk predictions when the mechanical state of the brain evolves with loading history.

\subsection{Multiaxial Simulations Examine Damage Effects in Realistic Loading Sequence}
To examine the progressive softening effects of Mullins damage under more realistic repeated head loading, the Mullins-based, HE, and LVHE models are subjected to the same sequence of twenty randomized multiaxial loading cycles. As detailed in Sec.~\ref{subsec:loadingconditions}, the peak angular velocity components are statistically sampled using the directional means and standard deviations reported for MMA head impacts in Table~\ref{tab:kinematicsvalues}, while the loading durations are sampled independently using the corresponding punch-duration statistics (also listed in Table~\ref{tab:kinematicsvalues}). The choice of \(N=20\) cycles is consistent with the number of head impacts recorded during MMA fights by \citet{OKeeffe2020}, such that the loading history incorporates realistic variation in impact intensity, directionality, duration, and exposure count. The HE model provides the direct damage-free counterpart to the Mullins-based model, while the LVHE model is included as an additional benchmark representative of the rate-dependent constitutive formulations most commonly used for brain tissue in computational head models \citep{Ji2022}.

\begin{figure}[t!]
    \centering    
    \includegraphics[width=4.997in]{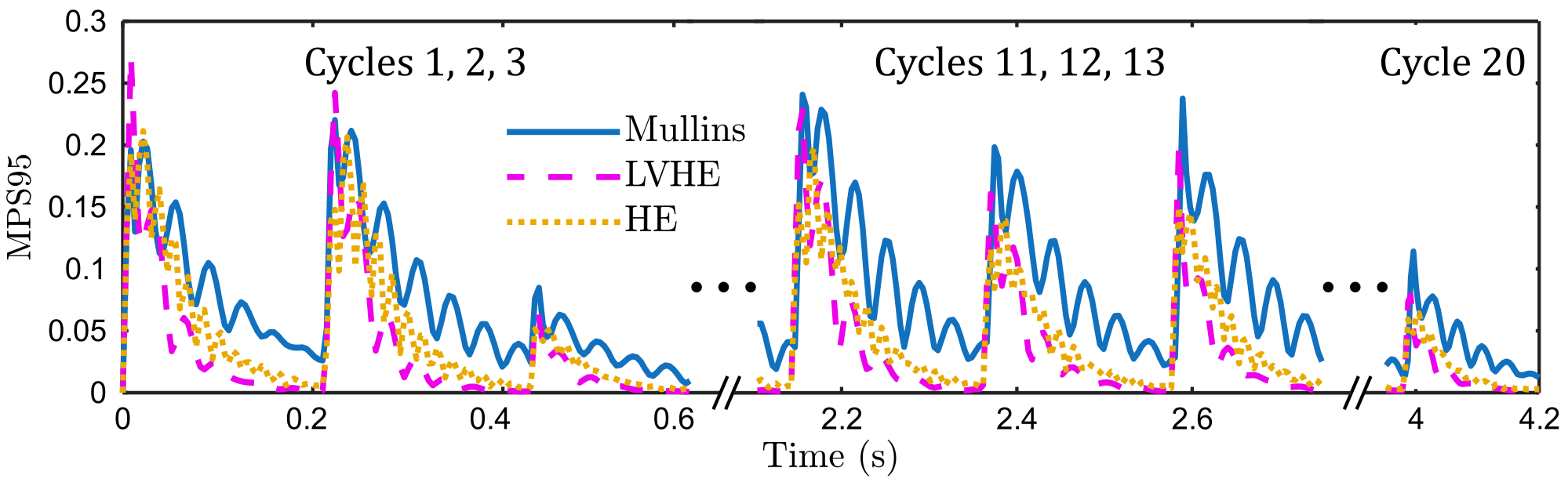}
    \caption{Abridged temporal evolution of the 95th-percentile maximum principal strain (MPS95) during the twenty-cycle multiaxial loading sequence. The figure shows the initial loading response (Cycles 1--3), an intermediate portion of the sequence (Cycles 11--13), and the final loading response (Cycle 20), with omitted intervals indicated by breaks in the time axis. Predictions from the Mullins-based, hyperelastic (HE), and linear visco-hyperelastic (LVHE) models are shown under the same prescribed loading history}
    \label{fig:multiMPS}
\end{figure}

\paragraph{Temporal evolution of MPS95} 
Figure~\ref{fig:multiMPS} presents an abridged view of the MPS95 time histories predicted by the Mullins-based, HE, and LVHE models during the twenty-cycle multiaxial loading sequence. The initial three cycles, three intermediate cycles, and the final cycle are shown for visual clarity. Because the loading intensity, directionality, and duration vary between cycles, the magnitude and temporal profile of MPS95 also vary throughout the sequence. Nevertheless, each cycle generally exhibits a rapid increase in MPS95 during the active loading pulse, followed by an oscillatory decay during the rest period.

During Cycles 1 and 2, the LVHE model predicts the largest peak MPS95 among the three models. By Cycle 3, the Mullins-based model predicts the largest response, and inspection of the complete twenty-cycle history shows that it continues to predict the highest peak MPS95 in Cycles 3--9 and 11--20; Cycle 10 is the only later exception. The increasing separation between the Mullins-based response and the two damage-free model responses is especially apparent in the intermediate and final cycles shown in Fig.~\ref{fig:multiMPS}. Thus, although the randomized loading prevents a monotonic increase in MPS95 with cycle number, the results show that accumulated Mullins-type softening progressively alters the tissue response and generally produces greater deformation during later loading cycles. Unlike the single-axis cases, in which the loading direction is fixed, the randomized variation in directionality and intensity exposes a broader and spatially evolving set of brain-bulk elements to deformation histories that produce Mullins-type softening. Because the Mullins formulation retains the maximum previously attained distortional strain energy at each element, softening generated during one cycle influences that element's response during subsequent cycles. The multiaxial response is therefore cumulative not only with respect to the number of loading cycles, but also through the progressively expanding spatial distribution of tissue with a prior softening history.

The LVHE model predicts higher peak MPS95 values than the HE model throughout the loading sequence, but its response decays more rapidly during the rest periods. The larger LVHE deformation is consistent with stress relaxation reducing the effective tissue stiffness during loading, whereas the more rapid attenuation of the post-loading response is consistent with viscous dissipation in the LVHE formulation \citep{Upadhyay2022}.

\begin{figure}[t!]
    \centering    
    \includegraphics[width=3.582in]{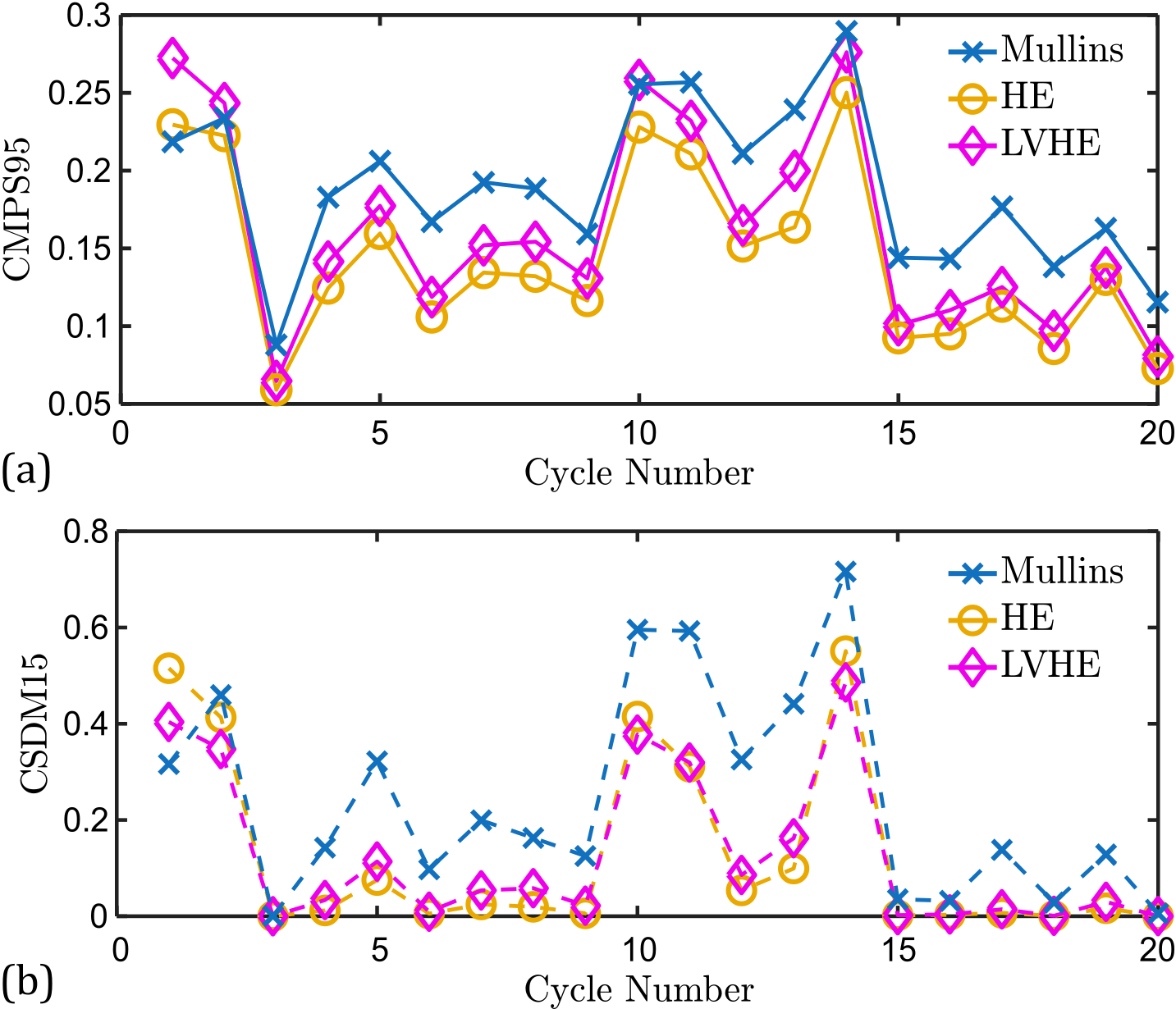}
    \caption{Cycle-wise evolution of tissue-strain-based injury predictors during the twenty-cycle randomized multiaxial loading sequence: (a) 95th-percentile cumulative maximum principal strain (CMPS95) and (b) cumulative strain damage measure with a 0.15 strain threshold (CSDM15). Predictions from the Mullins-based, hyperelastic (HE), and linear visco-hyperelastic (LVHE) models are shown under the same prescribed loading history}
    \label{fig:multiaxTBIP}
\end{figure}

\paragraph{Cycle-wise evolution of CMPS95 and CSDM15}
Figure~\ref{fig:multiaxTBIP} compares the cycle-wise evolution of CMPS95 and CSDM15 predicted by the Mullins-based, HE, and LVHE models during the twenty-cycle multiaxial loading sequence. Because the loading intensity and directionality vary between cycles, neither metric increases monotonically with cycle number. Nevertheless, the relative ordering of the model predictions changes as Mullins-type softening evolves. For CMPS95, the LVHE model predicts the largest values in Cycles 1, 2, and 10, whereas the Mullins-based model predicts the largest values in Cycles 3--9 and 11--20 (Fig.~\ref{fig:multiaxTBIP}(a)). For CSDM15, the Mullins-based model predicts the largest value in every cycle after Cycle 1 (Fig.~\ref{fig:multiaxTBIP}(b)). The LVHE model's CMPS95 predictions are generally greater than the corresponding HE predictions, consistent with the trends observed in the MPS95 histories.

The influence of loading history is particularly evident when comparing Cycles 1 and 14, which have similar kinematic intensities. In Cycle 1, the initially undamaged Mullins-based model predicts lower CMPS95 and CSDM15 values than the HE and LVHE models. By Cycle 14, prior loading has produced substantial cumulative Mullins-type softening, and the Mullins-based model predicts larger values of both metrics than the two damage-free models. Thus, even for cycles with similar kinematic intensity, the predicted brain deformation depends on the prior loading history and the resulting mechanical state of the tissue \citep{Franceschini2006,DeRooij2016}.

\begin{figure}[t!]
    \centering    
    \includegraphics[width=5.148in]{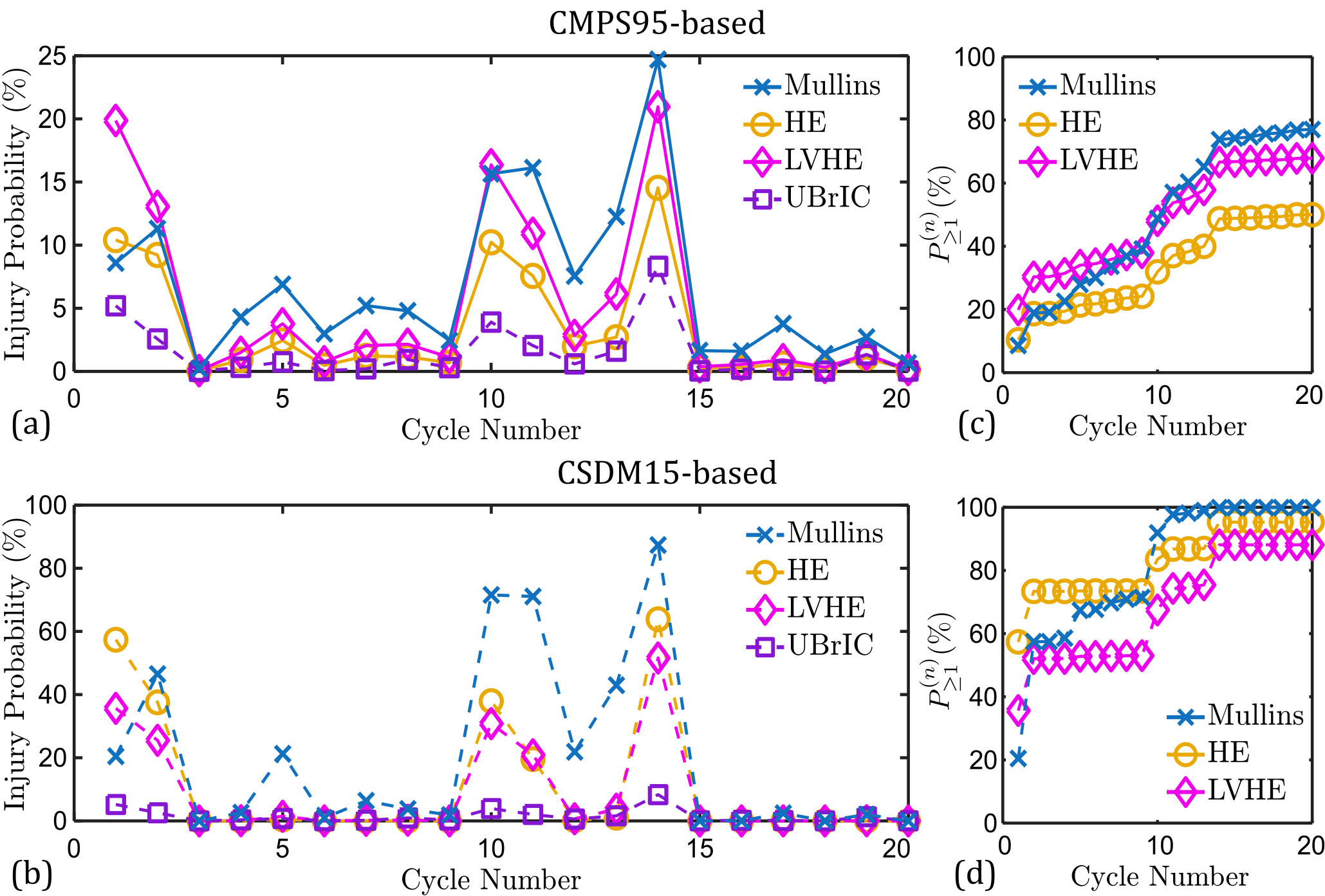}
    \caption{Cycle-wise and sequence-level AIS2 injury probability estimates during the twenty-cycle randomized multiaxial loading sequence. Left column: cycle-wise injury probabilities based on (a) the 95th-percentile cumulative maximum principal strain (CMPS95) and (b) the cumulative strain damage measure with a 0.15 strain threshold (CSDM15). Tissue-strain-based predictions are shown for the Mullins-based, hyperelastic (HE), and linear visco-hyperelastic (LVHE) models, with predictions based on the universal brain injury criterion (UBrIC) included as a kinematics-based reference. Right column: probability of at least one AIS2 injury \((P_{\geq 1}^{(n)})\) occurring through loading cycle \(n\), calculated from the corresponding cycle-wise (c) CMPS95-based and (d) CSDM15-based probabilities under a Bernoulli independence approximation. All cycle-wise probabilities are calculated using the IRFs developed by \citet{Wu2022}}
    \label{fig:multiinjury}
\end{figure}

\paragraph{Accumulation of AIS2 injury probability over the multiaxial sequence}
Figure~\ref{fig:multiinjury} presents cycle-wise and sequence-level estimates of AIS2 (i.e., mTBI/concussion) injury probability during the twenty-cycle randomized multiaxial loading history. The cycle-wise probabilities are obtained from the CMPS95-, CSDM15-, and UBrIC-based IRFs developed by \citet{Wu2022}.

Consistent with the corresponding tissue-strain-based injury metrics, the LVHE model produces the highest CMPS95-based injury probability in Cycles 1, 2, and 10, whereas the Mullins-based model produces the highest probability in Cycles 3--9 and 11--20 (Fig.~\ref{fig:multiinjury}(a)). For the CSDM15-based predictions, the Mullins-based model produces the highest injury probability in Cycles 2--20 (Fig.~\ref{fig:multiinjury}(b)). During the more kinematically intense portions of the sequence, including Cycles 1--2 and 10--14, the CSDM15-based IRF generally predicts higher probabilities than the CMPS95-based IRF. The UBrIC-based probabilities are generally lower in magnitude than the tissue-strain-based estimates, although absolute differences among IRFs should be interpreted cautiously because the functions are predictor-specific and may depend on the computational models and datasets used for their calibration.

The UBrIC-, HE-, and LVHE-based estimates generally follow the same cycle-to-cycle pattern, increasing and decreasing with the intensity of the current loading pulse. The Mullins-based predictions can deviate from this pattern because they additionally depend on the prior deformation history and evolving mechanical state of the brain. For example, the UBrIC-based probability decreases from Cycle 1 to Cycle 2 and from Cycle 10 to Cycle 11, indicating lower kinematic severity in the latter cycle of each pair. In contrast, the CMPS95-based probability predicted by the Mullins-based model increases across both pairs. Thus, progressive tissue softening can increase predicted injury probability even when the intensity of the current head kinematics decreases. As the loading sequence progresses, kinematics-only criteria and damage-free constitutive models may therefore increasingly underestimate injury probability relative to a damage-aware model.

To estimate the probability that at least one AIS2 injury occurs during the loading sequence, each loading cycle is treated as a Bernoulli trial with cycle-specific injury probability \(p_j\), obtained from Eq.~\eqref{eq:IRFprob} using the corresponding CMPS95 or CSDM15 value for Cycle \(j\). Assuming that injury occurrences across loading cycles are independent conditional on these cycle-specific probabilities, the probability of at least one injury occurring through Cycle \(n\) is \citep{blitzstein2019}
\begin{equation}
    \label{eq:cumulativeinjury}
    P_{\geq 1}^{(n)}
    =
    1-
    \prod_{j=1}^{n}
    \left(
    1-p_j
    \right).
\end{equation}
Equation~\eqref{eq:cumulativeinjury} is evaluated separately using the CMPS95- and CSDM15-based probabilities predicted by each computational model. Importantly, the independence approximation pertains to the occurrence of injury across loading cycles, not to the underlying mechanical response. In the Mullins-based model, \(p_j\) remains history dependent because softening accumulated during preceding cycles alters the injury metrics entering Eq.~\eqref{eq:IRFprob}.

From Figs.~\ref{fig:multiinjury}(c,d), the HE and LVHE models initially predict higher probabilities of at least one injury because they produce larger cycle-wise probabilities during the beginning of the sequence. As Mullins-type softening accumulates, however, the sequence-level probability predicted by the Mullins-based model increases more rapidly and surpasses the LVHE prediction by Cycle~9 for the CMPS95-based IRF and the HE prediction by Cycle~10 for the CSDM15-based IRF. By the end of the twenty-cycle sequence, the CMPS95-based probabilities of at least one injury are \(77.0\%\), \(50.0\%\), and \(67.9\%\) for the Mullins-based, HE, and LVHE models, respectively. The corresponding CSDM15-based probabilities are \(99.9\%\), \(95.3\%\), and \(88.1\%\).

These results demonstrate that loading-history-dependent softening can alter not only the predicted injury probability associated with an individual loading cycle, but also the estimated probability of injury accumulated across an MMA-derived sequence of repeated impacts. The mechanical consequences of a given loading pulse therefore depend on both its instantaneous kinematic characteristics and the evolving mechanical state produced by the preceding loading history.

\subsection{Brain Geometry Shapes Regional and Cortical-Fold Responses}
The brain's complex geometry governs how shear waves generated by impulsive head motion propagate and reflect through the tissue, producing deformation fields that evolve non-uniformly in space and time and vary with loading direction. Consequently, although the seven parenchymal brain substructures are assigned a common brain-bulk material response, their distinct anatomical locations within this complex geometry can lead to different deformation histories and patterns of Mullins-type softening. To examine this spatial variability, region-wise CMPS95 evolution is analyzed in the Mullins-based single-axis simulations, and cycle-wise CMPS is compared across 75 manually selected pairs of gyral and sulcal grey-matter elements. 

\begin{figure}[hbt!]
    \centering    
    \includegraphics[width=2.466in]{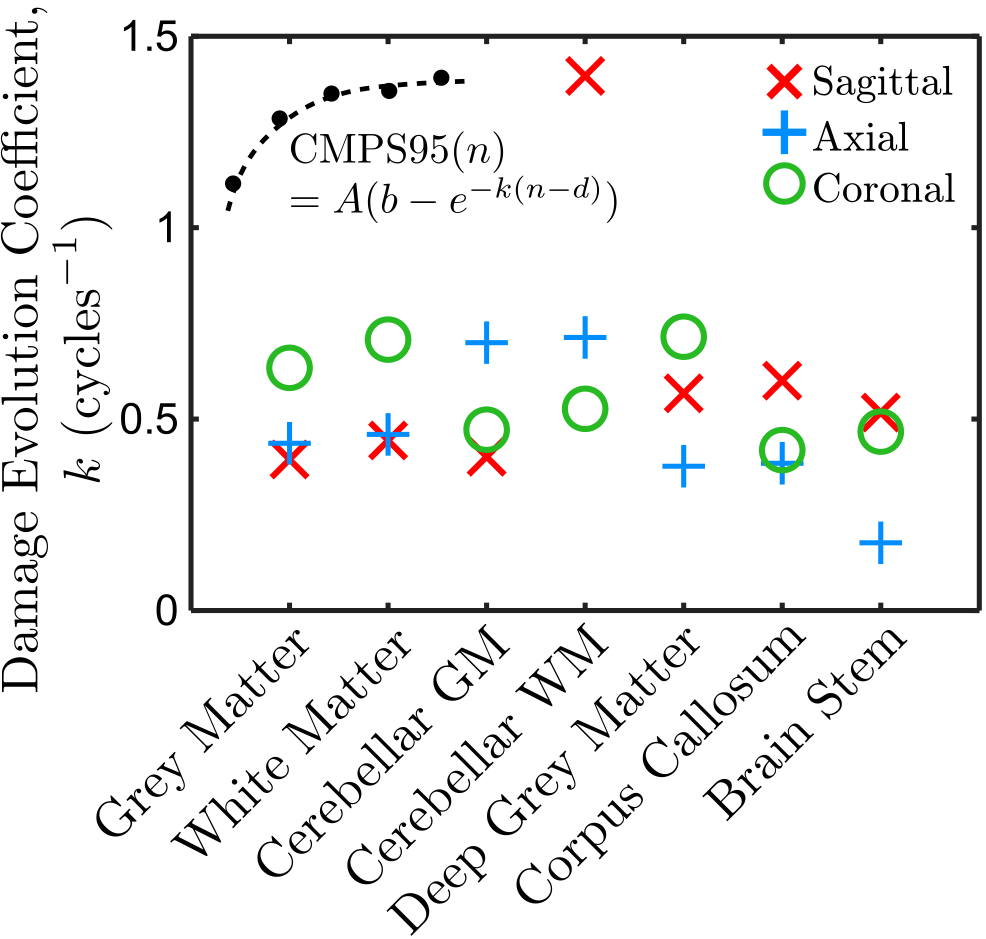}
    \caption{Regional damage evolution coefficients \(k\) obtained from exponential fits to the 95th-percentile cumulative maximum principal strain (CMPS95) versus loading-cycle responses of seven parenchymal brain substructures under sagittal, axial, and coronal single-axis rotation. Larger values of \(k\) indicate a more rapid approach toward the fitted asymptotic CMPS95 value and, consequently, greater damage-mediated deformation amplification during the earlier loading cycles. The inset shows the exponential relation used to determine \(k\)}
    \label{fig:regions}
\end{figure}

\paragraph{Deformation amplification across brain substructures}
Figure~\ref{fig:regions} compares the regional damage evolution coefficients \(k\) obtained by fitting the CMPS95-versus-cycle responses of the seven parenchymal brain substructures to the exponential relation defined in Eq. (\ref{eq:exp_k}). Within this relation, \(k\), with units of \(\mathrm{cycles}^{-1}\), controls the rate at which regional CMPS95 approaches its fitted asymptotic value. A larger \(k\) therefore indicates that a greater fraction of the total damage-mediated deformation amplification occurs during the earlier loading cycles. The term damage evolution coefficient is used because \(k\) summarizes the cycle-wise evolution of the regional deformation response as Mullins-type softening accumulates; it characterizes the rate of deformation amplification rather than the absolute CMPS95 magnitude and is not itself the local Mullins damage variable.

The damage evolution coefficients vary with both brain substructure and loading direction. Under sagittal rotation, the regional \(k\) values are comparatively closely grouped, with the notable exception of the cerebellar white matter, which exhibits a markedly larger \(k\) than all other substructures and loading directions shown in Fig.~\ref{fig:regions}. Axial rotation produces relatively large coefficients in the cerebellar grey and white matter and a substantially smaller coefficient in the brain stem. Under coronal rotation, comparatively large coefficients are observed in the grey matter, white matter, and deep grey matter. One direction-independent trend is that the white matter exhibits a larger \(k\) than the grey matter under all three rotation directions. Beyond this trend, however, the regional ordering changes substantially with loading direction, and no general direction-independent hierarchy of damage-mediated deformation amplification is evident. Because the seven substructures share the same brain-bulk constitutive properties, these differences arise from the interaction among the brain's complex geometry, the anatomical location of each substructure, and the applied loading direction.

\begin{figure}[hbt!]
    \centering    
    \includegraphics[width=5.105in]{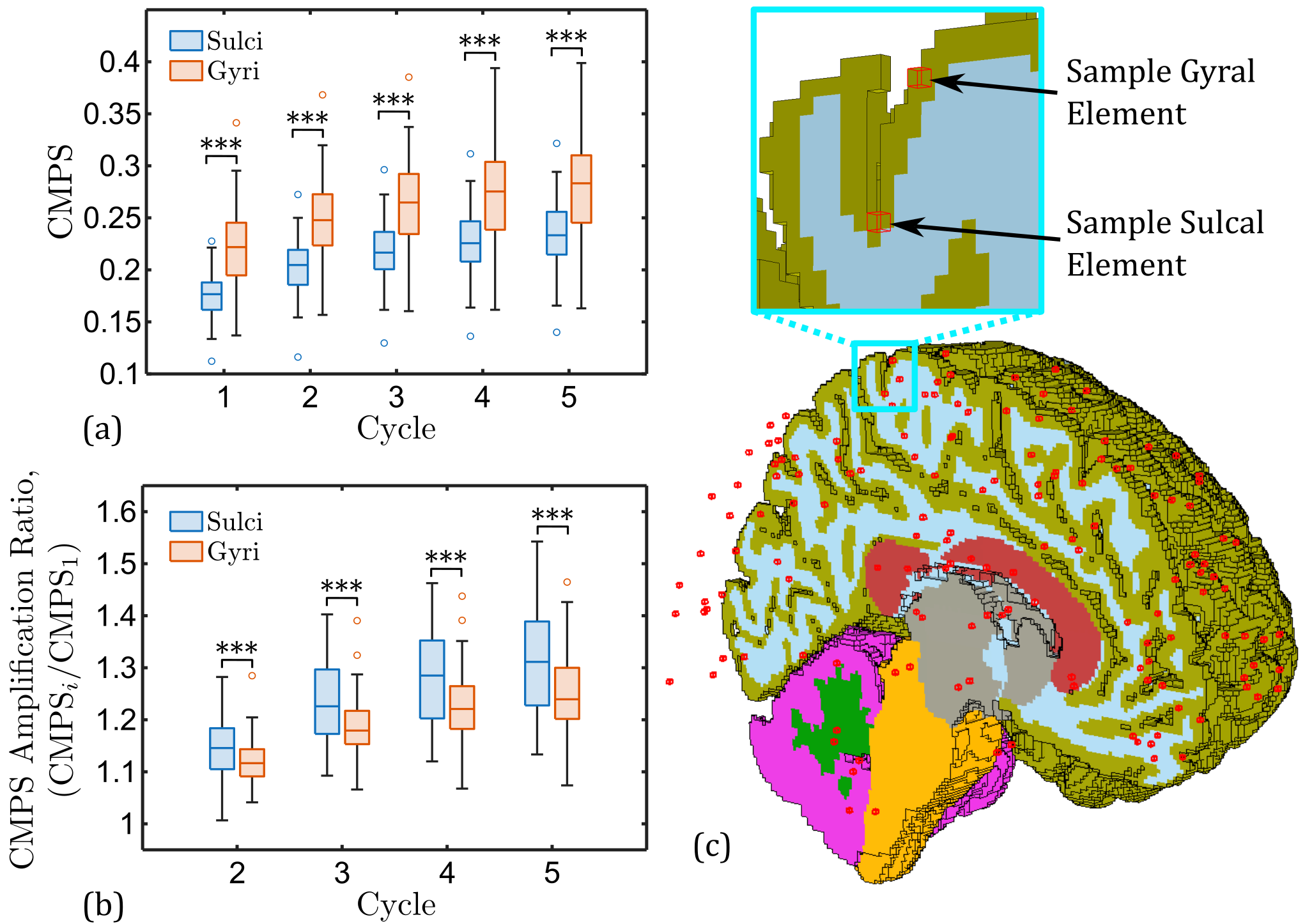}
    \caption{Comparison of cumulative maximum principal strain (CMPS) at 75 spatially paired gyral and sulcal grey-matter elements. Results in panels (a) and (b) are averaged across the sagittal, axial, and coronal single-axis simulations. (a) Box plots of cycle-wise absolute CMPS for the gyral and sulcal elements. (b) Box plots of the CMPS amplification ratio, \(\mathrm{CMPS}_{i}/\mathrm{CMPS}_{1}\), for Cycles 2--5; Cycle 1 is omitted because the ratio is identically unity. In each box plot, the box denotes the interquartile range, the central line denotes the median, the whiskers extend to the most extreme non-outlier values, and individual markers denote outliers. Statistical comparisons are performed using paired \(t\)-tests; \(***\) denotes \(p<0.001\). (c) Spatial distribution of the 75 gyral and 75 sulcal sampling locations in the computational head model, with the inset illustrating one representative paired selection}
    \label{fig:gyrisulci}
\end{figure}

\paragraph{Deformation amplification in cerebral gyri and sulci}
Figure~\ref{fig:gyrisulci} compares the cycle-wise CMPS responses of 75 spatially paired gyral and sulcal grey-matter elements. The pairs are selected approximately uniformly across the cerebral cortex near sulcal fissures, with each pair comprising one element located on a gyrus and one nearby element located within the adjacent sulcus. The analysis therefore includes 75 gyral and 75 sulcal sampling locations. For each element, absolute CMPS and the CMPS amplification ratio, \(\mathrm{CMPS}_{i}/\mathrm{CMPS}_{1}\), are calculated separately for the sagittal, axial, and coronal simulations and then averaged across the three loading directions. The direction-averaged distributions are shown in Figs.~\ref{fig:gyrisulci}(a,b); the same qualitative gyral--sulcal differences are also observed in the individual direction-specific datasets.

As shown in Fig.~\ref{fig:gyrisulci}(a), the gyral elements exhibit significantly higher absolute CMPS values than their paired sulcal elements in every loading cycle (\(p<0.001\)). Conversely, the sulcal elements exhibit significantly larger CMPS amplification ratios during Cycles 2--5 (\(p<0.001\); Fig.~\ref{fig:gyrisulci}(b)), indicating greater proportional amplification relative to their respective Cycle 1 responses. However, no statistically significant difference is detected between the gyral and sulcal elements in their absolute cycle-to-cycle CMPS increments, \(\mathrm{CMPS}_{i}-\mathrm{CMPS}_{i-1}\). Because the amplification ratio is normalized by \(\mathrm{CMPS}_{1}\), the lower Cycle 1 CMPS values at the sulcal elements likely contribute to their larger normalized ratios. The normalized results therefore should not, by themselves, be interpreted as evidence that sulcal elements experience larger absolute increases in deformation. Rather, the present results show that gyral elements experience greater absolute deformation, whereas sulcal elements exhibit greater deformation amplification relative to their lower initial response.

\section{Discussion}\label{chap:discussion}

The present study demonstrates that short-term Mullins-type softening can make the mechanical response of the brain during repeated head loading depend on both the current head kinematics and the tissue's prior deformation history. Under controlled single-axis loading, identical kinematic inputs produced progressively greater tissue deformation in the Mullins-based model, whereas the damage-free HE response remained nearly unchanged. This history dependence persisted under the randomized multiaxial sequence despite variations in loading intensity, directionality, and duration. The resulting amplification affected strain and strain rate metrics, tissue-strain-based injury probability estimates, and deformation across parenchymal brain substructures and cortical folds. These findings provide a first organ-scale demonstration that treating each head loading as an isolated event may omit an important source of variability in repeated-loading scenarios.

\paragraph{History-dependent deformation under repeated loading}

The single-axis simulations provide the clearest demonstration of damage-mediated deformation amplification because the same loading pulse is prescribed in every cycle. The largest cycle-to-cycle increases in CMPS95 and CSDM15 generally occur between Cycles 1 and 2, followed by progressively smaller increases. This rapid-then-gradual evolution is consistent with the bounded Ogden--Roxburgh formulation and with the pronounced preconditioning observed during the initial cycles of brain-tissue experiments \citep{Franceschini2006,DeRooij2016,Budday2020}. The first loading establishes a maximum distortional strain energy history at each material point, and subsequent unloading and reloading below that maximum follow a softened response. As the allowable softening approaches its limiting magnitude, additional identical cycles produce diminishing changes in global deformation.

The multiaxial results show that this history dependence persists when successive loadings are not identical. Global strain metrics do not increase monotonically because each response is governed jointly by the current loading pulse and the mechanical state established by preceding cycles. Random variations in directionality also expose a broader and spatially evolving set of brain-bulk elements to deformation histories that produce softening. Because the maximum previously attained distortional strain energy is retained locally, softening generated at an element during one cycle influences its response in later cycles. The multiaxial response is therefore cumulative both temporally, through the number and ordering of loadings, and spatially, through the evolving distribution of tissue with a prior softening history.

The Mullins-based model is nevertheless not expected to predict greater deformation than the HE or LVHE models during every loading. While reloading below the historical maximum, \(W_{\mathrm{dev}}<W_{\mathrm{dev,max}}\), the local response is softened and \(\eta<1\). When the current distortional strain energy reaches the historical maximum, \(W_{\mathrm{dev}}=W_{\mathrm{dev,max}}\), the response returns to the intact loading envelope and \(\eta=1\); further loading then establishes a new maximum. Mullins-type effects should therefore be most evident when a subsequent loading produces deformation within or near the previously experienced range. This helps explain why the damage-free LVHE model predicts the largest deformation during the initial cycles, before substantial Mullins-type softening has accumulated, and may remain dominant during occasional later cycles when its rate-dependent response produces greater deformation than the amplification caused by prior softening. In contrast, the Mullins-based model generally predicts greater deformation during later loadings of similar or lower severity. Because the global response depends on the spatial distribution of prior softening, attainment of \(\eta=1\) at newly maximized elements does not imply that the brain as a whole responds as an intact material.

Mullins-type softening also increases the strain rate metrics examined in the single-axis simulations. Because the formulation contains no intrinsic strain rate dependence, these increases arise from changes in the transient deformation response caused by tissue softening rather than from an explicitly modeled rate-dependent damage mechanism. Nevertheless, \citet{Elkin2007} reported that rate-dependent embrittlement of neurons may lower the strain threshold for tissue injury at higher strain rates. The simultaneous increases in strain and strain rate may therefore have important consequences during successive loadings and demonstrate the need for experimental characterization of rate-dependent damage evolution in brain tissue.

The term damage is used here in its constitutive-mechanics sense and should not be interpreted as a direct measure of irreversible biological injury. Cyclic softening of brain tissue can recover after sufficient rest, and mechanisms such as fluid redistribution and microstructural reorganization may contribute to the measured response \citep{Budday2020}. The Ogden--Roxburgh variable \(\eta\) therefore represents a short-term, history-dependent reduction in mechanical resistance, not the remaining functional integrity of neural tissue.

\paragraph{Implications for injury prediction}

The divergence among the Mullins-based, HE, LVHE, and UBrIC-based injury estimates illustrates the consequences of loading history. UBrIC depends only on the prescribed kinematics of the current cycle, while the HE and LVHE models retain the same brain-bulk constitutive properties throughout the sequence. Their injury estimates therefore tend to rise and fall with the severity of the current loading pulse. The Mullins-based predictions additionally depend on the evolving material state of the brain, such that kinematically similar loadings can produce markedly different tissue deformations depending on whether the tissue is initially intact or has been softened by previous loading.

This distinction is especially evident when the Mullins-based and UBrIC predictions change in opposite directions. From Cycle 1 to Cycle 2 and from Cycle 10 to Cycle 11 in the multiaxial simulation, the UBrIC-based probability decreases, indicating lower kinematic severity in the latter cycle of each pair, whereas the Mullins-based CMPS95 injury probability increases. Thus, a less intense loading pulse can produce a greater tissue-strain-based injury estimate when applied after sufficient prior softening. The results demonstrate a dual dependence on current loading and prior mechanical history, rather than diminishing the importance of current kinematics. Kinematics-only criteria and intact constitutive models may consequently underestimate tissue deformation and associated injury probability during short-term repeated loading. These observations provide mechanistic support for the suggestion by \citet{Carlsen2021} that whether a head loading occurs as a first isolated event or within a repeated loading history may contribute to uncertainty in injury prediction, including real-world observations that nominally similar kinematics can produce injury in one instance but not another.

The probability of at least one AIS2 injury over the MMA-derived sequence further illustrates how cycle-wise differences accumulate. By Cycle 20, the CMPS95-based sequence-level probabilities are \(77.0\%\), \(50.0\%\), and \(67.9\%\) for the Mullins-based, HE, and LVHE models, respectively. The corresponding CSDM15-based values are \(99.9\%\), \(95.3\%\), and \(88.1\%\). Although the intact models initially produce larger sequence-level estimates, the Mullins-based CMPS95 prediction surpasses the LVHE prediction by Cycle 9, while the Mullins-based CSDM15 prediction surpasses the HE prediction by Cycle 10. The CMPS95-based IRF provides greater separation among the models by the end of the sequence, whereas the CSDM15-based probabilities approach saturation and become less discriminating.

These sequence-level values are comparative model estimates rather than validated predictions of cumulative clinical risk. The Bernoulli calculation assumes conditionally independent injury events with cycle-specific probabilities \(p_j\), even though the Mullins-based mechanical response is intentionally history dependent. It also does not represent biological changes following an actual injury that could modify susceptibility to later impacts. Nevertheless, it provides a consistent comparison of how different constitutive descriptions translate the same loading sequence into the probability of at least one injury.

The absolute probabilities are additionally uncertain because the IRFs were calibrated using other computational head models. Tissue-strain metrics depend on model geometry, material properties, and interface conditions, so transferring an IRF between models introduces model-form uncertainty \citep{Gabler2018a,Wu2022,Ji2022}. The probabilities are therefore most informative as illustrations of how predicted risk changes when identical IRFs are applied to intact and history-dependent constitutive models under the same loading conditions.

The differences between the CMPS95- and CSDM15-based predictions also demonstrate sensitivity to the selected injury metric. Although CSDM15 has shown favorable injury-prediction performance in previous datasets \citep{Gabler2018a}, the present model reproduced the experimentally measured evolution of MPS percentiles more accurately than that of MPS volume fractions during validation. This may support greater confidence in the relative CMPS95 trends than in the corresponding CSDM15 trends, but it does not remove uncertainty in the absolute CMPS95-based probabilities because the associated IRF was not calibrated specifically for the present model.

\paragraph{Spatial heterogeneity across brain substructures and cortical folds}

The regional damage evolution coefficients demonstrate that deformation amplification is spatially heterogeneous and depends strongly on loading direction. Because the seven parenchymal brain substructures are assigned the same brain-bulk material properties, differences in \(k\) arise from the interaction among loading direction, anatomical location, and the complex geometry governing shear-wave propagation and reflection through the brain. White matter exhibits a larger \(k\) than grey matter under all three rotation directions, but the ordering of the remaining substructures changes substantially with direction. The results therefore do not support a general direction-independent hierarchy of regional deformation amplification.

This direction dependence is consistent with experimental and computational studies showing that different anatomical planes of head motion can produce distinct regional injury patterns \citep{Gennarelli1987,Kleiven2007}. For example, \citet{Gennarelli1987} observed brain-stem axonal damage under coronal loading that was not present under sagittal loading. The present results suggest that these relationships may be further modified during repeated loading by nonuniform softening: a region's later response may depend not only on the direction of the current loading, but also on whether it was preferentially deformed during previous events.

The gyral--sulcal analysis reveals different trends for absolute and normalized deformation. The sampled gyral elements experience significantly greater absolute CMPS than their paired sulcal elements, whereas the sulcal elements exhibit larger amplification ratios relative to their respective Cycle 1 responses. No significant difference is detected in the absolute cycle-to-cycle CMPS increments, however, and the lower Cycle 1 CMPS values at the sulcal elements likely contribute to their larger normalized ratios. The ratios therefore do not by themselves demonstrate greater absolute deformation growth or more extensive Mullins-type softening in sulcal tissue.

The higher absolute gyral CMPS differs from previous computational studies reporting elevated deformation at sulcal depths \citep{Ghajari2017,Fagan2020}. Sulcal response is highly sensitive to the representation of cortical geometry, cerebrospinal fluid, brain--CSF interaction, and the simulated time interval. For example, \citet{Fagan2020} used an Arbitrary Lagrangian--Eulerian formulation in a two-dimensional head model, with cerebrospinal fluid represented as an Eulerian fluid, and observed elevated shear strains within the sulci. Their simulation was limited to the first \(1~\mathrm{ms}\) after impact, considerably shorter than the response windows on the order of tens of milliseconds recommended or employed in previous head-modeling studies \citep{Takhounts2008,Carlsen2021,Ji2022}. The authors noted that features of the subsequent response may therefore have been missed. Fluid--structure interactions within narrow sulcal spaces remain a complex modeling problem and an important area for continued investigation.

Establishing that sulcal tissue experiences disproportionately greater cyclic softening would require direct examination of the internal damage or history variables. Relating such constitutive softening to increased injury propensity would additionally require experimental evidence connecting the chosen variable to tissue function. \citet{Nol2019} investigated an internal damage variable in an idealized two-dimensional sulcal geometry and related it directly to tissue functional integrity using guinea-pig optic-nerve injury data from \citet{Bain2000}. This approach provides a promising framework for studying cyclic response across brain substructures and gyral and sulcal locations. In the present study, however, there is insufficient evidence to relate the Ogden--Roxburgh parameter \(\eta\) directly to remaining neuronal function, axonal integrity, or tissue injury severity.

\paragraph{Current scope and future directions}

The current constitutive description is necessarily constrained by the available cyclic brain-tissue data. Experimental measurements are not yet available to determine how Mullins-type softening varies with strain rate, anatomical region, deformation mode, loading amplitude, or inter-cycle recovery time. The rate-independent Ogden--Roxburgh formulation and common Mullins parameters assigned to the seven parenchymal brain substructures therefore reflect the available calibration evidence \citep{Franceschini2006}, rather than an assumption that softening is inherently rate or region independent.

Systematic repeated-loading experiments should span impact-relevant strain rates, deformation modes, anatomical regions, loading amplitudes, and controlled rest intervals while measuring viscoelastic relaxation, cyclic softening, and recovery. Brain tissue is known to exhibit nonlinear viscoelasticity, stress relaxation, and viscous dissipation \citep{DeRooij2016,Atsumi2018,Upadhyay2022}, and recovery of cyclic softening has been reported over longer rest periods \citep{Budday2020}; however, their coupled influence during short-term repeated loading remains insufficiently characterized. Such data would enable calibration of nonlinear visco-hyperelastic damage models incorporating rate-dependent softening, partial recovery, and region-specific behavior. Direct full-field measurements under repeated head motion would also allow validation of the cycle-to-cycle deformation amplification predicted here.

The computational analysis is limited to one subject-specific anatomy and one statistically generated twenty-cycle MMA-derived loading sequence. These choices enable controlled comparison among the Mullins-based, HE, and LVHE models, but multiple anatomies and ensembles of randomized histories are needed to quantify sensitivity to anatomical and sequence variability. The transferred IRFs and Bernoulli independence approximation also limit the generality and absolute interpretation of the sequence-level probabilities, but do not alter the controlled finding that intact and history-dependent constitutive models can produce progressively divergent predictions under the same prescribed loading history.

Further extensions may include atlas-based regional parcellation, region-specific strained-volume analyses, and improved treatment of cortical sulci and brain--CSF interactions. Diffusion-tensor-imaging-based fiber orientations or explicitly embedded axonal tracts would additionally enable evaluation of axonal fiber strain, an important predictor of diffuse axonal injury \citep{Wu2022,Upadhyay2022,Gerber346700}. Together, these developments would link loading-history-dependent tissue softening more directly to regional deformation, axonal injury, and repeated-impact risk.

\section{Conclusion}\label{chap:conclusion}

This study implemented an Ogden--Roxburgh Mullins damage formulation in a high-fidelity computational head model to investigate how short-term tissue softening affects brain deformation and injury-risk prediction during repeated head loading. The model was validated against subject-specific full-field strain measurements obtained from tagged MRI and then subjected to controlled five-cycle single-axis loading and a randomized twenty-cycle multiaxial loading sequence. Under identical single-axis loading, Mullins-type softening progressively increased strain and strain rate metrics relative to the damage-free HE model. The largest cycle-to-cycle increase generally occurred between Cycles 1 and 2, followed by diminishing increases as the bounded softening response approached a plateau. Consequently, tissue-strain-based injury probabilities increased from cycle to cycle and progressively diverged from the unchanged kinematics-based predictions. The multiaxial simulations demonstrated that this history dependence persists when loading intensity, directionality, and duration vary between cycles. The Mullins-based model differed most strongly from the damage-free HE and LVHE models during later loadings of similar or lower severity, and kinematically similar cycles produced different mechanical responses depending on the previously accumulated softening. In some cases, the Mullins-based injury probability increased even when UBrIC indicated a reduction in current loading severity. Regional deformation amplification also depended strongly on loading direction, with no general direction-independent hierarchy among the parenchymal brain substructures. Gyral elements exhibited greater absolute CMPS than paired sulcal elements, whereas sulcal elements exhibited greater amplification relative to their lower Cycle 1 responses. Overall, these findings demonstrate that current head kinematics alone may not uniquely determine brain deformation or injury-risk estimates during repeated loading; the evolving mechanical state produced by prior deformation can also play an important role.

\backmatter

\bmhead{Supplementary Information}
Supplementary material accompanies this manuscript.

\bmhead{Acknowledgements}
This research was conducted in part using high performance computational resources provided by Louisiana State University (http://www.hpc.lsu.edu).

\bmhead{Author contribution}
Carson Cooper: Conceptualization, Methodology, Software, Validation, Formal analysis, Visualization, Writing --- original draft, Writing --- review \& editing.
Anu Tripathi: Methodology, Software, Visualization, Writing --- review \& editing.
Genevieve Palardy: Methodology, Resources, Writing --- review \& editing, Supervision, Project administration, Funding acquisition.
Kshitiz Upadhyay: Conceptualization, Methodology, Validation, Resources, Writing --- original draft, Writing --- review \& editing, Supervision, Project administration, Funding acquisition

\section*{Declarations}

\bmhead{Funding}
This material is based upon work supported by the National Science Foundation under Grant No. 2623467, awarded to the University of Minnesota, and by the Our Lady of the Lake (OLOL) Health - Collaboration in Action Program (CAP), administered through the LSU Foundation.

\bmhead{Conflict of interest}
The authors declare no competing interests. 

\bmhead{Data availability} 
All data generated or analyzed during this study are
included in the manuscript and the supplementary files. The raw data
will be made available upon reasonable request.


\bibliography{cas-refs}

@article{Upadhyay2022,
   author = {Kshitiz Upadhyay and Ahmed Alshareef and Andrew K. Knutsen and Curtis L. Johnson and Aaron Carass and Philip V. Bayly and Dzung L. Pham and Jerry L. Prince and K. T. Ramesh},
   doi = {10.1098/rsif.2022.0561},
   issn = {17425662},
   issue = {195},
   journal = {Journal of the Royal Society Interface},
   month = {10},
   publisher = {Royal Society Publishing},
   title = {Development and validation of subject-specific 3D human head models based on a nonlinear visco-hyperelastic constitutive framework},
   volume = {19},
   year = {2022}
}

@article{Upadhyay2022b,
author = {Upadhyay, Kshitiz and Giovanis, Dimitris G. and Alshareef, Ahmed and Knutsen, Andrew K. and Johnson, Curtis L. and Carass, Aaron and Bayly, Philip V. and Shields, Michael D. and Ramesh, K.T.},
doi = {10.1016/j.cma.2022.115108},
issn = {00457825},
journal = {Computer Methods in Applied Mechanics and Engineering},
month = {aug},
pages = {115108},
publisher = {Elsevier B.V.},
title = {{Data-driven uncertainty quantification in computational human head models}},
volume = {398},
year = {2022}
}

@article{Ogden1998,
   author = {R W Ogden and D G Roxburgh},
   issue = {1999},
   journal = {Proceedings of the Royal Society of London, Series A},
   month = {08},
   doi = {https://doi.org/10.1098/rspa.1999.0431},
   pages = {2861-2877},
   title = {A pseudo-elastic model for the Mullins effect in filled rubber},
   volume = {455},
   url = {http://royalsocietypublishing.org/rspa/article-pdf/455/1988/2861/634690/rspa.1999.0431.pdf},
   year = {1999}
}

@manual{abaqus2004,
  title        = {ABAQUS Theory Manual, Version 6.5},
  author       = {{Dassault Syst\`emes}},
  organization = {Dassault Syst\`emes Simulia Corp.},
  year         = {2004},
  address      = {Providence, RI}
}

@book{Holzapfel2000,
   author = {Gerhard A. Holzapfel},
   publisher = {John Wiley \& Sons Ltd},
   title = {Nonlinear Solid Mechanics},
   year = {2000},
   address = {Chichester, UK}
}

@article{DeRooij2016,
   author = {Rijk De Rooij and Ellen Kuhl},
   doi = {10.1115/1.4032436},
   issn = {00036900},
   issue = {1},
   journal = {Applied Mechanics Reviews},
   month = {1},
   publisher = {American Society of Mechanical Engineers (ASME)},
   title = {Constitutive Modeling of Brain Tissue: Current Perspectives},
   volume = {68},
   year = {2016}
}

@book{Rudnicki2015,
   author = {John W Rudnicki},
   edition = {1},
   publisher = {John Wiley \& Sons, Ltd},
   title = {Fundamentals of Continuum Mechanics},
   address = {Chichester, UK},
   year = {2015}
}

@article{Franceschini2006,
   author = {G. Franceschini and D. Bigoni and P. Regitnig and G. A. Holzapfel},
   doi = {10.1016/j.jmps.2006.05.004},
   issn = {00225096},
   issue = {12},
   journal = {Journal of the Mechanics and Physics of Solids},
   month = {12},
   pages = {2592-2620},
   title = {Brain tissue deforms similarly to filled elastomers and follows consolidation theory},
   volume = {54},
   year = {2006}
}

@article{Ericksen1954,
   author = {J L Ericksen and R S Rivlin},
   journal = {Journal of Rational Mechanics and Analysis},
   pages = {281-301},
   title = {Large Elastic Deformations of Homogeneous Anisotropic Materials},
   volume = {3},
   year = {1954}
}

@article{Madhukar2019,
   author = {Amit Madhukar and Martin Ostoja-Starzewski},
   doi = {10.1007/s10439-019-02205-4},
   issn = {15739686},
   issue = {9},
   journal = {Annals of Biomedical Engineering},
   month = {9},
   pages = {1832-1854},
   pmid = {30693442},
   publisher = {Springer New York LLC},
   title = {Finite Element Methods in Human Head Impact Simulations: A Review},
   volume = {47},
   year = {2019}
}

@article{Budday2020,
   author = {Silvia Budday and Timothy C. Ovaert and Gerhard A. Holzapfel and Paul Steinmann and Ellen Kuhl},
   doi = {10.1007/s11831-019-09352-w},
   issn = {18861784},
   issue = {4},
   journal = {Archives of Computational Methods in Engineering},
   month = {9},
   pages = {1187-1230},
   publisher = {Springer Science and Business Media B.V.},
   title = {Fifty Shades of Brain: A Review on the Mechanical Testing and Modeling of Brain Tissue},
   volume = {27},
   year = {2020}
}

@article{Mullins1947,
   author = {L Mullins},
   issue = {12},
   journal = {Journal of Rubber Research},
   month = {12},
   pages = {275-289},
   title = {Effect of Stretching on the Properties of Rubber},
   volume = {16},
   year = {1947}
}

@article{Fischl2004,
   author = {Bruce Fischl and André Van Der Kouwe and Christophe Destrieux and Eric Halgren and Florent Ségonne and David H. Salat and Evelina Busa and Larry J. Seidman and Jill Goldstein and David Kennedy and Verne Caviness and Nikos Makris and Bruce Rosen and Anders M. Dale},
   doi = {10.1093/cercor/bhg087},
   issn = {10473211},
   issue = {1},
   journal = {Cerebral Cortex},
   month = {1},
   pages = {11-22},
   pmid = {14654453},
   title = {Automatically Parcellating the Human Cerebral Cortex},
   volume = {14},
   year = {2004}
}

@article{Fischl2012,
   author = {Bruce Fischl},
   doi = {10.1016/j.neuroimage.2012.01.021},
   issn = {10538119},
   issue = {2},
   journal = {NeuroImage},
   month = {8},
   pages = {774-781},
   pmid = {22248573},
   title = {FreeSurfer},
   volume = {62},
   year = {2012}
}

@article{Sgonne2004,
   author = {F. Ségonne and A. M. Dale and E. Busa and M. Glessner and D. Salat and H. K. Hahn and B. Fischl},
   doi = {10.1016/j.neuroimage.2004.03.032},
   issn = {10538119},
   issue = {3},
   journal = {NeuroImage},
   month = {7},
   pages = {1060-1075},
   pmid = {15219578},
   title = {A hybrid approach to the skull stripping problem in MRI},
   volume = {22},
   year = {2004}
}

@ARTICLE{py06nimg,
  author = {Paul A. Yushkevich and Joseph Piven and Cody Hazlett, Heather and
    Gimpel Smith, Rachel and Sean Ho and James C. Gee and Guido Gerig},
  title = {User-Guided {3D} Active Contour Segmentation of
    Anatomical Structures: Significantly Improved Efficiency and Reliability},
  journal = {Neuroimage},
  doi = {10.1016/j.neuroimage.2006.01.015},
  year = {2006},
  volume = {31},
  number = {3},
  pages = {1116--1128},
}

@article{Shiino2017,
   author = {Akihiko Shiino and Yen Wei Chen and Kenji Tanigaki and Atsushi Yamada and Piers Vigers and Toshiyuki Watanabe and Ikuo Tooyama and Ichiro Akiguchi},
   doi = {10.1038/srep39818},
   issn = {20452322},
   journal = {Scientific Reports},
   month = {1},
   pmid = {28045130},
   publisher = {Nature Publishing Group},
   title = {Sex-related difference in human white matter volumes studied: Inspection of the corpus callosum and other white matter by VBM},
   volume = {7},
   year = {2017}
}

@Inbook{Kikinis2014,
author="Kikinis, Ron
and Pieper, Steve D.
and Vosburgh, Kirby G.",
title="3D Slicer: A Platform for Subject-Specific Image Analysis, Visualization, and Clinical Support",
bookTitle="Intraoperative Imaging and Image-Guided Therapy",
year="2014",
publisher="Springer New York",
address="New York, NY",
pages="277--289",
isbn="978-1-4614-7657-3",
doi="10.1007/978-1-4614-7657-3_19"
}

@inproceedings{Glaister2017a,
   author = {Jeffrey Glaister and Aaron Carass and Dzung L. Pham and John A. Butman and Jerry L. Prince},
   address = {Quebec City, QC},
   doi = {10.1007/978-3-319-66182-7},
   isbn = {978-3-319-66181-0},
   booktitle = {Med Image Comput Comput Assist Interv},
   month = {9},
   publisher = {Springer International Publishing},
   title = {Falx Cerebri Segmentation via Multi-atlas Boundary Fusion},
   year = {2017}
}

@inproceedings{Glaister2017b,
   author = {Jeffrey Glaister and Aaron Carass and Dzung L. Pham and John A. Butman and Jerry L. Prince},
   address = {Orlando, FL},
   doi = {10.1117/12.2255640},
   isbn = {9781510607194},
   issn = {16057422},
   booktitle = {Proc SPIE Int Soc Opt Eng},
   month = {3},
   publisher = {SPIE},
   title = {Automatic falx cerebri and tentorium cerebelli segmentation from magnetic resonance images},
   volume = {10137},
   year = {2017}
}

@article{Ghajari2017,
   author = {Mazdak Ghajari and Peter J. Hellyer and David J. Sharp},
   doi = {10.1093/brain/aww317},
   issn = {14602156},
   issue = {2},
   journal = {Brain},
   month = {2},
   pages = {333-343},
   pmid = {28043957},
   publisher = {Oxford University Press},
   title = {Computational modelling of traumatic brain injury predicts the location of chronic traumatic encephalopathy pathology},
   volume = {140},
   year = {2017}
}

@article{Takhounts2008,
   author = {Erik G Takhounts and Stephen A Ridella and Vikas Hasija and Rabih E Tannous and J Quinn Campbell and Dan Malone and Kerry Danelson and Joel Stitzel and Steve Rowson and Stefan Duma},
   journal = {Stapp Car Crash Journal},
   doi = {10.4271/2008-22-0001},
   title = {Investigation of Traumatic Brain Injuries Using the Next Generation of Simulated Injury Monitor (SIMon) Finite Element Head Model},
   volume = {52},
   year = {2008}
}

@article{Ganpule2018,
   author = {S. Ganpule and N. P. Daphalapurkar and M. P. Cetingul and K. T. Ramesh},
   doi = {10.1007/s00193-017-0791-z},
   issn = {09381287},
   issue = {1},
   journal = {Shock Waves},
   month = {1},
   pages = {127-139},
   publisher = {Springer New York LLC},
   title = {Effect of bulk modulus on deformation of the brain under rotational accelerations},
   volume = {28},
   year = {2018}
}

@article{Fagan2020,
   author = {Brian T. Fagan and Sikhanda S. Satapathy and J. Neal Rutledge and Steven E. Kornguth},
   doi = {10.3389/fneur.2020.00998},
   issn = {16642295},
   journal = {Frontiers in Neurology},
   month = {9},
   publisher = {Frontiers Media S.A.},
   title = {Simulation of the Strain Amplification in Sulci Due to Blunt Impact to the Head},
   volume = {11},
   year = {2020}
}

@article{Alshareef2021,
   author = {Ahmed Alshareef and Andrew K. Knutsen and Curtis L. Johnson and Aaron Carass and Kshitiz Upadhyay and Philip V. Bayly and Dzung L. Pham and Jerry L. Prince and K. T. Ramesh},
   doi = {10.1016/j.brain.2021.100038},
   issn = {26665220},
   journal = {Brain Multiphysics},
   month = {1},
   publisher = {Elsevier Ltd},
   title = {Integrating material properties from magnetic resonance elastography into subject-specific computational models for the human brain},
   volume = {2},
   year = {2021}
}

@article{Cloots2008,
   author = {R. J.H. Cloots and H. M.T. Gervaise and J. A.W. Van Dommelen and M. G.D. Geers},
   doi = {10.1007/s10439-008-9510-3},
   issn = {00906964},
   issue = {7},
   journal = {Annals of Biomedical Engineering},
   month = {7},
   pages = {1203-1215},
   pmid = {18465248},
   title = {Biomechanics of traumatic brain injury: Influences of the morphologic heterogeneities of the cerebral cortex},
   volume = {36},
   year = {2008}
}

@article{Duckworth2021,
   author = {Harry Duckworth and David J. Sharp and Mazdak Ghajari},
   doi = {10.1002/cnm.3440},
   issn = {20407947},
   issue = {4},
   journal = {International Journal for Numerical Methods in Biomedical Engineering},
   month = {4},
   pmid = {33480161},
   publisher = {John Wiley and Sons Inc},
   title = {Smoothed particle hydrodynamic modelling of the cerebrospinal fluid for brain biomechanics: Accuracy and stability},
   volume = {37},
   year = {2021}
}

@article{Takhounts2013,
   author = {Erik G Takhounts and Matthew J Craig and Kevin Moorhouse and Joe Mcfadden and Vikas Hasija},
   journal = {Stapp Car Crash Journal},
   doi = {10.4271/2013-22-0010},
   pages = {243-266},
   title = {Development of Brain Injury Criteria (BrIC)},
   volume = {57},
   year = {2013}
}

@techReport{Taylor2006,
   author = {Paul A Taylor and Corey C Ford},
   institution = {Computational Shock \& Multiphysics Sandia National Laboratories Albuquerque, NM 87185},
   month = nov,
   title = {Simulation of head impact leading to traumatic brain injury},
   year = {2006}
}

@article{Moerman2018, doi = {10.21105/joss.00506}, url = {https://doi.org/10.21105/joss.00506}, year = {2018}, publisher = {The Open Journal}, volume = {3}, number = {22}, pages = {506}, author = {Moerman, Kevin M.}, title = {GIBBON: The Geometry and Image-Based Bioengineering add-On}, journal = {Journal of Open Source Software} }

@article{Tripathi2026,
   author = {Anu Tripathi and Alison Brooks and Traci Snedden and Peter Ferrazzano and Christian Franck and Rika Wright Carlsen},
   doi = {10.1007/s10237-026-02095-1},
   issn = {16177940},
   issue = {4},
   journal = {Biomechanics and Modeling in Mechanobiology},
   month = {8},
   pmid = {42321586},
   publisher = {Springer Science and Business Media Deutschland GmbH},
   title = {Effect of modeling subject-specific cortical folds on brain injury risk prediction under blunt impact loading},
   volume = {25},
   year = {2026}
}

@article{Zhou2025,
   author = {Zhou Zhou and Xiaogai Li and Svein Kleiven},
   doi = {10.1007/s10237-025-01940-z},
   issn = {16177940},
   issue = {3},
   journal = {Biomechanics and Modeling in Mechanobiology},
   month = {6},
   pages = {845-864},
   pmid = {40067579},
   publisher = {Springer Science and Business Media Deutschland GmbH},
   title = {Surface-based versus voxel-based finite element head models: comparative analyses of strain responses},
   volume = {24},
   year = {2025}
}

@article{Zhang2013,
   author = {Liying Zhang and Rahul Makwana and Sumit Sharma},
   doi = {10.3389/fneur.2013.00088},
   issn = {16642295},
   journal = {Frontiers in Neurology},
   title = {Brain response to primary blast wave using validated finite element models of human head and advanced combat helmet},
   volume = {4},
   year = {2013}
}

@article{Wright2013,
   author = {Rika M. Wright and Andrew Post and Blaine Hoshizaki and Kaliat T. Ramesh},
   doi = {10.1089/neu.2012.2418},
   issn = {08977151},
   issue = {2},
   journal = {Journal of Neurotrauma},
   month = {1},
   pages = {102-118},
   pmid = {22992118},
   title = {A multiscale computational approach to estimating axonal damage under inertial loading of the head},
   volume = {30},
   year = {2013}
}

@article{Mao2013,
   author = {Haojie Mao and Liying Zhang and Binhui Jiang and Vinay V. Genthikatti and Xin Jin and Feng Zhu and Rahul Makwana and Amandeep Gill and Gurdeep Jandir and Amrinder Singh and King H. Yang},
   doi = {10.1115/1.4025101},
   issn = {01480731},
   issue = {11},
   journal = {Journal of Biomechanical Engineering},
   pmid = {24065136},
   title = {Development of a finite element human head model partially validated with thirty five experimental cases},
   volume = {135},
   year = {2013}
}

@article{Ji2015,
   author = {Songbai Ji and Wei Zhao and James C. Ford and Jonathan G. Beckwith and Richard P. Bolander and Richard M. Greenwald and Laura A. Flashman and Keith D. Paulsen and Thomas W. McAllister},
   doi = {10.1089/neu.2013.3268},
   issn = {15579042},
   issue = {7},
   journal = {Journal of Neurotrauma},
   month = {4},
   pages = {441-454},
   pmid = {24735430},
   publisher = {Mary Ann Liebert Inc.},
   title = {Group-wise evaluation and comparison of white matter fiber strain and maximum principal strain in sports-related concussion},
   volume = {32},
   year = {2015}
}

@article{Bain2000,
   author = {Allison C Bain and David F Meaney},
   journal = {Journal of Biomechanical Engineering},
   doi = {10.1115/1.1324667},
   month = {12},
   pages = {615-622},
   title = {Tissue-Level Thresholds for Axonal Damage in an Experimental Model of Central Nervous System White Matter Injury},
   volume = {122},
   url = {http://biomechanical.asmedigitalcollection.asme.org/},
   year = {2000}
}

@article{Gabler2018a,
   author = {Lee F. Gabler and Jeff R. Crandall and Matthew B. Panzer},
   doi = {10.1007/s10439-018-2015-9},
   issn = {15739686},
   issue = {7},
   journal = {Annals of Biomedical Engineering},
   month = {7},
   pages = {972-985},
   pmid = {29594689},
   publisher = {Springer New York LLC},
   title = {Development of a Metric for Predicting Brain Strain Responses Using Head Kinematics},
   volume = {46},
   year = {2018}
}

@article{Zhao2017,
   author = {Wei Zhao and Yunliang Cai and Zhigang Li and Songbai Ji},
   doi = {10.1007/s10237-017-0915-5},
   issn = {16177940},
   issue = {5},
   journal = {Biomechanics and Modeling in Mechanobiology},
   month = {10},
   pages = {1709-1727},
   pmid = {28500358},
   publisher = {Springer Verlag},
   title = {Injury prediction and vulnerability assessment using strain and susceptibility measures of the deep white matter},
   volume = {16},
   year = {2017}
}

@article{Knutsen2014,
   author = {Andrew K. Knutsen and Elizabeth Magrath and Julie E. McEntee and Fangxu Xing and Jerry L. Prince and Philip V. Bayly and John A. Butman and Dzung L. Pham},
   doi = {10.1016/j.jbiomech.2014.09.010},
   issn = {18732380},
   issue = {14},
   journal = {Journal of Biomechanics},
   pages = {3475-3481},
   pmid = {25287113},
   publisher = {Elsevier Ltd},
   title = {Improved measurement of brain deformation during mild head acceleration using a novel tagged MRI sequence},
   volume = {47},
   year = {2014}
}

@article{Laksari2020,
   author = {Kaveh Laksari and Michael Fanton and Lyndia C. Wu and Taylor H. Nguyen and Mehmet Kurt and Chiara Giordano and Eoin Kelly and Eoin O'Keeffe and Eugene Wallace and Colin Doherty and Matthew Campbell and Stephen Tiernan and Gerald Grant and Jesse Ruan and Saeed Barbat and David B. Camarillo},
   doi = {10.1089/neu.2018.6340},
   issn = {15579042},
   issue = {7},
   journal = {Journal of Neurotrauma},
   month = {4},
   pages = {982-993},
   pmid = {31856650},
   publisher = {Mary Ann Liebert Inc.},
   title = {Multi-Directional Dynamic Model for Traumatic Brain Injury Detection},
   volume = {37},
   year = {2020}
}

@article{OKeeffe2020,
   author = {Eoin O'Keeffe and Eoin Kelly and Yuzhe Liu and Chiara Giordano and Eugene Wallace and Mark Hynes and Stephen Tiernan and Aidan Meagher and Chris Greene and Stephanie Hughes and Tom Burke and John Kealy and Niamh Doyle and Alison Hay and Michael Farrell and Gerald A. Grant and Alon Friedman and Ronel Veksler and Michael G. Molloy and James F. Meaney and Niall Pender and David Camarillo and Colin P. Doherty and Matthew Campbell},
   doi = {10.1089/neu.2019.6483},
   issn = {15579042},
   issue = {2},
   journal = {Journal of Neurotrauma},
   month = {1},
   pages = {347-356},
   pmid = {31702476},
   publisher = {Mary Ann Liebert Inc.},
   title = {Dynamic Blood-Brain Barrier Regulation in Mild Traumatic Brain Injury},
   volume = {37},
   year = {2020}
}

@article{Adamec2020,
   author = {Jiri Adamec and Peter Hofer and Stefan Pittner and Fabio Monticelli and Matthias Graw and Jutta Schöpfer},
   doi = {10.1007/s00414-020-02440-8},
   issue = {3},
   journal = {International Journal of Legal Medicine},
   month = {10},
   pages = {853-859},
   title = {Biomechanical assessment of various punching techniques},
   volume = {135},
   url = {https://doi.org/10.1007/s00414-020-02440-8},
   year = {2020}
}

@article{Carlsen2021,
   author = {Rika Wright Carlsen and Alice Lux Fawzi and Yang Wan and Haneesh Kesari and Christian Franck},
   doi = {10.1016/j.brain.2021.100024},
   issn = {26665220},
   journal = {Brain Multiphysics},
   month = {1},
   publisher = {Elsevier Ltd},
   title = {A quantitative relationship between rotational head kinematics and brain tissue strain from a 2-D parametric finite element analysis},
   volume = {2},
   year = {2021}
}

@article{Wu2022,
   author = {Taotao Wu and Fusako Sato and Jacobo Antona-Makoshi and Lee F. Gabler and J. Sebastian Giudice and Ahmed Alshareef and Masayuki Yaguchi and Mitsutoshi Masuda and Susan S. Margulies and Matthew B. Panzer},
   doi = {10.1115/1.4053209},
   issn = {15288951},
   issue = {7},
   journal = {Journal of Biomechanical Engineering},
   month = {7},
   pmid = {34897386},
   publisher = {American Society of Mechanical Engineers (ASME)},
   title = {Integrating Human and Nonhuman Primate Data to Estimate Human Tolerances for Traumatic Brain Injury},
   volume = {144},
   year = {2022}
}

@book{aaam_ais1998,
  title        = {The Abbreviated Injury Scale-1990 Revision, Update 1998},
  author       = {{Association for the Advancement of Automotive Medicine (AAAM)}},
  year         = {1998},
  address      = {Des Plaines, IL},
  publisher    = {Association for the Advancement of Automotive Medicine (AAAM)}
}

@techreport{J211/1_202208,
  author = {{Safety Test Instrumentation Standards Committee}},
  title = {Instrumentation for Impact Test Part 1 - Electronic Instrumentation},
  type        = {SAE Recommended Practice},
  institution = {SAE International},
  number      = {SAE Standard J211/1\_202208},
  month       = aug,
  year        = {2022},
  doi         = {10.4271/J211/1_202208},
}

@techreport{ISO_TR_9790_1999,
  author       = {{The International Organization for Standardization (ISO)}},
  title        = {Road vehicles---anthropomorphic side impact dummy---lateral impact response requirements to assess the biofidelity of the dummy},
  institution  = {International Organization for Standardization},
  type         = {ISO Technical Report},
  number       = {ISO/TR 9790},
  year         = {1999},
  month        = dec
}

@techreport{CDC_2003,
  author       = {{National Center for Injury Prevention and Control (U.S.)}},
  title        = {Report to Congress on mild traumatic brain injury in the United States: steps to prevent a serious public health problem},
  institution  = {Centers for Disease Control and Prevention},
  url = {https://stacks.cdc.gov/view/cdc/6544},
  year         = {2003},
  month        = sep
}

@article{Knutsen2020,
   author = {Andrew K. Knutsen and Arnold D. Gomez and Mihika Gangolli and Wen Tung Wang and Deva Chan and Yuan Chiao Lu and Eftychios Christoforou and Jerry L. Prince and Philip V. Bayly and John A. Butman and Dzung L. Pham},
   doi = {10.1016/j.brain.2020.100015},
   issn = {26665220},
   journal = {Brain Multiphysics},
   month = {11},
   publisher = {Elsevier B.V.},
   title = {In vivo estimates of axonal stretch and 3D brain deformation during mild head impact},
   volume = {1},
   year = {2020}
}

@inproceedings{Gehre2009,
   author = {Christian Gehre and Heinrich Gades and Philipp Wernicke},
   address = {Stuttgart, Germany},
   booktitle = {21st International Technical Conference on the Enhanced Safety of Vehicles},
   month = {6},
   pages = {1-8},
   title = {OBJECTIVE RATING OF SIGNALS USING TEST AND SIMULATION RESPONSES},
   year = {2009}
}

@article{Atsumi2018,
   author = {Noritoshi Atsumi and Yuko Nakahira and Eiichi Tanaka and Masami Iwamoto},
   doi = {10.1007/s10439-018-1988-8},
   issn = {15739686},
   issue = {5},
   journal = {Annals of Biomedical Engineering},
   month = {5},
   pages = {736-748},
   pmid = {29404847},
   publisher = {Springer New York LLC},
   title = {Human Brain Modeling with Its Anatomical Structure and Realistic Material Properties for Brain Injury Prediction},
   volume = {46},
   year = {2018}
}

@article {Giordano2016,
	Title = {Development of an Unbiased Validation Protocol to Assess the Biofidelity of Finite Element Head Models used in Prediction of Traumatic Brain Injury},
	Author = {Giordano, Chiara and Kleiven, Svein},
	DOI = {10.4271/2016-22-0013},
	Volume = {60},
	Month = {November},
	Year = {2016},
	Journal = {Stapp car crash journal},
	ISSN = {1532-8546},
	Pages = {363—471},
	URL = {https://doi.org/10.4271/2016-22-0013},
}

@article{Upadhyay2024,
   author = {Kshitiz Upadhyay and Roshan Jagani and Dimitris G. Giovanis and Ahmed Alshareef and Andrew K. Knutsen and Curtis L. Johnson and Aaron Carass and Philip V. Bayly and Michael D. Shields and K. T. Ramesh},
   doi = {10.1093/milmed/usae199},
   issn = {1930613X},
   issue = {Supplement_3},
   journal = {Military Medicine},
   month = {9},
   pages = {608-617},
   pmid = {38739497},
   publisher = {Oxford University Press},
   title = {Effect of Human Head Shape on the Risk of Traumatic Brain Injury: A Gaussian Process Regression-Based Machine Learning Approach},
   volume = {189},
   year = {2024}
}

@article{Begonia2021,
   author = {Mark T. Begonia and Alexander M. Knapp and R. K. Prabhu and Jun Liao and Lakiesha N. Williams},
   doi = {10.1016/j.jbiomech.2021.110260},
   issn = {18732380},
   journal = {Journal of Biomechanics},
   month = {3},
   pmid = {33515903},
   publisher = {Elsevier Ltd},
   title = {Shear-deformation based continuum-damage constitutive modeling of brain tissue},
   volume = {117},
   year = {2021}
}

@article{Nol2019,
   author = {L. Noël and E. Kuhl},
   doi = {10.1007/s00466-019-01717-z},
   issn = {14320924},
   issue = {5},
   journal = {Computational Mechanics},
   month = {11},
   pages = {1375-1387},
   publisher = {Springer Verlag},
   title = {Modeling neurodegeneration in chronic traumatic encephalopathy using gradient damage models},
   volume = {64},
   year = {2019}
}

@article{Greenwald2008,
   author = {Richard M Greenwald and Joseph T Gwin and Jeffrey J Chu and Joseph J Crisco},
   doi =  {10.1227/01.neu.0000318162.67472.ad},
   journal = {Neurosurgery},
   month = {1},
   pages = {789-798},
   title = {HEAD IMPACT SEVERITY MEASURES FOR EVALUATING MILD TRAUMATIC BRAIN INJURY RISK EXPOSURE},
   volume = {62},
   url = {www.neurosurgery-online.com},
   year = {2008}
}

@article{Margulies1992,
   author = {Susan Sheps Margulies and Lawrence E Thibault},
   isbn = {00219290/92},
   issue = {8},
   doi = {10.1016/0021-9290(92)90231-o},
   journal = {J. Biomechanics},
   pages = {917-923},
   title = {A Proposed Tolerance Criterion for Diffuse Axonal Injury in Man},
   volume = {25},
   year = {1992}
}

@article{Kleiven2007,
   author = {Svein Kleiven},
   journal = {Stapp Car Crash Journal},
   doi = {10.4271/2007-22-0003},
   month = {10},
   title = {Predictors for Traumatic Brain Injuries Evaluated through Accident Reconstructions},
   volume = {51},
   year = {2007}
}

@article{Cloots2011,
   author = {R. J.H. Cloots and J. A.W. Van Dommelen and T. Nyberg and S. Kleiven and M. G.D. Geers},
   doi = {10.1007/s10237-010-0243-5},
   issn = {16822978},
   issue = {3},
   journal = {Biomechanics and Modeling in Mechanobiology},
   pages = {413-422},
   pmid = {20635116},
   publisher = {Springer Verlag},
   title = {Micromechanics of diffuse axonal injury: Influence of axonal orientation and anisotropy},
   volume = {10},
   year = {2011}
}

@article{Giza2014,
   author = {Christopher C. Giza and David A. Hovda},
   doi = {10.1227/NEU.0000000000000505},
   issn = {15244040},
   journal = {Neurosurgery},
   pages = {S24-S33},
   pmid = {25232881},
   publisher = {Lippincott Williams and Wilkins},
   title = {The new neurometabolic cascade of concussion},
   volume = {75},
   year = {2014}
}

@article{Smith2000,
   author = {Douglas H Smith and David F Meaney},
   doi = {https://doi.org/10.1177/107385840000600611},
   issue = {6},
   journal = {The Neuroscientist},
   month = {12},
   pages = {483-495},
   title = {Axonal Damage in Traumatic Brain Injury},
   volume = {6},
   year = {2000}
}

@article{Elkin2007,
   author = {Benjamin S Elkin and Morrison, III, Barclay},
   journal = {Stapp Car Crash Journal},
   doi={10.4271/2007-22-0005},
   month = {10},
   title = {Region-Specific Tolerance Criteria for the Living Brain},
   volume = {51},
   year = {2007}
}

@article{Ji2022,
   author = {Songbai Ji and Mazdak Ghajari and Haojie Mao and Reuben H. Kraft and Marzieh Hajiaghamemar and Matthew B. Panzer and Remy Willinger and Michael D. Gilchrist and Svein Kleiven and Joel D. Stitzel},
   doi = {10.1007/s10439-022-02999-w},
   issn = {15739686},
   issue = {11},
   journal = {Annals of Biomedical Engineering},
   month = {11},
   pages = {1389-1408},
   pmid = {35867314},
   publisher = {Springer},
   title = {Use of Brain Biomechanical Models for Monitoring Impact Exposure in Contact Sports},
   volume = {50},
   year = {2022}
}

@article{Zuidema2023,
   author = {Taylor R. Zuidema and Jeffrey J. Bazarian and Kyle A. Kercher and Rebekah Mannix and Reuben H. Kraft and Sharlene D. Newman and Keisuke Ejima and Devin J. Rettke and Jonathan T. Macy and Jesse A. Steinfeldt and Keisuke Kawata},
   doi = {10.1001/jamanetworkopen.2023.16601},
   issn = {25743805},
   issue = {5},
   journal = {JAMA network open},
   month = {5},
   pages = {e2316601},
   pmid = {37252737},
   publisher = {NLM (Medline)},
   title = {Longitudinal Associations of Clinical and Biochemical Head Injury Biomarkers With Head Impact Exposure in Adolescent Football Players},
   volume = {6},
   year = {2023}
}

@inproceedings{Zhou1995,
   author = {Chun Zhou and Tawfik B Khalil and Albert I King},
   booktitle = {39th Stapp Car Crash Conference},
   address = {San Diego, CA},
   doi = {https://doi.org/10.4271/952714},
   title = {A New Model Comparing Impact Responses of the Homogeneous and Inhomogeneous Human Brain},
   publisher = {SAE},
   year = {1995}
}

@article{Li2016,
   author = {Wenguang Li},
   doi = {10.1007/s40846-016-0132-1},
   issn = {21994757},
   issue = {3},
   journal = {Journal of Medical and Biological Engineering},
   month = {6},
   pages = {285-307},
   publisher = {Springer Berlin Heidelberg},
   title = {Damage Models for Soft Tissues: A Survey},
   volume = {36},
   year = {2016}
}

@article {Gerber346700,
	author = {Gerber, Jesse I. and Garimella, Harsha T. and Kraft, Reuben H.},
	title = {Computation of history-dependent mechanical damage of axonal fiber tracts in the brain: towards tracking sub-concussive and occupational damage to the brain},
	elocation-id = {346700},
	year = {2018},
	doi = {10.1101/346700},
	publisher = {Cold Spring Harbor Laboratory},
	journal = {bioRxiv}
}

@article{Sumelka2017,
   author = {Wojciech Sumelka and George Z. Voyiadjis},
   doi = {10.1016/j.ijsolstr.2017.06.024},
   issn = {00207683},
   journal = {International Journal of Solids and Structures},
   month = {10},
   pages = {151-160},
   publisher = {Elsevier Ltd},
   title = {A hyperelastic fractional damage material model with memory},
   volume = {124},
   year = {2017}
}

@article{Voyiadjis2019,
   author = {George Z. Voyiadjis and Wojciech Sumelka},
   doi = {10.1016/j.jmbbm.2018.09.029},
   issn = {18780180},
   journal = {Journal of the Mechanical Behavior of Biomedical Materials},
   month = {1},
   pages = {209-216},
   pmid = {30292967},
   publisher = {Elsevier Ltd},
   title = {Brain modelling in the framework of anisotropic hyperelasticity with time fractional damage evolution governed by the Caputo-Almeida fractional derivative},
   volume = {89},
   year = {2019}
}

@misc{CDC_TBI_Data_death,
  author       = {{Centers for Disease Control and Prevention (CDC), National Center for Injury Prevention and Control}},
  title        = {Web-based Injury Statistics Query and Reporting System (WISQARS)},
  howpublished = {\url{https://wisqars.cdc.gov/fatal-injury-trends/}},
  note         = {Accessed: 2026-06-01},
  institution  = {Centers for Disease Control and Prevention}
}

@misc{CDC_TBI_Data,
  author       = {{Centers for Disease Control and Prevention (CDC)}},
  title        = {TBI Data},
  howpublished = {\url{https://www.cdc.gov/traumatic-brain-injury/data-research/index.html}},
  note         = {Accessed: 2026-02-24},
  institution  = {Centers for Disease Control and Prevention}
}

@article{Meaney2014,
   author = {David F. Meaney and Barclay Morrison and Cameron Dale Bass},
   doi = {10.1115/1.4026364},
   issn = {01480731},
   issue = {2},
   journal = {Journal of Biomechanical Engineering},
   month = {2},
   pmid = {24384610},
   title = {The mechanics of traumatic brain injury: A review of what we know and what we need to know for reducing its societal burden},
   volume = {136},
   year = {2014}
}

@book{Brasure2012PostacuteTBI,
  author       = {Brasure, M. and Lamberty, G. J. and Sayer, N. A. and others},
  title        = {Multidisciplinary Postacute Rehabilitation for Moderate to Severe Traumatic Brain Injury in Adults},
  series       = {Comparative Effectiveness Reviews},
  edition      = {No. 72},
  publisher    = {Agency for Healthcare Research and Quality (US)},
  address      = {Rockville, MD, USA},
  year         = {2012},
  month        = {June},
  url          = {https://www.ncbi.nlm.nih.gov/books/NBK98993/},
}

@article{McKee2013,
   author = {Ann C. McKee and Thor D. Stein and Christopher J. Nowinski and Robert A. Stern and Daniel H. Daneshvar and Victor E. Alvarez and Hyo Soon Lee and Garth Hall and Sydney M. Wojtowicz and Christine M. Baugh and David O. Riley and Caroline A. Kubilus and Kerry A. Cormier and Matthew A. Jacobs and Brett R. Martin and Carmela R. Abraham and Tsuneya Ikezu and Robert Ross Reichard and Benjamin L. Wolozin and Andrew E. Budson and Lee E. Goldstein and Neil W. Kowall and Robert C. Cantu},
   doi = {10.1093/brain/aws307},
   issn = {14602156},
   issue = {1},
   journal = {Brain},
   pages = {43-64},
   pmid = {23208308},
   publisher = {Oxford University Press},
   title = {The spectrum of disease in chronic traumatic encephalopathy},
   volume = {136},
   year = {2013}
}

@article{Longhi2005,
   author = {Luca Longhi and Kathryn E. Saatman and Scott Fujimoto and Ramesh Raghupathi and David F. Meaney and Jason Davis and Asenia McMillan and Valeria Conte and Helmut L. Laurer and Sherman Stein and Nino Stocchetti and Tracy K. McIntosh},
   doi = {10.1227/01.NEU.0000149008.73513.44},
   issn = {0148396X},
   issue = {2},
   journal = {Neurosurgery},
   month = {2},
   pages = {364-373},
   pmid = {15670384},
   title = {Temporal window of vulnerability to repetitive experimental concussive brain injury},
   volume = {56},
   year = {2005}
}

@article{Meaney2011,
   author = {David F. Meaney and Douglas H. Smith},
   doi = {10.1016/j.csm.2010.08.009},
   issn = {02785919},
   issue = {1},
   journal = {Clinics in Sports Medicine},
   month = {1},
   pages = {19-31},
   pmid = {21074079},
   title = {Biomechanics of Concussion},
   volume = {30},
   year = {2011}
}

@article{Nakarmi2025,
   author = {Sushan Nakarmi and Yaohui Wang and Alice Lux Fawzi and Christian Franck and Rika Wright Carlsen},
   doi = {10.1093/milmed/usae309},
   issn = {1930613X},
   issue = {1-2},
   journal = {Military Medicine},
   month = {1},
   pages = {e202-e210},
   pmid = {38877897},
   publisher = {Oxford University Press},
   title = {Estimating Brain Injury Risk from Shipborne Underwater Blasts Using a High-fidelity Finite Element Head Model},
   volume = {190},
   year = {2025}
}

@article{Zhan2022,
   author = {Xianghao Zhan and Anna Oeur and Yuzhe Liu and Michael M Zeineh and Gerald A Grant and Susan S Margulies and David B Camarillo},
   doi = {https://doi.org/10.1016/j.cobme.2022.100422},
   journal = {Current Opinion in Biomedical Engineering},
   month = {12},
   title = {Translational models of mild traumatic brain injury tissue biomechanics},
   volume = {24},
   year = {2022}
}

@article{Hajiaghamemar2021,
   author = {Marzieh Hajiaghamemar and Susan S. Margulies},
   doi = {10.1089/neu.2019.6791},
   issn = {15579042},
   issue = {1},
   journal = {Journal of Neurotrauma},
   month = {1},
   pages = {144-157},
   pmid = {32772838},
   publisher = {Mary Ann Liebert Inc.},
   title = {Multi-Scale White Matter Tract Embedded Brain Finite Element Model Predicts the Location of Traumatic Diffuse Axonal Injury},
   volume = {38},
   year = {2021}
}

@inproceedings{Gennarelli1987,
   author = {Thomas A Gennarelli and Lawrence E Thibault and G Tomei and R Wiser and D Graham and J Adams},
   address = {New Orleans, LA},
   doi = {https://doi.org/10.4271/872197},
   issn = {0148-7191},
   booktitle = {31st Stapp Car Crash Conference},
   month = {10},
   publisher = {SAE},
   title = {Directional Dependence of Axonal Brain Injury due to Centroidal and Non-Centroidal Acceleration},
   year = {1987}
}

@article{Bayly2021,
   author = {Philip V. Bayly and Ahmed Alshareef and Andrew K. Knutsen and Kshitiz Upadhyay and Ruth J. Okamoto and Aaron Carass and John A. Butman and Dzung L. Pham and Jerry L. Prince and K. T. Ramesh and Curtis L. Johnson},
   doi = {10.1007/s10439-021-02820-0},
   issn = {15739686},
   issue = {10},
   journal = {Annals of Biomedical Engineering},
   month = {10},
   pages = {2677-2692},
   pmid = {34212235},
   publisher = {Springer},
   title = {MR Imaging of Human Brain Mechanics In Vivo: New Measurements to Facilitate the Development of Computational Models of Brain Injury},
   volume = {49},
   year = {2021}
}

@article{Chandrashekar2026,
author = {Chandrashekar, Yogesh C. and Upadhyay, Kshitiz},
doi = {10.1016/j.jmps.2026.106700},
issn = {00225096},
journal = {Journal of the Mechanics and Physics of Solids},
month = {sep},
pages = {106700},
publisher = {Elsevier Ltd},
title = {{A constitutive framework for distortional-mode-dependent failure in soft materials: Tension–compression asymmetry and beyond}},
volume = {215},
year = {2026}
}

@book{blitzstein2019,
  title={Introduction to Probability, Second Edition},
  author={Blitzstein, J.K. and Hwang, J.},
  address={Boca Raton, FL},
  isbn={9780429428357},
  lccn={2020693004},
  series={Chapman and Hall/CRC Texts in Statistical Science Series},
  year={2019},
  publisher={CRC Press}
}

@article{Simo1987,
   author = {J C Simo},
   journal = {Computer Methods in Applied Mechanics and Engineering},
   doi = {10.1016/0045-7825(87)90107-1},
   pages = {153-173},
   title = {ON A FULLY THREE-DIMENSIONAL FINITE-STRAIN VISCOELASTIC DAMAGE MODEL: FORMULATION AND COMPUTATIONAL ASPECTS},
   volume = {60},
   year = {1987}
}

\end{document}


\renewcommand{\thesection}{S\arabic{section}}
\renewcommand{\thesubsection}{S\arabic{section}.\arabic{subsection}}

\renewcommand{\thefigure}{S\arabic{figure}}
\renewcommand{\thetable}{S\arabic{table}}
\renewcommand{\theequation}{S\arabic{equation}}

\setcounter{section}{0}
\setcounter{figure}{0}
\setcounter{table}{0}
\setcounter{equation}{0}

\title[Supplementary Material for History Matters: Damage-Mediated Amplification of Brain Deformation and Injury Risk under Repeated Head Impacts
]{\begin{center} \textit{Supplementary Material} \end{center} History Matters: Damage-Mediated Amplification of Brain Deformation and Injury Risk under Repeated Head Impacts
}


\author[1]{\fnm{Carson} \sur{Cooper}}\email{ccooper7003@gmail.com}
\equalcont{Current email: ccoope@umich.edu}

\author[1]{\fnm{Genevieve} \sur{Palardy}}\email{gpalardy@lsu.edu}

\author[2]{\fnm{Anu} \sur{Tripathi}}\email{tripathia@rmu.edu}

\author*[3]{\fnm{Kshitiz} \sur{Upadhyay}}\email{kshitizu@umn.edu}

\affil[1]{\orgdiv{Department of Mechanical and Industrial Engineering}, \orgname{Louisiana State University}, \orgaddress{\city{Baton Rouge}, \state{LA}, \country{USA}}}

\affil[2]{\orgdiv{Department of Engineering}, \orgname{Robert Morris University}, \orgaddress{\city{Moon Township}, \state{PA}, \country{USA}}}

\affil*[3]{\orgdiv{Department of Aerospace Engineering and Mechanics}, \orgname{University of Minnesota}, \orgaddress{\city{Minneapolis}, \state{MN}, \country{USA}}}

\maketitle

\section{FE-Based Head Model Validation}\label{chap:valmethods}
\subsection{Validation Methods}
\begin{figure}[hbt!]
    \centering
    \includegraphics[width=2.4645in]{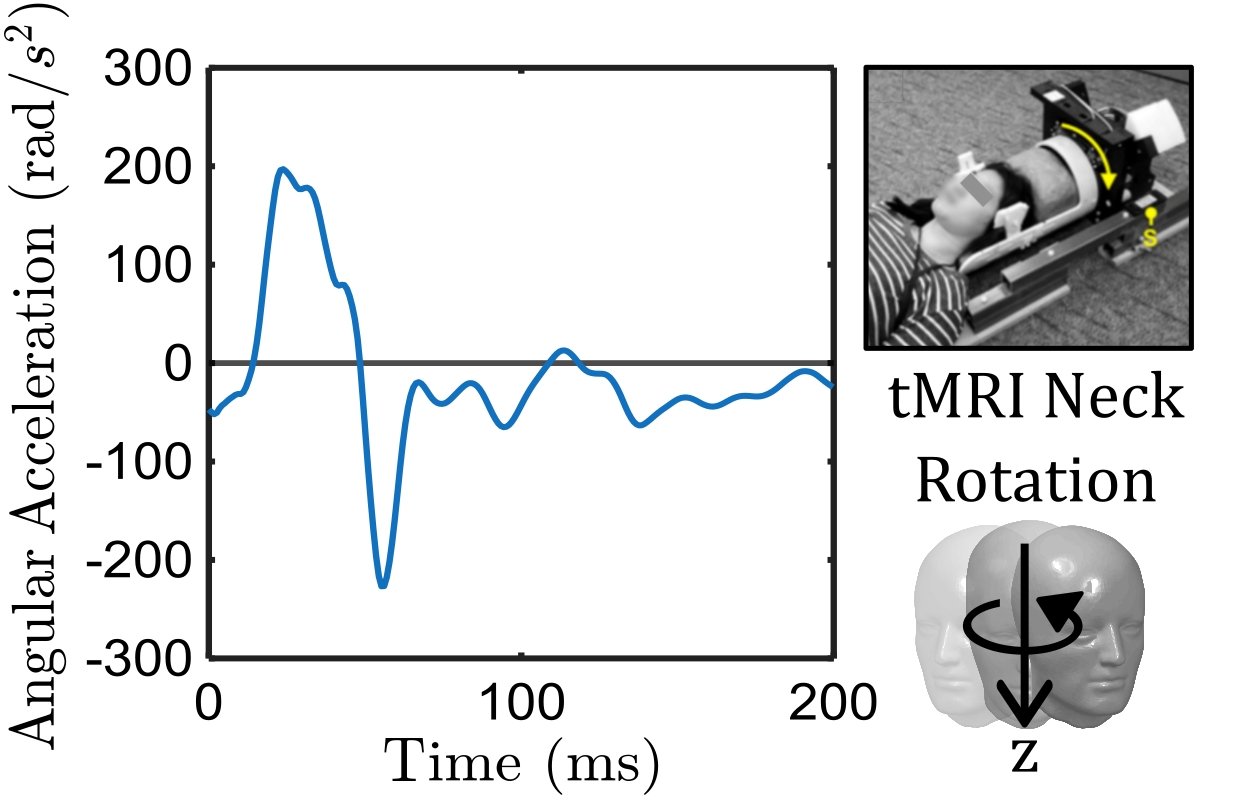}
    \caption{Kinematic history for validation load case. Modified from \cite{Upadhyay2022}}
    \label{fig:validation_loading2}
\end{figure}

The Mullins damage-based head model is validated against subject-specific experimental tagged MRI (tMRI) strain data from the Neuroimaging Tools and Resources Collaboratory (NITRC) Brain Biomechanics Imaging Resources (BBIR) database \citep{Knutsen2020, Bayly2021}.
In this test case, the head is loaded in rotation about the superior--inferior axis before impacting a stopper, which causes a mild deceleration (Fig.~\ref{fig:validation_loading2}).
In tagged MRI, tag lines are imposed over the brain \citep{Knutsen2014}, which deform with the brain tissue. 
Strain is estimated from the displacement of the tag lines from a sequence of images, providing a full-field spatiotemporal strain measurement.

The experimental loading conditions are applied to the present computational model, and the simulation's predicted strains are compared to those obtained from tMRI.
Instantaneous peak (95th percentile) and average (75th percentile) values of the strain components (e.g., $\varepsilon_{xx}, \varepsilon_{yy}, \varepsilon_{xy}$) are computed according to Eqs.~(\ref{eq:comp95}) and (\ref{eq:comp75}).
\begin{equation}
\label{eq:comp95}
    \varepsilon_{ij}\mathrm{95}(t)
    =
    P_{95}
    \left(
    \left\{
    \varepsilon_{ij,e}(t)
    \right\}_{e\in\mathcal{B}}
    \right)
\end{equation}
\begin{equation}
\label{eq:comp75}
    \varepsilon_{ij}\mathrm{75}(t)
    =
    P_{75}
    \left(
    \left\{
    \varepsilon_{ij,e}(t)
    \right\}_{e\in\mathcal{B}}
    \right)
\end{equation}
Here, $\varepsilon_{ij}$ is a component of the logarithmic strain tensor $\boldsymbol{\varepsilon}^{L}$, with indices \(i = x,y,z\), and \(j=x,y,z\).
Additionally, \(P_{95}(\cdot)\) denotes the 95th percentile, \(P_{75}(\cdot)\) is the 75th percentile, \(e\) denotes a brain-bulk element (i.e., within the seven parenchymal segmented regions), and \(\mathcal{B}\) is the set of all brain-bulk elements.

The 95th percentile is chosen over the 100th percentile value to avoid numerical outliers in the data set \citep{Gabler2018a}. The symmetry in positively and negatively straining regions in the brain under this loading causes the 50th percentile values to be approximately equal to zero. Instead, the 75th percentile is taken as the average strain \citep{Upadhyay2022}. Strain components adhere to the SAE J211 coordinate system \citep{J211/1_202208}.

The maximum principal strain, or the first eigenvalue $(\lambda_{\mathrm{max}})$ of the strain tensor $(\boldsymbol{\varepsilon}^{L})$, has been shown to be closely tied to tissue injury \citep{Bain2000}. 
The 95th percentile value of the maximum principal strain (MPS95), defined in Eq.~(15) of the main manuscript, is often used to characterize the global extent of deformation within the brain and predict risk of injury~\citep{Takhounts2008, Zhao2017, Gabler2018a}. 
Therefore, accurate prediction of MPS95 is considered an important indicator of the model's suitability for strain-based injury prediction.
Since maximum principal strain is a strictly positive quantity (for an incompressible material), the average MPS value is taken at 50th percentile. 
Time histories of 95th and 75th percentile strain components, as well as MPS95 and MPS50, are compared.
Additionally, to spatially validate the model, strained volume fractions ($\mathrm{SVF}$) are computed based on voxels whose deformation exceeds a given threshold \citep{Wright2013},
\begin{equation}
    \mathrm{SVF}_{ij}(t)
    =
    \frac{
    \sum\limits_{e\in\mathcal{B}}
    V_e\,
    H
    \left(
    \varepsilon_{ij,e}-
    \varepsilon_{\mathrm{thresh}}
    \right)
    }{
    \sum\limits_{e\in\mathcal{B}} V_e
    },
\end{equation}
where \(V_e\) is the volume of element \(e\), \(H(\cdot)\) is the Heaviside step function, and \(\varepsilon_{\mathrm{thresh}}\) is the selected strain threshold. 

The correlation and analysis objective rating (CORA) \citep{Gehre2009} is used to quantify the degree of agreement between simulation and experiment.
CORA is widely used in validation of computational head models~\citep{Upadhyay2022,Wright2013,Duckworth2021,Tripathi2026,Atsumi2018}. CORA outputs a score between 0 and 1, and a rating based on the scale defined in Technical Report ISO/TR 9790 \citep{ISO_TR_9790_1999} as: ``Excellent": 0.86--1.0, ``Good": 0.65--0.86, ``Fair": 0.44--0.65, ``Marginal": 0.26--0.44, and ``Unacceptable": 0.0--0.26 \citep{Atsumi2018}. CORA scores are calculated for both percentile strains and strained volume fractions. 
Input parameters for CORA are taken from the recommendations in \cite{Giordano2016} for computational head model validation.

\subsection{Validation Results}\label{chap:valresults}
\begin{figure}[hbt!]
    \centering    
    \includegraphics[width=4.993in]{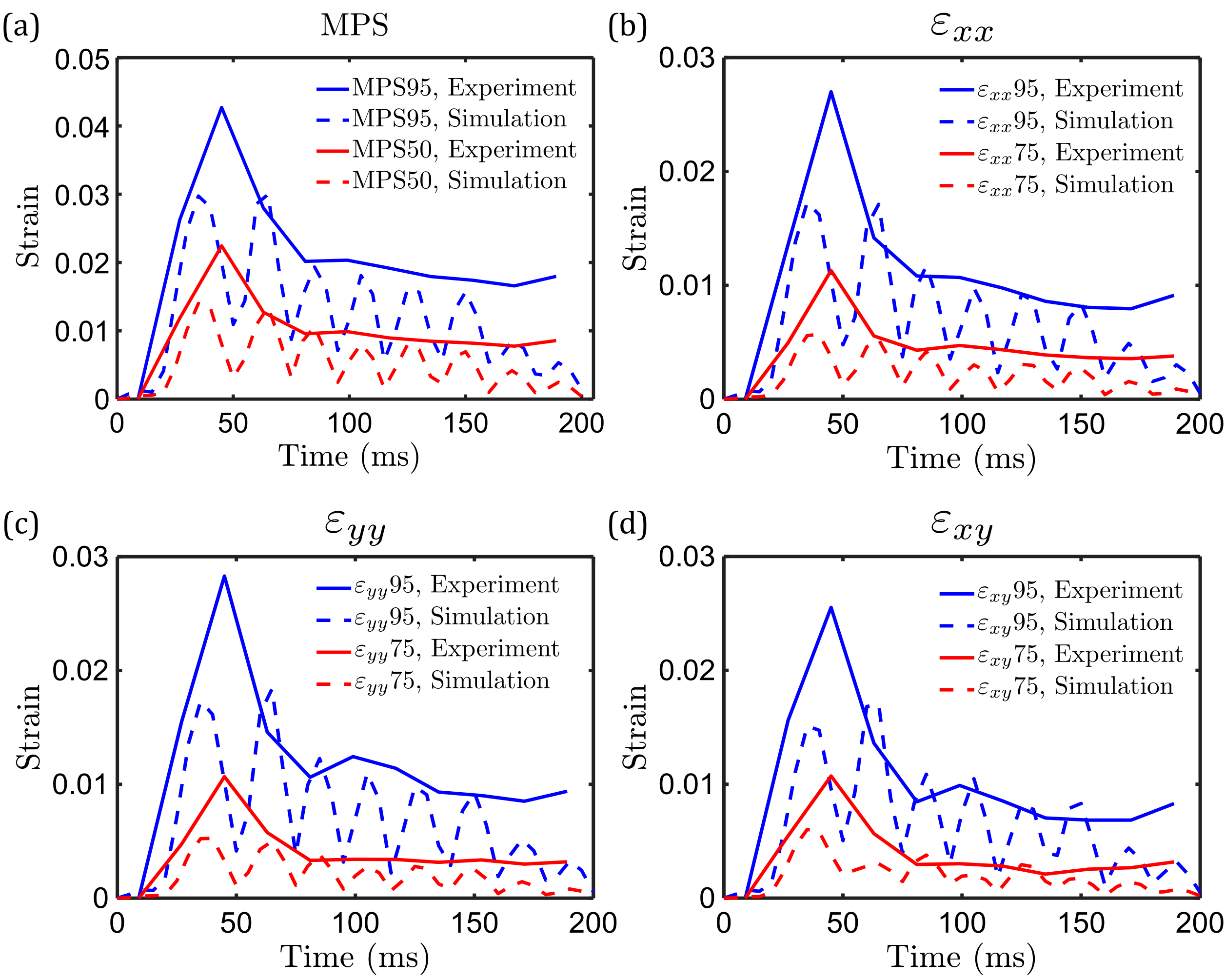}
    \caption{Comparison of experimental and simulated scalar percentile strain histories in validation load case for (a) maximum principal strain (MPS), (b) $\varepsilon_{xx}$, (c) $\varepsilon_{yy}$, (d) $\varepsilon_{xy}$. Mullins-based head model is used for simulations. Experimental values are taken from \cite{Knutsen2020}}
    \label{fig:valpercentiles}
\end{figure}

\begin{figure}[hbt!]
    \centering    
    \includegraphics[width=4.993in]{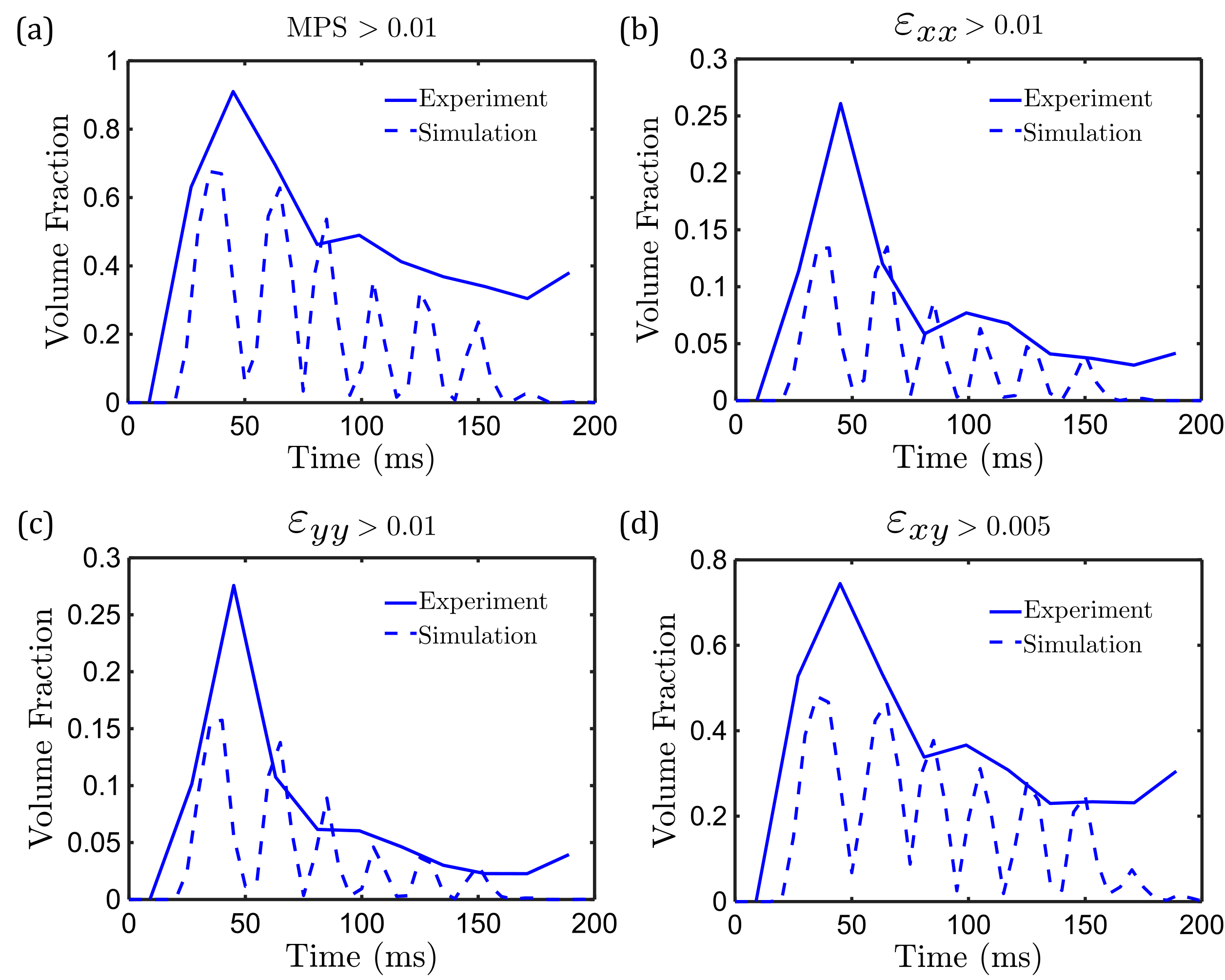}
    \caption{Comparison of experimental and simulated strained volume fraction ($\mathrm{SVF}$) histories in validation load case for (a) maximum principal strain (MPS), (b) $\varepsilon_{xx}$, (c) $\varepsilon_{yy}$, (d) $\varepsilon_{xy}$. Strain threshold for MPS, $\varepsilon_{xx}$, and $\varepsilon_{yy}$ is 0.01; strain threshold for $\varepsilon_{xy}$ is 0.005. Mullins-based head model is used for simulations. Experimental values are taken from \cite{Knutsen2020}}
    \label{fig:valvolfrac}
\end{figure}

The Mullins-damage based head model was validated against experimental tagged MRI strain data. 
Time histories of scalar percentile strains are plotted in Fig.~\ref{fig:valpercentiles}, and time histories of strained volume fractions are presented in Fig.~\ref{fig:valvolfrac}. 
Model CORA scores and ratings are given in Table~\ref{tab:CORA}. Comparison is restricted to 0--117 milliseconds, as the fading of tag lines causes erroneous strain measurement later in the loading, where, the experiment suggests permanent strains.

As shown in Fig.~\ref{fig:valpercentiles} and Fig.~\ref{fig:valvolfrac}, the Mullins-based model underpredicts strains and strained volume fractions for MPS and the in-plane strain components.
In tMRI, the strain history is built by taking successive images in many repeat loadings, so it is likely that the experimental response is that of a preconditioned, slightly softened brain. Therefore, the first-cycle simulation response may be expected to underpredict strains.
Volume fraction strain thresholds were chosen to be consistent within the two deformation modes (normal and shear) and as middle-ground values clearly exceeded in the percentile history plots. 
The model was found to better predict volume fractions for shear strains as opposed to normal strains, which is important as brain tissue is known to be more vulnerable to injury under shear deformation \citep{Ganpule2018, Begonia2021}. 

Oscillations also appear in the simulation output, which are absent in the experimental data. 
Two factors may contribute to this difference: (1) the Mullins model lacks viscoelasticity, which dissipates energy (i.e., dampens oscillations) in brain tissue, and (2) the low temporal resolution of 17 milliseconds in the experimental data, which may alias high-frequency oscillations. 
Strain oscillations under this load case are also observed in validation efforts for a number of computational models \citep{Upadhyay2022,Alshareef2021,Tripathi2026}.

The model received ``Marginal" to ``Good" CORA scores overall, with MPS95 predicted well (score of 0.724---``Good"). 
In general, the model better predicted peak (95th percentile) values than average (75th percentile) values, with in-plane shear ($\varepsilon_{xy}$) being the sole exception. 
The prediction of MPS volume fractions and in-plane normal strain volume fractions received ``Marginal" ratings, suggesting the model's prediction of CSDM may be subject to relatively high error. Therefore, more confidence may be placed in the model's MPS95-based injury predictions than CSDM-based ones.  

\begin{figure}[hbt!]
    \centering    
    \includegraphics[width=3.644in]{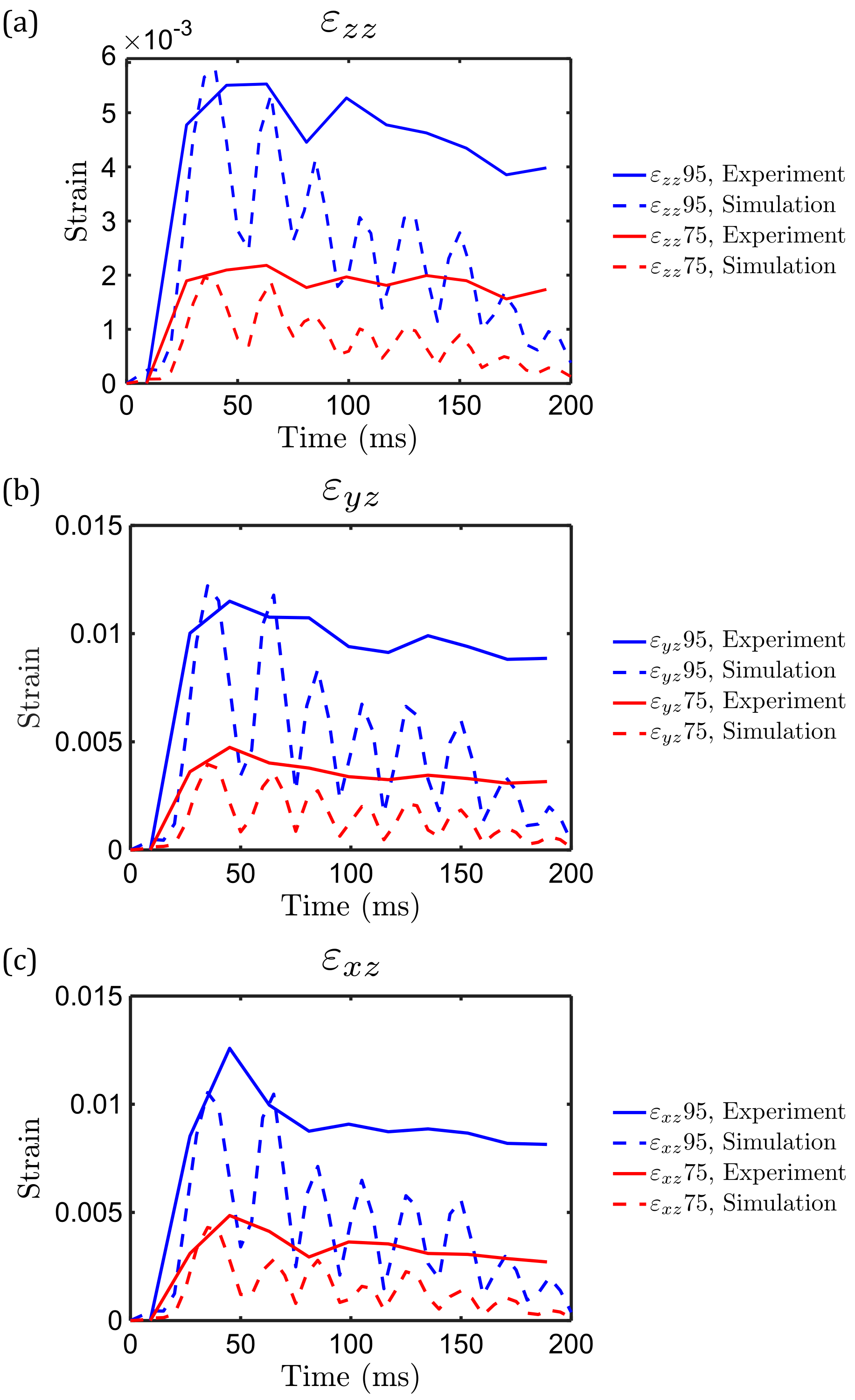}
    \caption{Comparison of experimental and simulated scalar percentile strain histories in validation load case for (a) $\varepsilon_{zz}$, (b) $\varepsilon_{yz}$, (c) $\varepsilon_{xz}$. Mullins-based head model is used for simulations. Experimental values are taken from \cite{Knutsen2020}}
    \label{fig:appPERC}
\end{figure}

\begin{figure}[hbt!]
    \centering    
    \includegraphics[width=4.889in]{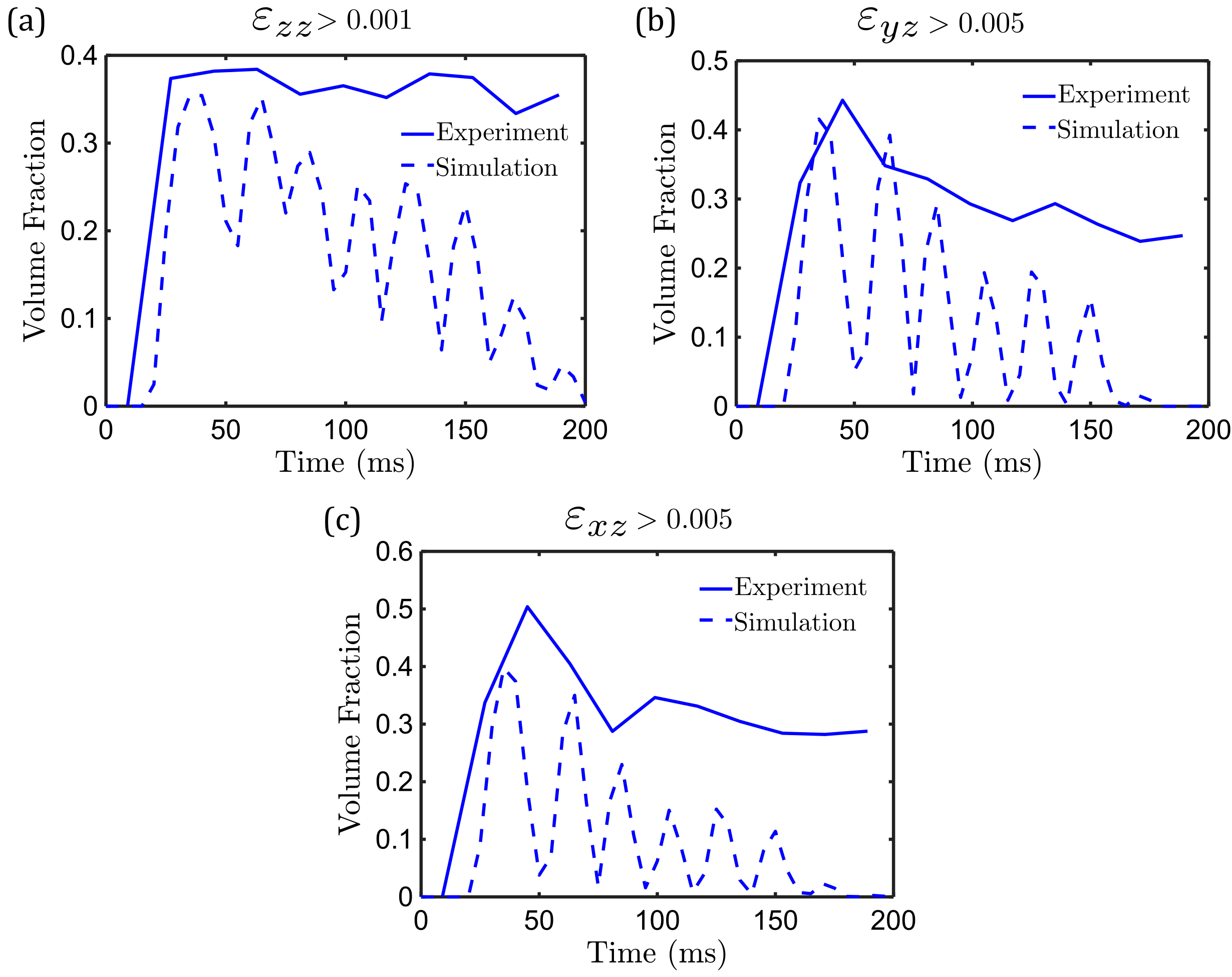}
    \caption{Comparison of experimental and simulated strained volume fraction histories in validation load case for (a) $\varepsilon_{zz}$, (b) $\varepsilon_{yz}$, (c) $\varepsilon_{xz}$. Strain threshold for $\varepsilon_{zz}$ is 0.001; strain threshold for $\varepsilon_{yz}$ and $\varepsilon_{xz}$ is 0.005. Mullins-based head model is used for simulations. Experimental values are taken from \cite{Knutsen2020}}
    \label{fig:appVOL}
\end{figure}

Comparisons of the percentile strain histories and strained volume fraction histories for the out-of-plane strain components are shown in Fig.~\ref{fig:appPERC} and Fig.~\ref{fig:appVOL}. The peak out-of-plane strains seen in Fig.~\ref{fig:appPERC} are approximately two orders of magnitude smaller than those observed in the in-plane strain components.
These results are shown for completeness, as the validation focuses primarily on larger deformations characteristic of mild head impacts.

\begin{table}[t]
\centering
\caption{CORA scores for temporal validation of the Mullins-based head model}
\label{tab:CORA}
\ \\
\begin{tabular}{ccccc}

    \multicolumn{5}{c}{\textbf{CORA Scores for Percentile Strains}} \\
    \toprule
    Strain & Peak Strain (95th) & Rating & Average Strain (50th) & Rating\\
    MPS & 0.724 & Good & 0.349 & Marginal \\
    \midrule
    Strain& Peak Strain (95th)&Rating & Average Strain (75th)&Rating \\
    $\varepsilon_{xx}$ & 0.363 & Marginal & 0.299 & Marginal\\
    $\varepsilon_{yy}$ & 0.362 & Marginal & 0.323 & Marginal \\
    $\varepsilon_{zz}$ & 0.743 & Good & 0.693 & Good\\
    $\varepsilon_{xy}$ & 0.401 & Marginal & 0.649 & Fair \\
    $\varepsilon_{yz}$ & 0.732 & Good & 0.669 & Good\\
    $\varepsilon_{xz}$ & 0.705 & Good & 0.667 & Good\\
    \midrule
    \multicolumn{5}{c}{\textbf{CORA Scores for 
    Volume Fractions}} \\
    \midrule
    Strain & Score & Rating \\
    MPS $>0.01$ & 0.283 & Marginal \\
    $\varepsilon_{xx}>0.01$& 0.315 & Marginal \\
    $\varepsilon_{yy}>0.01$& 0.307 & Marginal \\
    $\varepsilon_{zz}>0.001$& 0.752 & Good \\
    $\varepsilon_{xy}>0.005$& 0.647 & Fair \\
    $\varepsilon_{yz}>0.005$& 0.637 & Fair\\
    $\varepsilon_{xz}>0.005$& 0.565 & Fair\\
    \bottomrule
\end{tabular}
\end{table}

\clearpage
\section{Constitutive Descriptions for Hyperelastic (HE) and Linear Visco-hyperelastic (LVHE) models}
As described in Sec. 2.1 of the main manuscript, a decoupled representation of the strain energy density function is utilized, in which the strain energy density $W$ in the material is decomposed additively into isochoric and volumetric components \citep{Holzapfel2000}. This yields an additive split for the Cauchy stress,
\begin{equation}
\label{eq:addsplit}
    \boldsymbol{\sigma}=
    \boldsymbol{\sigma}_{\mathrm{dev}} +
    \boldsymbol{\sigma}_{\mathrm{vol}} \ ,
\end{equation}
where $\boldsymbol{\sigma}_{\mathrm{dev}}$ represents the stress due to isochoric/distortional deformation and $\boldsymbol{\sigma}_{\mathrm{vol}}$ is the stress due to volumetric deformation. For a purely hyperelastic solid, these stresses describe the instantaneous elastic response with no rate dependence. For a viscoelastic solid, the same instantaneous hyperelastic response is supplemented by a history-dependent contribution accounting for time-dependent isochoric stress relaxation \citep{Simo1987,Holzapfel2000},
\begin{equation}
\label{eq:addsplitVISC}
    \boldsymbol{\sigma}=
    \boldsymbol{\sigma}_{0,\mathrm{dev}} +
    \boldsymbol{\sigma}_{0,\mathrm{vol}} + 
    \boldsymbol{\sigma}_{\mathrm{dev},v} \ 
\end{equation}
Here, \(\boldsymbol{\sigma}_{0,\mathrm{dev}}\) and \(\boldsymbol{\sigma}_{0,\mathrm{vol}}\) denote the instantaneous deviatoric and volumetric hyperelastic responses, respectively, and \(\boldsymbol{\sigma}_{\mathrm{dev},v}\) represents the isochoric viscoelastic history.

\paragraph{Hyperelastic (HE) Model}
Following Eq.~(\ref{eq:addsplit}), the Cauchy stress formulation for the neo-Hookean hyperelastic solid without damage is given as 
\begin{equation}
\label{eq:neohooke}
    \boldsymbol{\sigma}=
    \underbrace{\left(\frac{2}{J}\operatorname{dev}(C_{10}\mathbf{B}^{*})\right)}
    _{\boldsymbol{\sigma}_{\mathrm{dev}}}
    +\underbrace{\frac{2}{D_{1}}(J-1)\mathbf{I}}_{\boldsymbol{\sigma}_{\mathrm{vol}}} \ ,
\end{equation}
where $\mathbf{B}^*$ is the corrected left Cauchy-Green deformation tensor based on the deviatoric part of the deformation gradient, \(\mathbf{F}^{*}\);
\begin{equation}
    J=\det{\mathbf{F}},
\end{equation}
\begin{equation}
    \mathbf{F}^{*}=J^{-1/3}\mathbf{F}, \
\end{equation}
and
\begin{equation}
    \mathbf{B}^*=\mathbf{F}^{*}\cdot\mathbf{\left(F^{*}\right)}^{\mathrm{T}}.
\end{equation}
Additionally, $\mathbf{I}$ is the second-order identity tensor and the deviatoric operator is defined as \(\mathrm{dev}(\cdot)=(\cdot)-\frac{1}{3}\operatorname{tr}(\cdot)\mathbf{I}\), with \(\operatorname{tr}(\cdot)\) as the trace of the tensor.
The constants $C_{10}=\frac{\mu}{2}$ and $D_{1}=\frac{2}{K}$ are material parameters related to the shear modulus ($\mu$) and bulk modulus ($K$), respectively.

\paragraph{Linear Visco-hyperelastic (LVHE) Model}
Recall the definition of the Prony-series based time-dependent shear modulus from Eq. (10) of the main manuscript, reproduced here for convenience,
\begin{equation}
\label{eq:visco}
\mu(t)=\mu_{0}-\mu_{0}\sum^{n}_{i=1}g_{i}\left(1-\exp{\left(-t/\tau_{i}\right)}\right) \ .
\end{equation}
Here, \(\mu_0\) is the instantaneous shear modulus, \(g_i\) is the dimensionless relaxation coefficient associated with the \(i\)th Prony-series term, and \(\tau_i\) is the corresponding relaxation time. From Eq.~\eqref{eq:visco}, the normalized relaxation function is defined as
\begin{equation}
\label{eq:relax}
r(t)=\frac{\mu(t)}{\mu_0}=
1 - \sum^{n}_{i=1}g_{i}\left(1-
\exp{\left(-t/\tau_{i}\right)}\right) \ ,
\end{equation}
with first time derivative
\begin{equation}
\label{eq:relaxdot}
\dot{r}(t)=
-\sum^{n}_{i=1}\left(\frac{g_{i}}{\tau_{i}}
\exp{\left(-t/\tau_{i}\right)}\right) \ .
\end{equation}

The deviatoric viscous stress contribution (with viscous denoted by $v$) is given as
\begin{equation}
    \boldsymbol{\tau}_{\mathrm{dev},v} =
    \operatorname{dev}\left[\int^{t}_{0}
    \dot{r}(s)
    \mathcal{P}_{t-s\to t}[\boldsymbol{\tau}_{0}(t-s)]
    ds\right] \ ,
\end{equation}
where $\boldsymbol{\tau}$ is the Kirchhoff stress, $\boldsymbol{\tau} = J\boldsymbol{\sigma}$ \citep{abaqus2004}.
Here, $\mathcal{P}_{t-s\to t}[\cdot]$ is a push-forward operation that transforms the instantaneous deviatoric Kirchhoff stress $\boldsymbol{\tau}_{0}(t-s)$ from the spatial configuration at time $t-s$ to the spatial configuration at time $t$, defined as
\begin{equation}
    \mathcal{P}_{t-s\to t}[\cdot]=
    \mathbf{F}^{*-1}_{t}(t-s)\cdot[\cdot]\cdot
    \mathbf{F}^{*-T}_{t}(t-s) \ ,
\end{equation}
where $\mathbf{F}^{*}_{t}(t-s)$ is the corrected relative deformation gradient, 
\begin{equation}
    \mathbf{F}^{*}_{t}(t-s) =
    \left(\frac{J(t-s)}{J(t)}
    \right)^{-1/3}
    \frac{\partial\mathbf{x}(t-s)}
    {\partial\mathbf{x} (t)}=
    \left(\frac{J(t-s)}{J(t)}
    \right)^{-1/3}
    \mathbf{F}(t-s)
    \mathbf{F}^{-1}(t) \ .
\end{equation}

This yields the form
\begin{equation}
\label{eq:viscstress}
    \boldsymbol{\tau}_{\mathrm{dev},v} =
    \operatorname{dev}\left[
    \int^{t}_{0}
    \left(-\sum^{n}_{i=1}\frac{g_{i}}{\tau_{i}}\exp(-s/\tau_{i})\right)
    \mathbf{F}^{*-1}_{t}(t-s)\cdot
    [\boldsymbol{\tau}_{0}(t-s)]
    \cdot
    \mathbf{F}^{*-T}_{t}(t-s) 
    ds
    \right] \ ,
\end{equation}
where the instantaneous deviatoric Kirchhoff stress is obtained from the neo-Hookean formulation as
\begin{equation}
    \boldsymbol{\tau}_{0}(t-s) = 
    J(t-s)\boldsymbol{\sigma}_{\mathrm{dev}}(t-s) =
    2\operatorname{dev}(C_{10}\mathbf{B}^{*}(t-s)) \ .
\end{equation}

The $\boldsymbol{\sigma}_{\mathrm{dev},v}$ term in Eq.~(\ref{eq:addsplitVISC}) is obtained by dividing Eq.~(\ref{eq:viscstress}) by $J$. 
\begin{equation}
\label{eq:viscstresssigma}
    \boldsymbol{\sigma}_{\mathrm{dev},v} =
    \frac{1}{J}
    \operatorname{dev}\left[
    \int^{t}_{0}
    \left(-\sum^{n}_{i=1}\frac{g_{i}}{\tau_{i}}\exp(-s/\tau_{i})\right)
    \mathbf{F}^{*-1}_{t}(t-s)\cdot
    [\boldsymbol{\tau}_{0}(t-s)]
    \cdot
    \mathbf{F}^{*-T}_{t}(t-s) 
    ds
    \right] \ .
\end{equation}
The instantaneous terms $\boldsymbol{\sigma}_{0,\mathrm{dev}}$ and $\boldsymbol{\sigma}_{0,\mathrm{vol}}$ in Eq.~(\ref{eq:addsplitVISC}) are taken to have the same respective forms as the deviatoric and volumetric terms in Eq.~(\ref{eq:neohooke}).

\clearpage
\backmatter

\bibliography{cas-refs}